\documentclass[12pt]{article}
\usepackage{amssym}
\usepackage{graphics}

\renewcommand{\theequation}{\arabic{section}.\arabic{equation}}

\def\bp{b^{\dagger}}
\def\ap{a^{\dagger}}
\def\Z{\mbox{$\Bbb Z$}}
\def\C{\mbox{$\Bbb C$}}
\def\N{\mbox{$\Bbb N$}}
\def\case#1#2{{\textstyle{#1\over #2}}}
\def\bbeta{\bar{\beta}}
\def\bn{\mathbf{n}}
\def\ss{\scriptstyle}

\title{
\hfill{\normalsize ULB/229/CQ/01/1}\\
\vspace{1cm}
Generalized Coherent States Associated with the $C_{\lambda}$-Extended Oscillator}
\author{C. Quesne\\ 
{\sl \small Physique Nucl\'eaire Th\'eorique et Physique
Math\'ematique,  Universit\'e Libre de Bruxelles,} \\ 
{\sl \small Campus de la Plaine CP229, Boulevard~du Triomphe, B-1050 Brussels,
Belgium}\\
{\small E-mail: cquesne@ulb.ac.be}}
\date{ }
\begin{document}
\baselineskip=20pt plus 1pt minus 1pt
%%%%%%%%%%%%%%%%%%%%%%%%%%%%%%%%%%%%%%%%%%%%%%%%%%%%%%%%%%
\maketitle

\begin{abstract}
Two new types of coherent states associated with the $C_{\lambda}$-extended
oscillator, where $C_{\lambda}$ is the cyclic group of order $\lambda$, are introduced.
The first ones include as special cases both the Barut-Girardello and the Perelomov
su(1,1) coherent states for $\lambda=2$, as well as the annihilation-operator coherent
states of the $C_{\lambda}$-extended oscillator spectrum generating algebra for higher
$\lambda$ values. The second ones, which are eigenstates of the
$C_{\lambda}$-extended oscillator annihilation operator, extend to higher $\lambda$
values the paraboson coherent states, to which they reduce for $\lambda=2$. All these
states satisfy a unity resolution relation in the $C_{\lambda}$-extended oscillator Fock
space (or in some subspace thereof). They give rise to Bargmann representations of
the latter, wherein the generators of the $C_{\lambda}$-extended oscillator algebra
are realized as differential-operator-valued matrices (or differential operators). The
statistical and squeezing properties of the new coherent states are investigated over a
wide range of parameters and some interesting nonclassical features are exhibited.   
\end{abstract}

\vfill
%73 pages, 8 figures 

\newpage
%
%======================================================================
%
%{\parindent = 0pt
%Running head: GENERALIZED COHERENT STATES
%
%Mailing address: C. Quesne, Physique Nucl\'eaire Th\'eorique et Physique
%Math\'ematique,  Universit\'e Libre de Bruxelles, Campus de la Plaine CP229,
%Boulevard du Triomphe, B-1050 Brussels, Belgium
%
%Tel.: +32-2-6505559
%
%Fax: +32-2-6505045
%
%E-mail: cquesne@ulb.ac.be}
%\newpage
%
%======================================================================
%
\section{INTRODUCTION}

Coherent states (CS) of the harmonic oscillator~\cite{glauber, klauder63, sudarshan},
dating back to the birth of quantum mechanics~\cite{schroedinger}, are known to have
properties similar to those of the classical radiation field. They are specific superpositions
$|z\rangle$, parametrized by a single complex number~$z$, of the eigenstates
$|n\rangle$ of the number operator
$N_b
\equiv
\bp b$ with
\begin{equation}
  \left[N_b, \bp\right] = \bp, \qquad \left[N_b, b\right] = - b, \qquad \left[b, \bp\right]
  = I.  \label{eq:osc-alg}
\end{equation}
They may be constructed in three equivalent ways: (i) by defining them as eigenstates of
the annihilation operator~$b$; (ii) by applying a unitary displacement operator on the
vacuum state $|0\rangle$ (such that $b |0\rangle = 0$); (iii) by considering them as
quantum states with a minimum uncertainty relationship.\par
%
%-------------------------------------------------------------------------------------------------------------
%
These three construction methods become inequivalent when considering generalized CS
associated with other algebras than the oscillator one, defined in~(\ref{eq:osc-alg}). One
may therefore distinguish between annihilation-operator CS (often called Barut-Girardello
CS~\cite{barut}), displacement-operator CS (often called Perelomov
CS~\cite{perelomov72}), and minimum-uncertainty or ``intelligent'' CS~\cite{aragone},
but this in no way exhausts all the possibilities of defining generalized CS.\par
%
%-------------------------------------------------------------------------------------------------------
% 
Such states provide a fundamental framework wherein global dynamical properties of
quantum systems can be interpreted~\cite{klauder85, perelomov86, zhang}. They are
commonly used for instance in connection with variational and path-integral methods or to
study the classical and thermodynamic limits of quantum mechanics.\par
%
%-------------------------------------------------------------------------------------------------------------
%
They have also found considerable applications in quantum optics, because they may
exhibit some nonclassical properties, such as photon antibunching~\cite{kimble}, or
sub-Poissonian photon statistics~\cite{short}, and squeezing~\cite{hong}. As examples of
generalized CS with nonclassical properties, we may quote the eigenstates of~$b^2$,
which were introduced as even and odd CS or cat states~\cite{dodonov}, and are a
special case of Barut-Girardello CS associated with the Lie algebra
su(1,1)~\cite{barut}. We may also mention the eigenstates of $b^{\lambda}$ ($\lambda
> 2$) or kitten states~\cite{buzek}, which are an example of multiphoton
CS~\cite{dariano}.\par
%
%-----------------------------------------------------------------------------------------------------------
% 
Recently there has been much interest in the study of nonlinear CS associated with a
deformed oscillator (or $f$-oscillator). The latter is defined in terms of creation,
annihilation, and number operators, $\ap = f(N_b) \bp$, $a = b f(N_b)$, $N = N_b$,
satisfying the commutation relations~\cite{daska, solomon, katriel, manko}
\begin{equation}
  \left[N, \ap\right] = \ap, \qquad [N, a] = - a, \qquad \left[a, \ap\right] = G(N),
  \label{eq:gdoa}
\end{equation}
where $f$ is some Hermitian operator-valued function of the number operator and $G(N)
= (N+1) f^2(N+1) - N f^2(N)$. Nonlinear CS defined as eigenstates of
$a$~\cite{solomon, manko, shanta, matos, fernandez}, of $a^2$~\cite{mancini}, or of
an arbitrary power $a^{\lambda}$ ($\lambda > 2$)~\cite{liu} have, for instance, been
considered in connection with nonclassical properties. For a particular class of
nonlinearities, the first ones were shown to be realized physically as the stationary states
of the centre-of-mass motion of a trapped ion~\cite{matos}. It should be noted that the
algebra~(\ref{eq:gdoa}) may also be seen as a deformed su(1,1) algebra. Hence the
eigenstates of~$a$ may be considered as a deformation of the Barut-Girardello su(1,1)
CS~\cite{junker}.\par  
%
%-----------------------------------------------------------------------------------------------------------
% 
The purpose of the present paper is to study some mathematical and physical properties
of various CS associated with the recently introduced $C_{\lambda}$-extended oscillator
(where $C_{\lambda} = \Z_{\lambda}$ denotes the cyclic group of
order~$\lambda$)~\cite{cq98}. The latter, which shares some properties with a
fractional supersymmetric oscillator~\cite{daoud}, has proved very useful in the context
of supersymmetric quantum mechanics and some of its variants~\cite{cq98,cq99,cq00a}.
It may be considered as a deformed oscillator with a \Z$_{\lambda}$-graded Fock space
(with components labelled by $\mu = 0$, 1, \ldots,~$\lambda - 1$), generalizing the
Calogero-Vasiliev~\cite{vasiliev} or modified~\cite{brze} oscillator, to which it reduces
for $\lambda = 2$. Its coherent states are therefore special cases of the nonlinear CS
considered in~\cite{solomon, manko, shanta, matos, fernandez, mancini, liu, junker}.\par
%
%-----------------------------------------------------------------------------------------------------------
% 
In~\cite{cq00b}, we already studied one type of CS $|z; \mu\rangle$, associated with the
$C_{\lambda}$-extended oscillator, namely annihilation-operator CS for its spectrum
generating algebra (SGA), a $C_{\lambda}$-extended polynomial deformation of
su(1,1). Since such states are eigenstates of $J_- = a^{\lambda}/\lambda$,
corresponding to the eigenvalue $z$, they fall into the class of multiphoton nonlinear
CS~\cite{liu}, and reduce to the Barut-Girardello su(1,1) CS for $\lambda = 2$ and to
the standard multiphoton CS~\cite{buzek} (equivalent to some Mittag-Leffler
CS~\cite{sixdeniers99}) for arbitrary $\lambda > 2$ and vanishing
$C_{\lambda}$-extended oscillator parameters. Their main mathematical feature is that
they satisfy Klauder's minimal set of conditions for generalized CS~\cite{klauder63},
including the existence of a unity resolution relation, property that is not shared by
generic multiphoton nonlinear CS. In addition, from a physical viewpoint, they were shown
to exhibit, especially for $\mu = 0$, strong nonclassical properties, such as antibunching
and quadrature squeezing.\par
%
%-------------------------------------------------------------------------------------------------------------
% 
Here we will extend the results of~\cite{cq00b} in two different directions. Firstly, we will
consider a whole family of CS, each member of the family being labelled by some index
$\alpha \in \{0, 1, \ldots, [\lambda/2]\}$, where $[\lambda/2]$ denotes the largest
integer contained in $\lambda/2$. To each $\alpha$ value will be associated some CS
$|z; \mu; \alpha\rangle$, $\mu = 0$, 1, \ldots,~$\lambda - \alpha - 1$, satisfying
Klauder's conditions~\cite{klauder63} in some subspace and for some appropriate values
of the $C_{\lambda}$-extended oscillator parameters. This family will include not only the
multiphoton nonlinear CS $|z, \mu\rangle$ of~\cite{cq00b}, corresponding to $\alpha =
0$, but also the Perelomov su(1,1) CS~\cite{perelomov72}, corresponding to $\lambda
= 2$ and $\alpha = 1$.\par
%
%----------------------------------------------------------------------------------------------------------
%
Secondly, we will use the CS $|z; \mu; 0\rangle$, with $z$ replaced by $\omega =
z^{\lambda}$, to obtain the eigenstates $|z\rangle$ of the $C_{\lambda}$-extended
oscillator annihilation operator $a$. For $\lambda \ge 3$, such states will generalize the
paraboson CS~\cite{sharma}, to which they will reduce for $\lambda = 2$ owing to the
connection~\cite{chaturvedi} between the Calogero-Vasiliev algebra~\cite{vasiliev, brze}
and the paraboson one~\cite{ohnuki}.\par
%
%----------------------------------------------------------------------------------------------------------
%
Both types of CS $|z; \mu; 0\rangle$ and $|z\rangle$ providing us with a Bargmann
representation~\cite{bargmann} of the $C_{\lambda}$-extended oscillator Fock space,
we will use them to derive differential-operator-valued matrix realizations of the
$C_{\lambda}$-extended oscillator algebra and of its SGA. In such a context, the $\mu$
index will play a role similar to that of the discrete label appearing in vector
CS~\cite{zhang, deenen}.\par
%
%-----------------------------------------------------------------------------------------------------------
%
This paper is organized as follows. The $C_{\lambda}$-extended oscillator and the
corresponding algebras are briefly reviewed in Section~2. Generalized CS associated with
this oscillator are introduced and proved to satisfy unity resolution relations in
Sections~3 and 4, respectively. They are then used in Section~5 to construct Bargmann
representations. Some physical applications are discussed in Section~6. Finally, Section~7
contains the conclusion.\par
%
%=============================================================
%
\section{\boldmath THE $C_{\lambda}$-EXTENDED OSCILLATOR}

\setcounter{equation}{0}

The $C_{\lambda}$-extended oscillator algebra was introduced as a generalization of the
Calogero-Vasiliev~\cite{vasiliev} or modified oscillator~\cite{brze} algebra, defined by
\begin{equation}
  \left[N, \ap\right] = \ap, \qquad \left[a, \ap\right] = I + \alpha_0 K, \qquad
  [K, N] = \left\{K, \ap\right\} = 0,  \label{eq:calogero}
\end{equation}
and their Hermitian conjugates, where $\alpha_0$ is some real parameter subject
to the condition $\alpha_0 > -1$, and $N$, $K$ are some Hermitian operators. The latter
may be realized as $K = (-1)^N$, so that the second equation in~(\ref{eq:calogero})
becomes equivalent to $\left[a, \ap\right] = I + \alpha_0 P_0 + \alpha_1 P_1$,
where $\alpha_0 + \alpha_1 = 0$ and $P_0 = \frac{1}{2} \left[I + (-1)^N\right]$,
$P_1 = \frac{1}{2} \left[I - (-1)^N\right]$ project on the even and odd subspaces, 
${\cal F}_0 \equiv \{\, |2k\rangle \mid k = 0, 1, \ldots\,\}$, ${\cal F}_1 \equiv \{\,
|2k+1\rangle \mid k = 0, 1, \ldots\,\}$, of the Fock space ${\cal F} = {\cal F}_0 \oplus
{\cal F}_1 = \{\, |n\rangle \mid n = 0, 1, \ldots\,\}$, respectively. The oscillator
Hamiltonian associated with the algebra~(\ref{eq:calogero}),
\begin{equation}
  H_0 = \case{1}{2} \left\{a, \ap\right\},  \label{eq:Ham}  
\end{equation}
provides an algebraic formulation of the two-particle Calogero problem (see~\cite{brze}
and references quoted therein).\par
%
%----------------------------------------------------------------------
%
When partitioning $\cal F$ into $\lambda$ subspaces ${\cal F}_{\mu} \equiv
\{\, |k\lambda + \mu\rangle \mid k = 0, 1, \ldots\,\}$, $\mu=0$, 1,
\ldots,~$\lambda-1$, instead of two, the Calogero-Vasiliev algebra~(\ref{eq:calogero}) is
replaced by the $C_{\lambda}$-extended oscillator algebra, satisfying the relations
\begin{equation}
  \left[N, \ap\right] = \ap, \qquad \left[a, \ap\right] = I + \sum_{\mu=0}^{\lambda-1}
  \alpha_{\mu} P_{\mu}, \qquad \left[N, P_{\mu}\right] = 0, \qquad \ap P_{\mu} =
  P_{\mu+1} \ap,  \label{eq:alg-def}
\end{equation}
and their Hermitian conjugates~\cite{cq98,cq00a}. Here the $\alpha_{\mu}$'s are
some real parameters subject to the conditions 
\begin{eqnarray}
  \sum_{\mu=0}^{\lambda-1} \alpha_{\mu} & = & 0, \\
  \sum_{\nu=0}^{\mu-1} \alpha_{\nu} & > & - \mu, \qquad \mu = 1, 2, \ldots,\lambda-1.
         \label{eq:Fock-cond}
\end{eqnarray}
The operators
\begin{equation}
  P_{\mu} = \frac{1}{\lambda} \sum_{\nu=0}^{\lambda-1} e^{2\pi {\rm i} \nu
  (N-\mu)/\lambda},  \label{eq:Pmu}
\end{equation}
which are linear combinations of the operators of a cyclic group $C_{\lambda}$, project
on the subspaces ${\cal F}_{\mu}$, and satisfy the relations $P_{\mu} P_{\nu} =
\delta_{\mu, \nu} P_{\mu}$ and $\sum_{\mu=0}^{\lambda-1} P_{\mu} = I$.
Throughout this paper, except otherwise stated, we use the convention $P_{\mu} =
P_{\mu'}$ if $\mu' - \mu = 0\, {\rm mod} \lambda$ (and similarly for other operators or
parameters labelled by $\mu$, $\mu'$).\par
%
%-----------------------------------------------------------------------
%
When we insert (\ref{eq:Pmu}) into the second equation of~(\ref{eq:alg-def}), it
becomes clear that the $C_{\lambda}$-extended oscillator algebra is a special case of
deformed oscillator algebra, as defined in~(\ref{eq:gdoa}). Here $f(N) =
[F(N)/N]^{1/2}$, where the so-called structure function $F(N)$ is given by
\begin{equation}
  F(N) = N + \sum_{\mu=0}^{\lambda-1} \beta_{\mu} P_{\mu}, \qquad \beta_{\mu} 
  \equiv \sum_{\nu=0}^{\mu-1} \alpha_{\nu}.  \label{eq:F}
\end{equation}
In the Fock space representation, the operators $N$, $\ap$, $a$ are related to each
other through the function $F(N)$,
\begin{equation}
  \ap a = F(N), \qquad a \ap = F(N+1).
\end{equation}
\par
%
%---------------------------------------------------------------------------------------------------------
%
The Fock space basis states
\begin{equation}
  |n\rangle = |k \lambda + \mu\rangle = \left[\lambda^{k\lambda+\mu} k!
  \left(\prod_{\nu=1}^{\mu} (\bbeta_{\nu})_{k+1}\right)
  \left(\prod_{\nu=\mu+1}^{\lambda-1}
  (\bbeta_{\nu})_k\right)\right]^{-1/2} \left(\ap\right)^{k\lambda+\mu} 
  |0\rangle,  \label{eq:Fock-state}
\end{equation}
where $n$, $k=0$, 1,~\ldots, $\mu = 0$, 1, \ldots,~$\lambda-1$, $a |0\rangle = 0$,
$\bbeta_{\nu}
\equiv (\beta_{\nu} + \nu)/\lambda$, and $(\bbeta_{\nu})_k$ denotes
Pochhammer's symbol, satisfy the relations
\begin{equation}
  N |n\rangle = n |n\rangle, \qquad \ap |n\rangle = \sqrt{F(n+1)}\, |n+1\rangle, \qquad
  a |n\rangle = \sqrt{F(n}\, |n-1\rangle.  \label{eq:alg-action}  
\end{equation}
Due to the restriction~(\ref{eq:Fock-cond}) on the range of the
parameters~$\alpha_{\mu}$, $F(\mu) = \beta_{\mu} + \mu = \lambda \bbeta_{\mu} >
0$ so that all the states~$|n\rangle$ are well defined.\par
%
%--------------------------------------------------------------------------------------------------------
% 
The $C_{\lambda}$-extended oscillator Hamiltonian is still defined by~(\ref{eq:Ham}). Its
eigenstates are the states $|n\rangle = |k\lambda + \mu\rangle$ and their eigenvalues
are given by
\begin{equation}
  E_{k\lambda + \mu} = k \lambda + \mu + \gamma_{\mu} + \case{1}{2}, \qquad 
  \gamma_{\mu} \equiv \case{1}{2} (\beta_{\mu} + \beta_{\mu+1}). \label{eq:E}
\end{equation}
In each ${\cal F}_{\mu}$ subspace of $\cal F$, the spectrum of $H_0$ is harmonic, but
the $\lambda$ infinite sets of equally spaced energy levels, corresponding to $\mu=0$,
1,~\ldots, $\lambda-1$, are shifted with respect to each other by some amounts
depending upon the parameters $\alpha_0$, $\alpha_1$,~\ldots,
$\alpha_{\lambda-1}$ through their linear combinations~$\gamma_{\mu}$.\par
%
%-------------------------------------------------------------------------------------------------------
%
The whole spectrum of~$H_0$ can be generated from the lowest eigenstates
$|\mu\rangle$, $\mu = 0$, 1, \ldots,~$\lambda-1$, belonging to ${\cal F}_{\mu}$,
$\mu = 0$, 1, \ldots,~$\lambda-1$, by using the operators~\cite{cq00b}
\begin{equation}
  J_+ = \frac{1}{\lambda} \left(\ap\right)^{\lambda}, \qquad J_- =
  \frac{1}{\lambda} a^{\lambda}, \qquad J_0 = \frac{1}{\lambda} H_0 =
  \frac{1}{2\lambda} \left\{a, \ap\right\}. \label{eq:SGA-gen}
\end{equation}
Equation~(\ref{eq:Fock-state}) can indeed be rewritten as
\begin{equation}
  |n\rangle = |k \lambda + \mu\rangle = \left[\lambda^{k(\lambda-2)} k!
  \left(\prod_{\nu=1}^{\mu} (\bbeta_{\nu} + 1)_k\right)
  \left(\prod_{\nu=\mu+1}^{\lambda-1}
  (\bbeta_{\nu})_k\right)\right]^{-1/2} J_+^k |\mu\rangle. 
  \label{eq:Fock-state-bis}
\end{equation}
The operators~(\ref{eq:SGA-gen}) satisfy the Hermiticity properties
\begin{equation}
  J_+^{\dagger} = J_-, \qquad J_0^{\dagger} = J_0,  \label{eq:SGA-Herm}
\end{equation}
and the commutation relations
\begin{equation}
  [J_0, J_{\pm}] = \pm J_{\pm}, \qquad [J_+, J_-] = f(J_0, P_{\mu}), \qquad 
  [J_0, P_{\mu}] = [J_{\pm}, P_{\mu}] = 0,  \label{eq:SGA-com}
\end{equation}
where the function $f(J_0, P_{\mu})$ (not to be confused with the function $f(N)$
of~(\ref{eq:gdoa})) is a ($\lambda-1$)th-degree polynomial in $J_0$ with 
$P_{\mu}$-dependent coefficients, given by
\begin{eqnarray}
  f(J_0, P_{\mu}) & = & \sum_{i=0}^{\lambda-1} s_i(P_{\mu}) J_0^i \nonumber \\
  & = & - \frac{1}{\lambda} \Biggl\{\prod_{l=0}^{\lambda-2} \left(
         \lambda J_0 + \case{1}{2} \sum_{\mu} \left(2l + 1 + \alpha_{\mu} + 2
         \sum_{m=1}^l \alpha_{\mu+m}\right) P_{\mu}\right) \nonumber \\
  && \mbox{} + \sum_{i=1}^{\lambda-1} \left(\lambda J_0 - \case{1}{2}
         \sum_{\mu} (1 + \alpha_{\mu}) P_{\mu}\right) \nonumber \\
  && \mbox{} \times \left[\prod_{j=1}^{i-1}
         \left(\lambda J_0 + \case{1}{2} \sum_{\mu} \left(- 2j - 1 + \alpha_{\mu} + 2
         \sum_{k=1}^{\lambda-j-1} \alpha_{\mu+k}\right) P_{\mu}\right)\right]
         \nonumber \\
  && \mbox{} \times \left[\prod_{l=0}^{\lambda-i-2} \left(\lambda J_0 +
         \case{1}{2} \sum_{\mu} \left(2l + 1 + \alpha_{\mu} + 2 \sum_{m=1}^l
         \alpha_{\mu+m}\right) P_{\mu}\right)\right]\Biggr\}. \label{eq:SGA-f} 
\end{eqnarray}
\par
%
%---------------------------------------------------------------------------------------------------------
%
The SGA of the $C_{\lambda}$-extended oscillator is therefore a
$C_{\lambda}$-extended polynomial deformation of the Lie algebra su(1,1): in each 
${\cal F}_{\mu}$ subspace or for vanishing algebra parameters, it reduces to a standard
polynomial deformation of su(1,1)~\cite{poly}. Nonlinearities are only present for
$\lambda \ge 3$, since for $\lambda = 2$ the algebra reduces to the well-known su(1,1)
SGA of the Calogero-Vasiliev oscillator~\cite{brze}.\par
%
%-----------------------------------------------------------------------------------------------------------
%
The algebra (\ref{eq:SGA-Herm})--(\ref{eq:SGA-f}) has a Casimir operator, which can be
written as
\begin{equation}
  C = J_- J_+ + h(J_0, P_{\mu}) = J_+ J_- + h(J_0, P_{\mu}) - f(J_0, P_{\mu}),
\end{equation}
where $h(J_0, P_{\mu})$ is a $\lambda$th-degree polynomial in~$J_0$ with
$P_{\mu}$-dependent coefficients, $h(J_0, P_{\mu}) = \sum_{i=0}^{\lambda}
t_i(P_{\mu}) J_0^i$, whose explicit form is given in~\cite{cq00b}.\par
%
%---------------------------------------------------------------------------------------------------------
%
Each ${\cal F}_{\mu}$ subspace of~$\cal F$ carries a unitary irreducible representation
(unirrep) of the SGA, characterized by the eigenvalue $c_{\mu}$ of~$C$ and by the
lowest eigenvalue of~$J_0$, namely $\left(\mu + \gamma_{\mu} +
\case{1}{2}\right)/{\lambda}$.\par
%
%=============================================================
%
\section{\boldmath GENERALIZED COHERENT STATES ASSOCIATED WITH THE
$C_{\lambda}$-EXTENDED OSCILLATOR}

\setcounter{equation}{0}

\subsection{\boldmath Family of Coherent States $|z; \mu; \alpha\rangle$}

Let $|z; \mu; \alpha\rangle$ be a normalizable state of~${\cal F}_{\mu}$, satisfying the
equation
\begin{equation}
  \left[a^{\lambda - \alpha} - z \left(\ap\right)^{\alpha}\right] |z; \mu; \alpha\rangle
  = 0  \label{eq:CS1-def}
\end{equation}
for some $\alpha \in \{0, 1, \ldots, [\lambda/2]\}$ and some $z \in \C$. The set of
states $|z; \mu; \alpha\rangle$ is surely non-empty, since for $\alpha = 0$,
Eq.~(\ref{eq:CS1-def}) reduces to
\begin{equation}
  a^{\lambda} |z; \mu; 0\rangle = z |z, \mu; 0\rangle,
\end{equation}
showing that $|z; \mu; 0\rangle$ is an annihilation-operator CS $|z/\lambda;
\mu\rangle$ for the $C_{\lambda}$-extended oscillator SGA, corresponding to the
eigenvalue $z/\lambda$ of~$J_-$,
\begin{equation}
  |z; \mu; 0\rangle = |z/\lambda; \mu\rangle.  \label{eq:SGA-CS}
\end{equation}
The latter is known to be normalizable for any $z \in \C$ and any $\mu \in \{0, 1, \ldots,
\lambda-1\}$~\cite{cq00b}.\par
%
%---------------------------------------------------------------------------------------------------------
%
{}Furthermore, for $\lambda = 2$, $\mu = 0$, and $\alpha = 1$, Eq.~(\ref{eq:CS1-def})
becomes
\begin{equation}
  \left(a - z \ap\right) |z; 0; 1\rangle = 0.
\end{equation}
By using Baker-Campbell-Hausdorf formula, this equation can be rewritten as 
\begin{equation}
  e^{z J_+} a e^{-z J_+} |z; 0; 1\rangle = 0,
\end{equation}
where $J_+$ is given by~(\ref{eq:SGA-gen}) where $\lambda = 2$. It is therefore clear
that
\begin{equation}
  |z; 0; 1\rangle = \left(1 - |z|^2\right)^{\bbeta_1/2} e^{z J_+} |0\rangle
  \label{eq:Pere-CS}
\end{equation}
is a displacement-operator (or Perelomov~\cite{perelomov72}) CS for su(1,1),
corresponding to the unirrep characterized by $\frac{1}{2} \left(\gamma_0 +
\frac{1}{2}\right) = \frac{1}{2} \bbeta_1$. Such a state is defined on the unit disc $|z|
< 1$.\par
%
%--------------------------------------------------------------------------------------------------------
%
We may therefore say that the states~(\ref{eq:CS1-def}) in some way interpolate
between Barut-Girardello and Perelomov CS. We shall now proceed to construct them and
to determine under which conditions they are normalizable.\par
%
%-----------------------------------------------------------------------------------------------------
%
Let us expand $|z; \mu; \alpha\rangle$ in the number-state basis $\{\, |k \lambda +
\mu\rangle \mid k = 0, 1, \ldots\,\}$ of ${\cal F}_{\mu}$,
\begin{equation}
  |z; \mu; \alpha\rangle = \sum_{k=0}^{\infty} c_k(z, z^*; \mu; \alpha) z^k |k \lambda
  + \mu\rangle,  \label{eq:CS1-ansatz}  
\end{equation}
where $c_k(z, z^*; \mu; \alpha)$ are some yet unknown coefficients. By
iterating~(\ref{eq:alg-action}) and taking~(\ref{eq:F}) into account, we easily obtain
\begin{eqnarray}
  a^{\lambda - \alpha} |k \lambda + \mu\rangle & = & \left[\lambda^{\lambda - \alpha}
          k \left(\prod_{\nu=1}^{\mu} (\bbeta_{\nu} + k)\right)
          \left(\prod_{\nu=\mu + \alpha +1}^{\lambda-1} (\bbeta_{\nu} + k - 1)
          \right)\right]^{1/2} \nonumber \\
  && \times |(k-1) \lambda + \mu + \alpha\rangle \qquad {\rm if\ } \mu = 0, 1, \ldots,
          \lambda - \alpha - 1, \nonumber \\
  & = & \left[\lambda^{\lambda - \alpha} \left(\prod_{\nu=\mu - \lambda + \alpha +1}
          ^{\mu} (\bbeta_{\nu} + k)\right)\right]^{1/2} |k \lambda + \mu -
          \lambda +\alpha\rangle \nonumber \\
  && {\rm if\ } \mu = \lambda - \alpha, \lambda - \alpha +1, \ldots, \lambda - 1,
          \label{eq:CS1-inter1}
\end{eqnarray}
and
\begin{eqnarray}
  \left(\ap\right)^{\alpha} |k \lambda + \mu\rangle & = & \left[\lambda^{\alpha}
          \left(\prod_{\nu= \mu +1}^{\mu + \alpha} (\bbeta_{\nu} + k)\right)
          \right]^{1/2} |k \lambda + \mu + \alpha\rangle \qquad {\rm if\ } \mu = 0,
         1, \ldots, \lambda - \alpha - 1, \nonumber \\
  & = & \left[\lambda^{\alpha} (k+1) \left(\prod_{\nu=1}^{\mu - \lambda + \alpha}
          (\bbeta_{\nu} + k + 1)\right) \left(\prod_{\nu= \mu +1}^{\lambda
          -1} (\bbeta_{\nu} + k)\right)\right]^{1/2} \nonumber \\
  && \mbox{} \times |(k+1) \lambda + \mu - \lambda +\alpha\rangle \qquad {\rm if\ }
          \mu = \lambda - \alpha, \lambda - \alpha +1, \ldots, \lambda - 1. 
          \label{eq:CS1-inter2}
\end{eqnarray}
Inserting (\ref{eq:CS1-ansatz}), (\ref{eq:CS1-inter1}), and~(\ref{eq:CS1-inter2})
into~(\ref{eq:CS1-def}) leads to a recursion relation for $c_k(z, z^*; \mu; \alpha)$. For
$\mu \in \{0, 1, \ldots, \lambda - \alpha - 1\}$, the latter has a solution given by
\begin{equation}
  c_k(z, z^*; \mu; \alpha) = \left(\frac{\lambda^{k (2\alpha - \lambda)} 
  \left(\prod_{\nu=\mu+1}^{\mu+\alpha} (\bbeta_{\nu})_k\right)}{k! 
  \left(\prod_{\nu=1}^{\mu} (\bbeta_{\nu} + 1)_k\right)
  \left(\prod_{\nu=\mu+\alpha+1}^{\lambda-1} (\bbeta_{\nu})_k\right)}
  \right)^{1/2} c_0(z, z^*; \mu; \alpha),  \label{eq:CS1-coeff} 
\end{equation}
while for $\mu \in \{\lambda - \alpha, \lambda - \alpha + 1, \ldots, \lambda - 1\}$, it
has no solution apart from the trivial one, $c_k(z, z^*; \mu; \alpha) = 0$, $k = 0$,
1,~\ldots.\par
%
%------------------------------------------------------------------------------------------------------
%
By introducing (\ref{eq:CS1-coeff}) into~(\ref{eq:CS1-ansatz}), we obtain
\begin{eqnarray}
  |z; \mu; \alpha\rangle & = & \left[N^{(\alpha)}_{\mu}(|z|)\right]^{-1/2} \sum_{k=0}
          ^{\infty} \left(\frac{\prod_{\nu=\mu+1}^{\mu+\alpha}
          (\bbeta_{\nu})_k}{k! \left(\prod_{\nu=1}^{\mu}
          (\bbeta_{\nu} + 1)_k\right) \left(\prod_{\nu=\mu+\alpha+1}^{\lambda-1}
          (\bbeta_{\nu})_k\right)}\right)^{1/2} \nonumber \\
  && \times \left(\lambda^{- (\lambda - 2\alpha)/2} z\right)^k |k \lambda + \mu
          \rangle,  \label{eq:CS1-exp}
\end{eqnarray}
where the normalization coefficient $N^{(\alpha)}_{\mu}(|z|)$ can be expressed in terms
of a generalized hypergeometric function,
\begin{eqnarray}
  N^{(\alpha)}_{\mu}(|z|) & \equiv & \left[c_0(z, z^*; \mu; \alpha)\right]^{-2}
         \nonumber \\
  & = & {}_{\alpha}F_{\lambda - \alpha - 1} \left(\bbeta_{\mu+1}, \ldots,
         \bbeta_{\mu+\alpha}; \bbeta_1 + 1, \ldots, \bbeta_{\mu} + 1, 
         \bbeta_{\mu+\alpha+1}, \ldots, \bbeta_{\lambda-1}; y\right), \nonumber \\
  y & \equiv & \frac{|z|^2}{\lambda^{\lambda - 2\alpha}}.  \label{eq:CS1-norm}   
\end{eqnarray}
Since the latter converges for all finite $y$ if $\alpha \le \lambda - \alpha - 1$, and for
$|y| < 1$ if $\alpha = \lambda - \alpha$, we conclude that Eq.~(\ref{eq:CS1-def}) has
normalizable solutions defined on the complex plane for any $\alpha \in \{0, 1, \ldots,
[(\lambda - 1)/2]\}$ and any $\mu \in \{0, 1, \ldots, \lambda - \alpha - 1\}$, and
normalizable solutions defined on the unit disc for $\lambda$ even, $ \alpha =
\lambda/2$, and any $\mu \in \{0, 1, \ldots, (\lambda - 2)/2\}$.\par
%
%---------------------------------------------------------------------------------------------------------
% 
In the former subset, there are the CS~(\ref{eq:SGA-CS}) of the
$C_{\lambda}$-extended oscillator SGA. As new examples, we find, for instance, for
$\lambda = 3$ the states
\begin{eqnarray}
  |z; 0; 1\rangle & = & \left[{}_1F_1 \left(\bbeta_1; \bbeta_2; y\right)\right]^{-1/2}
         \sum_{k=0}^{\infty} \left(\frac{(\bbeta_1)_k}{k!\, (\bbeta_2)_k}\right)^{1/2}
         \left(\frac{z}{\sqrt{3}}\right)^k |3k\rangle,  \label{eq:CS1-ex1} \\
  |z; 1; 1\rangle & = & \left[{}_1F_1 \left(\bbeta_2; \bbeta_1+1; y\right)\right]^{-1/2}
         \sum_{k=0}^{\infty} \left(\frac{(\bbeta_2)_k}{k!\, (\bbeta_1+1)_k}\right)^{1/2}
         \left(\frac{z}{\sqrt{3}}\right)^k |3k+1\rangle,  \label{eq:CS1-ex2}
\end{eqnarray}  
with $y \equiv |z|^2/3$, which satisfy the equation $\left(a^2 - z \ap\right) |z; \mu;
1\rangle = 0$, $\mu = 0$,~1.\par
%
%---------------------------------------------------------------------------------------------------------
%
The Perelomov su(1,1) CS~(\ref{eq:Pere-CS}) belong to the latter subset, wherein we
also encounter, for instance, for $\lambda = 4$ the states
\begin{eqnarray}
  |z; 0; 2\rangle & = & \left[{}_2F_1 \left(\bbeta_1, \bbeta_2; \bbeta_3;
         y\right)\right]^{-1/2} \sum_{k=0}^{\infty} \left(\frac{(\bbeta_1)_k
         (\bbeta_2)_k}{k!\, (\bbeta_3)_k}\right)^{1/2} z^k\, |4k\rangle,  
         \label{eq:CS1-ex3} \\
  |z; 1; 2\rangle & = & \left[{}_2F_1 \left(\bbeta_2, \bbeta_3; \bbeta_1+1;
         y\right)\right]^{-1/2} \sum_{k=0}^{\infty} \left(\frac{(\bbeta_2)_k
         (\bbeta_3)_k}{k!\, (\bbeta_1+1)_k}\right)^{1/2} z^k |4k+1\rangle, 
         \label{eq:CS1-ex4}
\end{eqnarray}  
with $y \equiv |z|^2$, which are solutions of $\left[a^2 - z \left(\ap\right)^2\right] |z;
\mu; 2\rangle = 0$, $\mu = 0$,~1.\par
%
%---------------------------------------------------------------------------------------------------------
%
By using~(\ref{eq:Fock-state-bis}), Eq.~(\ref{eq:CS1-exp}) can be written in an
alternative form
\begin{equation}
  |z; \mu; \alpha\rangle = \left[N^{(\alpha)}_{\mu}(|z|)\right]^{-1/2} {}_0F_{\lambda -
  \alpha -1} \left(\bbeta_1+1, \ldots, \bbeta_{\mu}+1, \bbeta_{\mu+\alpha+1}, \ldots,
  \bbeta_{\lambda-1}; z J_+/\lambda^{\lambda - \alpha - 1}\right) |\mu\rangle
\end{equation}
in terms of the SGA generator $J_+$.\par
%
%-----------------------------------------------------------------------------------------------------------
%
The overlap of two states of type~(\ref{eq:CS1-exp}) can be easily calculated and is
given by
\begin{eqnarray}
  \lefteqn{\langle z'; \mu'; \alpha' | z; \mu; \alpha\rangle = \delta_{\mu',\mu}
          \left[N^{(\alpha)}_{\mu}(|z|) N^{(\alpha')}_{\mu}(|z|)\right]^{-1/2}
          \nonumber }\\
  && \mbox{} \times {}_{\alpha_m}F_{\lambda - \alpha_M -1} \Bigl(\bbeta_{\mu+1},
          \ldots, \bbeta_{\mu+\alpha_m}; \bbeta_1 + 1, \ldots, \bbeta_{\mu} + 1, 
         \bbeta_{\mu+\alpha_M+1}, \ldots, \bbeta_{\lambda-1}; \nonumber \\
  && \mbox{} \hphantom{\times} z^{\prime*} z/\lambda^{\lambda - \alpha -
         \alpha'}\Bigr),  \label{eq:CS1-overlap}
\end{eqnarray}
where $\alpha_m \equiv \min(\alpha, \alpha')$ and $\alpha_M \equiv \max(\alpha,
\alpha')$. From the continuity of the overlapping factor $\langle z'; \mu; \alpha | z; \mu;
\alpha\rangle$, it follows that the normalizable states $|z; \mu; \alpha\rangle$ are
continuous in the label~$z$, i.e.,
\begin{equation}
  |z - z'| \to 0 \quad \Rightarrow \quad \Bigl||z; \mu; \alpha\rangle - |z'; \mu;
  \alpha\rangle\Bigr|^2 \to 0.  \label{eq:continuity} 
\end{equation}
Hence, they satisfy two of the three conditions considered by Klauder~\cite{klauder63}
as minimal for the existence of generalized CS. We leave the discussion of the third one,
i.e., the existence of a unity resolution relation, for the next Section.\par
%
%xxxxxxxxxxxxxxxxxxxxxxxxxxxxxxxxxxxxxxxxxxxxxxxxxxxxxxxxxxxxxxxxxxxxxxxxxxx
%
\subsection{\boldmath Eigenstates $|z\rangle$ of the Annihilation Operator $a$}

Let us now look for the eigenstates $|z\rangle$ of the $C_{\lambda}$-extended
oscillator annihilation operator~$a$,
\begin{equation}
  a |z\rangle = z |z\rangle,  \label{eq:CS2-def}
\end{equation}
corresponding to some eigenvalues $z \in \C$.\par
%
%-----------------------------------------------------------------------------------------------------------
%
We shall try to construct them as linear combinations of the CS $|\omega; \mu;
0\rangle$, $\mu=0$, 1, \ldots,~$\lambda-1$, considered in the previous Subsection,
where we replace $z$ by $\omega \equiv z^{\lambda}$,
\begin{equation}
  |z\rangle = \sum_{\mu=0}^{\lambda-1} d_{\mu}\left(z, z^*\right)
  \left(\frac{z}{\sqrt{\lambda}}\right)^{\mu} |\omega; \mu; 0\rangle.
  \label{eq:CS2-ansatz} 
\end{equation}
Here $d_{\mu}\left(z, z^*\right)$ are some coefficients to be determined.\par
%
%---------------------------------------------------------------------------------------------------------
%
{}From (\ref{eq:F}), (\ref{eq:alg-action}), (\ref{eq:CS1-ansatz}),
and~(\ref{eq:CS1-coeff}), we easily obtain
\begin{eqnarray}
  a |\omega; 0; 0\rangle & = & \left[\lambda^{\lambda-1}
          \left(\prod_{\nu=1}^{\lambda-1} \bbeta_{\nu}\right)\right]^{-1/2} 
          \frac{c_0(\omega, \omega^*; 0; 0)}{c_0(\omega, \omega^*; \lambda-1; 0)}\,
          \omega |\omega; \lambda-1; 0\rangle,  \label{eq:CS2-inter1} \\
  a |\omega; \mu; 0\rangle & = & \sqrt{\lambda \bbeta_{\mu}}\, 
          \frac{c_0(\omega, \omega^*; \mu; 0)}{c_0(\omega, \omega^*; \mu-1; 0)}\,
          |\omega; \mu-1; 0\rangle, \qquad \mu=1, 2, \ldots,\lambda-1. 
          \label{eq:CS2-inter2}
\end{eqnarray}
Inserting (\ref{eq:CS2-ansatz}), (\ref{eq:CS2-inter1}), and (\ref{eq:CS2-inter2})
into~(\ref{eq:CS2-def}) leads to
\begin{equation}
  d_{\mu}\left(z, z^*\right) = \left(\prod_{\nu=1}^{\mu} \bbeta_{\nu}\right)^{-1/2} 
  \frac{c_0(\omega, \omega^*; 0; 0)}{c_0(\omega, \omega^*; \mu; 0)}\,
  d_0\left(z, z^*\right), \qquad \mu=1, 2, \ldots,\lambda-1.  
\end{equation}
\par
%
%--------------------------------------------------------------------------------------------------------
%
By taking~(\ref{eq:CS1-norm}) into account, the solutions of~(\ref{eq:CS2-def}) can
therefore be written as
\begin{equation}
  |z\rangle = [{\cal N}(|z|)]^{-1/2} \sum_{\mu=0}^{\lambda-1}
  \left(\frac{N^{(0)}_{\mu} (|\omega|)}{\prod_{\nu=1}^{\mu} \bbeta_{\nu}}\right)
  ^{1/2} \left(\frac{z}{\sqrt{\lambda}}\right)^{\mu} |\omega; \mu; 0\rangle,
  \label{eq:CS2-exp}
\end{equation}
where the normalization factor ${\cal N}(|z|)$ is given by a linear combination of
generalized hypergeometric functions,
\begin{eqnarray}
  {\cal N}(|z|) & \equiv & \left[c_0\left(z, z^*; 0; 0\right) d_0\left(z, z^*\right)\right]
         ^{-2} \nonumber \\
  & = & \sum_{\mu=0}^{\lambda-1} N^{(0)}_{\mu}
         \left(|z|^{\lambda}\right) \frac{\left(|z|^2/\lambda\right)^{\mu}}
         {\prod_{\nu=1}^{\mu} \bbeta_{\nu}} \nonumber \\
  & = & \sum_{\mu=0}^{\lambda-1} {}_0F_{\lambda-1} \left(\bbeta_1 + 1, \ldots,
         \bbeta_{\mu} + 1, \bbeta_{\mu+1}, \ldots, \bbeta_{\lambda-1}; t^{\lambda}
         \right) \frac{t^{\mu}}{\prod_{\nu=1}^{\mu} \bbeta_{\nu}},  \label{eq:CS2-norm} 
\end{eqnarray}
with $t \equiv |z|^2/\lambda$. As the CS $|\omega; \mu; 0\rangle$, the annihilation
operator eigenstates $|z\rangle$ are defined on the whole complex plane.\par
%
%----------------------------------------------------------------------------------------------------------
%
By using (\ref{eq:CS1-exp}), they can be rewritten in the number-state basis
$\{\, |k\lambda + \mu\rangle \mid k=0, 1, \ldots, \mu=0, 1, \ldots, \lambda -1\,\}$ of
$\cal F$ as
\begin{equation}
  |z\rangle = [{\cal N}(|z|)]^{-1/2} \sum_{\mu=0}^{\lambda-1} \sum_{k=0}^{\infty}
  \frac{\left(z/\sqrt{\lambda}\right)^{k\lambda + \mu}}{\left[k!
  \left(\prod_{\nu=1}^{\mu} (\bbeta_{\nu})_{k+1}\right)
  \left(\prod_{\nu=\mu+1}^{\lambda-1} (\bbeta_{\nu})_k\right)\right]^{1/2}}
  |k\lambda + \mu\rangle.
\end{equation}
\par
%
%-----------------------------------------------------------------------------------------------------------
%
{}From (\ref{eq:CS1-overlap}) and (\ref{eq:CS2-exp}), the overlap of two eigenstates
of~$a$ is obtained as
\begin{eqnarray}
  \langle z'|z \rangle & = & [{\cal N}(|z|) {\cal N}(|z'|)]^{-1/2}
        \sum_{\mu=0}^{\lambda-1} {}_0F_{\lambda-1} \left(\bbeta_1 + 1, \ldots,
        \bbeta_{\mu} + 1, \bbeta_{\mu+1}, \ldots, \bbeta_{\lambda-1}; 
        \left(z^{\prime*} z/\lambda\right)^{\lambda}\right) \nonumber \\
  && \mbox{} \times \frac{\left(z^{\prime*} z/\lambda\right)^{\mu}}
        {\prod_{\nu=1}^{\mu} \bbeta_{\nu}},  \label{eq:CS2-overlap}
\end{eqnarray}
which is a continuous function. Hence the normalizable states $|z\rangle$ are continuous
in~$z$, i.e., satisfy a property similar to~(\ref{eq:continuity}).\par
%
%---------------------------------------------------------------------------------------------------------
%
Before proceeding to the determination of a unity resolution relation in the next Section,
it is worth considering the example $\lambda = 2$, corresponding to the
Calogero-Vasiliev algebra~\cite{vasiliev,brze} or, equivalently, to the paraboson
one~\cite{ohnuki}. In such a case, the ${}_0F_1$ function appearing
in~(\ref{eq:CS2-norm}) and~(\ref{eq:CS2-overlap}) can be expressed in terms of a
modified Bessel function $I_{\nu}$. Hence we get
\begin{eqnarray}
  |z\rangle & = & [{\cal N}(|z|)]^{-1/2} \left\{\left[N^{(0)}_0 (|z|^2)\right]
        ^{1/2} \left|z^2; 0; 0\right\rangle + \left[N^{(0)}_1 (|z|^2)\right]
        ^{1/2} \frac{z}{\sqrt{2\bbeta_1}} \left|z^2; 1; 0\right\rangle\right\}
        \nonumber \\
  & = & \left(\frac{\Gamma(\bbeta_1)}{{\cal N}(|z|)}\right)^{1/2}
        \sum_{n=0}^{\infty} \frac{(z/\sqrt{2})^n}{\left\{[n/2]!\, \Gamma\left(\bbeta_1 +
        [(n+1)/2]\right)\right\}^{1/2}}\, |n\rangle, 
\end{eqnarray}
where
\begin{equation}
  N^{(0)}_{\mu}\left(|z|^2\right) = \Gamma(\bbeta_1 + \mu) \left(|z|^2/2
  \right)^{1 - \bbeta_1 - \mu} I_{\bbeta_1 - 1 + \mu} (|z|^2), \qquad \mu=0, 1,  
\end{equation}
and
\begin{equation}
  {\cal N}(|z|) = \Gamma(\bbeta_1) t^{1 - \bbeta_1} \left[I_{\bbeta_1 - 1}
  (2t) + I_{\bbeta_1}(2t)\right], \qquad t \equiv |z|^2/2.
\end{equation}
These results coincide with those previously obtained in~\cite{sharma} for
paraboson~CS.\par
%
%=============================================================
%
\section{UNITY RESOLUTION RELATIONS}

\setcounter{equation}{0}

\subsection{\boldmath Unity Resolution Relations for the Coherent States $|z; \mu;
\alpha\rangle$}

\subsubsection{\boldmath Unity Resolution Relations in ${\cal F}_{\mu}$}

Let us first restrict ourselves to a given subspace ${\cal F}_{\mu}$ of the Fock
space~$\cal F$. In such a subspace, the number states $|k\lambda + \mu\rangle$,
$k=0$, 1,~\ldots, form a complete set and their resolution of unity is given by a sum of
orthogonal projection operators, $\sum_{k=0}^{\infty} |k\lambda + \mu\rangle
\langle k\lambda + \mu| = I_{\mu}$, where $I_{\mu}$ denotes the unit operator
in~${\cal F}_{\mu}$.\par
%
%----------------------------------------------------------------------------------------------------------
%
{}For any $\alpha \le \lambda - \mu - 1$, the CS $|z; \mu; \alpha\rangle$ give a
resolution of unity in ${\cal F}_{\mu}$ with a positive measure
$d\rho^{(\alpha)}_{\mu} \left(z, z^*\right)$ if
\begin{equation}
  \int d\rho^{(\alpha)}_{\mu} \left(z, z^*\right) |z; \mu; \alpha\rangle \langle z; \mu;
  \alpha| = I_{\mu},  \label{eq:resol-mu} 
\end{equation}
where the integration is carried out over the complex plane or the unit disc according to
whether $\alpha \le [(\lambda - 1)/2]$ or $\alpha = \lambda/2$ and $\lambda$ even.
Making a polar decomposition $z = |z| \exp({\rm i} \phi)$ and the ansatz
\begin{eqnarray}
  d\rho^{(\alpha)}_{\mu} \left(z, z^*\right) & = & {}_{\alpha}F_{\lambda - \alpha - 1} 
         \left(\bbeta_{\mu+1}, \ldots, \bbeta_{\mu+\alpha}; \bbeta_1 + 1, \ldots,
         \bbeta_{\mu} + 1, \bbeta_{\mu+\alpha+1}, \ldots, \bbeta_{\lambda-1}; y\right)
         \nonumber \\
  && \mbox{} \times h^{(\alpha)}_{\mu}(y)\, d^2z, \qquad y \equiv |z|^2/\lambda^
         {\lambda - 2\alpha}, \qquad d^2z \equiv |z|\, d|z| d\phi,  \label{eq:CS1-drho}
\end{eqnarray}
where $h^{(\alpha)}_{\mu}(y)$ is a yet unknown weight function, we find that
Eq.~(\ref{eq:resol-mu}) is equivalent to
\begin{eqnarray}
  && \pi \lambda^{\lambda - 2\alpha} \sum_{k=0}^{\infty} \left(\frac{
        \prod_{\nu=\mu+1}^{\mu+\alpha} (\bbeta_{\nu})_k}{k!
        \left(\prod_{\nu=1}^{\mu} (\bbeta_{\nu} + 1)_k\right)
        \left(\prod_{\nu=\mu+\alpha+1}^{\lambda-1}
        (\bbeta_{\nu})_k\right)} \int_0^{y_{max}} dy\,  y^k
        h^{(\alpha)}_{\mu}(y)\right) \nonumber \\
  && \mbox{} \times |k\lambda + \mu\rangle \langle k\lambda + \mu| = I_{\mu},    
\end{eqnarray}
where $y_{max} = \infty$ for $\alpha \le [(\lambda-1)/2]$ and $y_{max} = 1$ for
$\alpha = \lambda/2$ and $\lambda$ even. Consequently, the requirement that for a
given $\alpha$, $|z; \mu; \alpha\rangle$ form a complete (in fact, an overcomplete)
set in~${\cal F}_{\mu}$ is equivalent to the resolution of a Stieltjes or a Hausdorff
power-moment problem~\cite{akhiezer}, namely determine a positive weight function
$h^{(\alpha)}_{\mu}(y)$ such that
\begin{eqnarray}
  \int_0^{y_{max}} dy\,  y^k h^{(\alpha)}_{\mu}(y) & = & A^{(\alpha)}_{\mu}
          \frac{\Gamma(k+1) \left(\prod_{\nu=1}^{\mu} \Gamma(\bbeta_{\nu} + k
          + 1)\right) \left(\prod_{\nu=\mu+\alpha+1}^{\lambda-1}
          \Gamma(\bbeta_{\nu} + k)\right)}{\prod_{\nu=\mu+1}^
          {\mu+\alpha} \Gamma(\bbeta_{\nu} + k)} \nonumber \\
  & \equiv & B^{(\alpha)}_{\mu}(k), \qquad k=0, 1, 2, \ldots,  \label{eq:moment}
\end{eqnarray}   
where
\begin{equation}
  A^{(\alpha)}_{\mu} \equiv \frac{\prod_{\nu=\mu+1}^{\mu+\alpha}
  \Gamma(\bbeta_{\nu})}{\pi \lambda^{\lambda - 2\alpha} 
  \left(\prod_{\nu=1}^{\mu} \Gamma(\bbeta_{\nu} + 1)\right)
  \left(\prod_{\nu=\mu+\alpha+1}^{\lambda-1} \Gamma(\bbeta_{\nu})\right)
  }  \label{eq:Aalphamu}
\end{equation}
is a positive constant.\par
%
%----------------------------------------------------------------------------------------------------------
%
The general theory of the power-moment problem~\cite{akhiezer} provides us with
conditions for the solvability of~(\ref{eq:moment}). In the case where $y_{max} =
\infty$, for instance, what is required is the positivity of the two series $\left\{
H^{(k)}_0\right\}$, $\left\{ H^{(k)}_1 \right\}$ ($k=1$, 2, 3,~\ldots) of so-called
Hankel-Hadamard matrices, defined by
\begin{equation}
  H^{(k)}_0(i, j) = B^{(\alpha)}_{\mu} (i+j-2), \qquad H^{(k)}_1(i, j) =
  B^{(\alpha)}_{\mu} (i+j-1), \qquad i, j = 1, 2, \ldots, k.   
\end{equation}
The proof of positivity being extremely difficult, we will instead directly construct a
solution of~(\ref{eq:moment}) in both cases $y_{max} = \infty$ and $y_{max} = 1$. For
such a purpose, we will interpret the latter as an inverse Mellin transform
problem~\cite{sneddon}, as done in related contexts before~\cite{fernandez, junker,
cq00b, sixdeniers99, penson, sixdeniers00}, by setting for complex $s$, $k \to s-1$ and
rewriting~(\ref{eq:moment}) as
\begin{equation}
  \int_0^{\infty} dy\, y^{s-1} g(y) = \frac{\prod_{\nu=1}^{r + \alpha} \Gamma(s +
  b_{\nu})}{\prod_{\nu=1}^{\alpha} \Gamma(s + a_{\nu})} \equiv g^*(s),  
  \label{eq:mellin}
\end{equation}
where for simplicity's sake we skipped the $(\alpha)$ and $\mu$ indices. The function
$g(y) \equiv \left(A^{(\alpha)}_{\mu}\right)^{-1} h^{(\alpha)}_{\mu}(y)$, which is the
inverse Mellin transform of $g^*(s)$, is assumed to vanish for $y > 1$ whenever
$y_{max} = 1$, and
$r$,
$a_{\nu}$, $b_{\nu}$ are defined by
\begin{eqnarray}
  r & = & \lambda - 2\alpha \ge 0, \\
  a_{\nu} & = & \bbeta_{\mu + \nu} - 1, \qquad \nu = 1, 2, \ldots, \alpha,  \label{eq:a}\\
  b_{\nu} & = & \left\{\begin{array}{ll}
        0, & \nu = 1, \\[0.2cm]
        \bbeta_{\nu-1}, & \nu = 2, 3, \ldots, \mu+1, \\[0.2cm]
        \bbeta_{\nu + \alpha - 1} - 1, & \nu = \mu + 2, \mu + 3, \ldots, r + \alpha.
     \end{array}\right.  \label{eq:b}
\end{eqnarray}
It is worth noting that from the restriction~(\ref{eq:Fock-cond}) and the definition of
$\bbeta_{\nu}$ given below~(\ref{eq:Fock-state}), it results that the parameters
$a_{\nu}$ and $b_{\nu}$ are greater than $-1$. In addition, $r > 0$ corresponds to
$y_{max} = \infty$, whereas $r = 0$ is associated with $y_{max} = 1$.\par
%
%----------------------------------------------------------------------------------------------------------
%
To ensure the positivity of $g(y)$ on $(0, y_{max})$, and hence that of
$h^{(\alpha)}_{\mu}(y)$, we shall proceed step by step by using the Mellin convolution
property of inverse Mellin transforms~\cite{sneddon}, which states that if $g^*(s) =
g^*_1(s) g^*_2(s)$ and $g_1(y)$, $g_2(y)$ exist, then the inverse Mellin transform of
$g^*(s)$ is 
\begin{equation}
  g(y) = \int_0^{\infty} g_1\left(\frac{y}{t}\right) g_2(t) \frac{dt}{t} =
  \int_0^{\infty} g_1(t) g_2\left(\frac{y}{t}\right) \frac{dt}{t}.  \label{eq:convolution} 
\end{equation}
In the present context, if we are sure that $g_1(y)$ and $g_2(y)$ are positive on $(0,
y_{max})$, then the same will be true for $g(y)$.\par
%
%------------------------------------------------------------------------------------------------------------
%
{}For $r> 0$ and $\alpha = 0$, it is known~\cite{fernandez, junker, cq00b} that
Eq.~(\ref{eq:mellin}) has a positive solution in terms of a Meijer 
$G$-function~\cite{erdelyi},
\begin{equation}
  g(y) = G^{r0}_{0r} (y | b_1, b_2, \ldots, b_r),  \label{eq:g-zero}
\end{equation}
for all allowed values of $b_{\nu}$. In Appendix A, we use this result as the starting point
of an induction process over $\alpha$, based upon~(\ref{eq:convolution}), to show that
for any $r > 0$ and any $0 \le \alpha \le \lambda - \mu - 1$, Eq.~(\ref{eq:mellin}) has a
positive solution, given by
\begin{equation}
  g(y) = G^{r + \alpha, 0}_{\alpha, r + \alpha} \left(y\, \Bigg| 
      \begin{array}{l}
         a_1, a_2, \ldots, a_{\alpha}\\
         b_1, b_2, \ldots, b_{r + \alpha}
      \end{array}\right),  \label{eq:g}
\end{equation}
if the set $\{b_1, b_2, \ldots, b_{r + \alpha}\}$ contains some subset of $\alpha$
elements $\{b_{i_1}, b_{i_2}, \ldots, b_{i_{\alpha}}\}$ such that
\begin{equation}
  a_1 > b_{i_1}, \qquad a_2 > b_{i_2}, \qquad \ldots, \qquad a_{\alpha} >
  b_{i_{\alpha}}.  \label{eq:pos-cond}
\end{equation}
We conclude that for any $\alpha \le \min(\lambda - \mu - 1, [(\lambda-1)/2])$, there
is a positive weight function $h^{(\alpha)}_{\mu}(y)$, solution of~(\ref{eq:moment})
and given by
\begin{equation}
  h^{(\alpha)}_{\mu}(y) = A^{(\alpha)}_{\mu} G^{\lambda - \alpha, 0}_{\alpha, \lambda
  - \alpha} \left(y\, \Bigg| 
      \begin{array}{l}
         \bbeta_{\mu+1} - 1, \ldots, \bbeta_{\mu+\alpha} - 1\\[0.1cm]
         0, \bbeta_1, \ldots, \bbeta_{\mu}, \bbeta_{\mu+\alpha+1} - 1, \ldots, 
              \bbeta_{\lambda-1} - 1
      \end{array}\right),  \label{eq:halphamu}
\end{equation}
if condition~(\ref{eq:pos-cond}) is satisfied by the sets $\{a_1, a_2, \ldots,
a_{\alpha}\}$ and $\{b_1, b_2, \ldots, b_{r + \alpha}\}$ defined in~(\ref{eq:a})
and~(\ref{eq:b}), respectively.\par
%
%-------------------------------------------------------------------------------------------------------------
%
As an example, let us consider the case of~${\cal F}_0$ and of the CS $|z; 0; 1\rangle$
corresponding to $\lambda = 3$, $\alpha = 1$, defined in~(\ref{eq:CS1-ex1}). From the
above results, it follows that if $\bbeta_1 > 1$ or $\bbeta_1 > \bbeta_2$ (i.e.,
$\alpha_0 > 2$ or $\alpha_1 < -1$), they satisfy the unity resolution
relation~(\ref{eq:resol-mu}) with
\begin{equation}
  h^{(1)}_0(y) = \frac{\Gamma(\bbeta_1)}{3\pi \Gamma(\bbeta_2)}\, 
  G^{20}_{12} \left(y\, \Bigg| 
      \begin{array}{l}
         \bbeta_1 - 1\\
         0, \bbeta_2 - 1
      \end{array}\right)
  = \frac{\Gamma(\bbeta_1)}{3\pi \Gamma(\bbeta_2)}\, e^{-y} 
  U\left(\bbeta_1 - \bbeta_2, 2 - \bbeta_2, y\right),  
\end{equation}
where $U(a, b, z) = \Psi(a, b; z)$ is Kummer's confluent hypergeometric
function~\cite{erdelyi}. As illustrated in Fig.~1, $h^{(1)}_0(y) \to 0$ for $y \to \infty$
as it should be, and for $y \to 0$, $h^{(1)}_0(y) \to + \infty$ or $(\bbeta_1 - 1)
[3\pi (\bbeta_2 - 1)]^{-1}$ according to whether $\bbeta_2 \le 1$ or $\bbeta_2 >
1$ (i.e., $\alpha_0 + \alpha_1 \le 1$ or $\alpha_0 + \alpha_1 > 1$).\par
%
%---------------------------------------------------------------------------------------------------------
%  
It now remains to consider the case where $r=0$, or $\alpha = \lambda/2$,
$\lambda$ even (and $\mu \le \alpha - 1$). For $\lambda = 2$, $\alpha = 1$ (and
$\mu=0$), corresponding to the Perelomov su(1,1) CS $|z; 0; 1\rangle$, defined
in~(\ref{eq:Pere-CS}), it is well known that there exists a positive weight function
$h^{(1)}_0(y)$ on $(0,1)$, solution of~(\ref{eq:moment}) and given
by~\cite{perelomov86}
\begin{equation}
  h^{(1)}_0(y) = \frac{\bbeta_1 - 1}{\pi}\, (1 - y)^{\bbeta_1 - 2} = 
  \frac{\bbeta_1 - 1}{\pi}\, {}_1F_0(2 - \bbeta_1; y),  \label{eq:h10} 
\end{equation}
provided the condition $\bbeta_1 > 1$ (i.e., $\alpha_0 > 1$) is satisfied. The problem
to be solved is therefore finding a solution of~(\ref{eq:mellin}), positive on $(0, 1)$ and
vanishing on $(1, \infty)$, for $r=0$ and $\alpha \ge 2$.\par
%
%------------------------------------------------------------------------------------------------------
% 
By using the convolution property~(\ref{eq:convolution}) again, it is shown in
Appendix~B that for $r=0$ and $\alpha=2$, Eq.~(\ref{eq:mellin}) has indeed a positive
solution~(\ref{eq:gB-1}) on $(0,1)$ (vanishing on $(1, \infty)$) if either $a_1 > b_1$,
$a_2 > b_2$ or $a_1 > b_2$, $a_2 > b_1$. It leads to the following result for the
weight function $h^{(2)}_0(y)$ of the CS $|z; 0; 2\rangle$, defined
in~(\ref{eq:CS1-ex3}),
\begin{equation}
  h^{(2)}_0(y) = A^{(2)}_0 \frac{(1 - y)^{\bbeta_1 + \bbeta_2 - \bbeta_3 - 2}}
  {\Gamma(\bbeta_1 + \bbeta_2 - \bbeta_3 - 1)}\, {}_2F_1\left(\bbeta_1 - \bbeta_3,
  \bbeta_2 - \bbeta_3; \bbeta_1 + \bbeta_2 - \bbeta_3 - 1; 1 - y\right), 
\label{eq:h20}
 \end{equation}
if either $\bbeta_1 > \bbeta_3$, $\bbeta_2 > 1$ or $\bbeta_1 > 1$, $\bbeta_2 >
\bbeta_3$ (i.e., $\alpha_0 + \alpha_1 > 2$, $\alpha_1 + \alpha_2 < -2$ or $\alpha_0 >
3$, $\alpha_2 < -1$). The weight function $h^{(2)}_1(y)$ of the CS $|z; 1; 2\rangle$,
defined in~(\ref{eq:CS1-ex4}), can be obtained from the latter by performing the
substitutions $\bbeta_1 \to \bbeta_2$, $\bbeta_2 \to \bbeta_3$, $\bbeta_3 \to
\bbeta_1 + 1$. As illustrated in Fig.~2, for $y \to 0$, $h^{(2)}_0(y) \to
+\infty$ or $(\bbeta_1 - 1)(\bbeta_2 - 1) [\pi (\bbeta_3 - 1)]^{-1}$ according to
whether $\bbeta_3 \le 1$ or $\bbeta_3 > 1$ (i.e., $\alpha_0 + \alpha_1 + \alpha_2 \le
1$ or $\alpha_0 + \alpha_1 + \alpha_2 > 1$), and for $y \to 1$, $h^{(2)}_0(y) \to
+\infty$, $\Gamma(\bbeta_1) \Gamma(\bbeta_2) [\pi \Gamma(\bbeta_3)]^{-1}$, or
0 according to whether $\bbeta_1 + \bbeta_2 - \bbeta_3 < 2$, $\bbeta_1 + \bbeta_2
- \bbeta_3 = 2$, or $\bbeta_1 + \bbeta_2 - \bbeta_3 > 2$ (i.e., $\alpha_0 -
\alpha_2 < 8$, $\alpha_0 - \alpha_2 = 8$, or $\alpha_0 - \alpha_2 > 8$).\par
%
%--------------------------------------------------------------------------------------------------------
%    
{}For $r=0$ and $\alpha > 2$, Eq.~(\ref{eq:h20}) can be generalized in the form of a
formal multiple series of hypergeometric functions (see Appendix~B). For $\mu=0$, the
result can be written as
\begin{eqnarray}
  h^{(\alpha)}_0(y) & = & \frac{A^{(\alpha)}_0}{\Gamma(\bbeta_1 -
        \bbeta_{\alpha+1}) \left(\prod_{p=1}^{\alpha-2} \Gamma(\bbeta_{p+1} -
        \bbeta_{\alpha+p})\right)} \nonumber \\
  && \mbox{} \times \sum_{n_1 n_2 \ldots n_{\alpha-2}} \frac{\prod_{p=1}^{\alpha-2} 
        [\Gamma(\bbeta_{p+1} - \bbeta_{\alpha+p} + n_p) \Gamma(\xi_{\alpha,
        p}(\bn))]}{\left(\prod_{p=1}^{\alpha-3} \Gamma(\eta_{\alpha, p}(\bn))\right)
        \Gamma(\zeta_{\alpha}(\bn)) \left(\prod_{p=1}^{\alpha-2} n_p!\right)}
        \nonumber \\
  && \mbox{} \times (1 - y)^{\zeta_{\alpha}(\bn) - 1} {}_2F_1\left(\bbeta_{\alpha} -
        \bbeta_{2\alpha -1}, \eta_{\alpha, \alpha-2}(\bn); \zeta_{\alpha}(\bn); 1-y
        \right),  \label{eq:halpha0}
\end{eqnarray}
where
\begin{eqnarray}
  \xi_{\alpha,p}(\bn) & = & \sum_{q=1}^p (\bbeta_q - \bbeta_{\alpha+q} + n_q),
         \qquad \eta_{\alpha,p}(\bn) = \sum_{q=1}^{p+1} (\bbeta_q - \bbeta_{\alpha+q})
         + \sum_{q=1}^p n_q, \nonumber \\
  \zeta_{\alpha}(\bn) & = & \sum_{p=1}^{\alpha} \bbeta_p - \sum_{p=1}^{\alpha-1}
         \bbeta_{\alpha+p} + \sum_{p=1}^{\alpha-2} n_p - 1,  \label{eq:halpha0-inter}
\end{eqnarray}
if there exists some permutation $(i_1, i_2, \ldots, i_{\alpha})$ of $(1, 2, \ldots,
\alpha)$ such that
\begin{equation}
  \bbeta_{i_1} > \bbeta_{\alpha+1}, \qquad \bbeta_{i_2} > \bbeta_{\alpha+2}, \qquad
  \ldots, \qquad \bbeta_{i_{\alpha-1}} > \bbeta_{2\alpha-1}, \qquad \bbeta_{i_{\alpha}}
  > 1. \label{eq:halpha0-cond}
\end{equation}
The other weight functions $h^{(\alpha)}_{\mu}(y)$, $\mu = 1$, 2, \ldots,~$(\lambda -
2)/2$, and their respective conditions of applicability can be found from
(\ref{eq:halpha0}), (\ref{eq:halpha0-inter}), and~(\ref{eq:halpha0-cond}) by making the
substitutions 
\begin{eqnarray}
  \bbeta_1 & \to & \bbeta_{\mu+1}, \qquad \bbeta_2 \to \bbeta_{\mu+2}, \qquad
        \ldots, \qquad \bbeta_{\lambda-\mu-1} \to \bbeta_{\lambda-1}, \qquad
        \bbeta_{\lambda-\mu} \to \bbeta_1 + 1, \nonumber \\
  \bbeta_{\lambda-\mu+1} &\to & \bbeta_2 + 1, \qquad \ldots, \qquad
        \bbeta_{\lambda-1} \to \bbeta_{\mu+1}.  \label{eq:substitution}
\end{eqnarray}
\par
%
%-----------------------------------------------------------------------------------------------------------
% 
Setting $\alpha=2$ in~(\ref{eq:halpha0}) and~(\ref{eq:halpha0-inter}) gives back
(\ref{eq:h20}) as it should be. For $\alpha = 3$, Eq.~(\ref{eq:halpha0}) contains a single
summation, which can be performed by using Eq.~(I.33) of~\cite{appell}. This leads to a
closed expression of $h^{(3)}_0(y)$ in terms of an Appell function $F_3(\alpha, \alpha';
\beta, \beta'; \gamma; x, y)$,
\begin{eqnarray}
  h^{(3)}_0(y) & = & A^{(3)}_0 \frac{y^{\bbeta_5 - \bbeta_3} (1-y)^{\bbeta_1 +
        \bbeta_2 + \bbeta_3 - \bbeta_4 - \bbeta_5 - 2}}{\Gamma(\bbeta_1 +
        \bbeta_2 + \bbeta_3 - \bbeta_4 - \bbeta_5 - 1)} F_3\Biggl(\bbeta_1 - \bbeta_4,
        \bbeta_3 - \bbeta_5; \bbeta_2 - \bbeta_4, \bbeta_3 - 1; \nonumber \\ 
  && \bbeta_1 + \bbeta_2 + \bbeta_3 - \bbeta_4 - \bbeta_5 - 1; 1 - y, 1 -
        \frac{1}{y}\Biggr).  \label{eq:halpha0-bis}  
\end{eqnarray}
It should be noted that this result can also be obtained by direct integration without
expanding the hypergeometric function into powers, as done in Appendix~B, by using
instead Eq.~(2.21.1.20) of~\cite{prudnikov}.\par
%
%---------------------------------------------------------------------------------------------------------
%
{}For higher $\alpha$ values, no closed expression of~(\ref{eq:halpha0}) has been
proved to be valid in general, but we conjecture that it is given in terms of a Meijer
$G$-function,
\begin{equation}
  h^{(\alpha)}_0(y) = A^{(\alpha)}_0 G^{\alpha, 0}_{\alpha, \alpha} \left(y\, \Bigg| 
      \begin{array}{l}
         \bbeta_1 - 1, \ldots, \bbeta_{\alpha} - 1\\
         0, \bbeta_{\alpha+1} - 1, \ldots,  \bbeta_{\lambda-1} - 1
      \end{array}\right).  \label{eq:conjecture}
\end{equation}
Equation~(\ref{eq:conjecture}) has been demonstrated only for $\alpha=1$, 2, and~3, for
which its equivalence to~(\ref{eq:h10}), (\ref{eq:h20}), and~(\ref{eq:halpha0-bis})
results from Eqs.~(8.4.2.3), (8.4.49.22), and~(8.4.51.3) of~\cite{prudnikov},
respectively.\par
%
%--------------------------------------------------------------------------------------------------------
%
By making the parameter substitution~(\ref{eq:substitution}) in~(\ref{eq:conjecture})
and~(\ref{eq:halpha0-cond}), we get for $h^{(\alpha)}_{\mu}(y)$ and the corresponding
applicability conditions the same results~(\ref{eq:halphamu}) and~(\ref{eq:pos-cond}) as
previously obtained when $\alpha \le [(\lambda-1)]/2$. In the present case, of course,
we have to set $\lambda = 2\alpha$ in~(\ref{eq:halphamu}) and~(\ref{eq:pos-cond})
and to restrict the range of~$y$ to the interval $(0,1)$. At this point, it is worth noting
that Eq.~(\ref{eq:halphamu}) would straightforwardly result from an unquestioning
inversion of~(\ref{eq:moment}), but that the convolution process adopted here is
essential to derive the positivity conditions~(\ref{eq:pos-cond}).\par
%
%----------------------------------------------------------------------------------------------------------
%
Having found a solution to the problem stated in~(\ref{eq:moment}), we may now ask
whether this solution is unique. An answer is provided by the (sufficient) condition of
Carleman~\cite{akhiezer}: if a solution exists and $S \equiv \sum_{k=1}^{\infty}
\left(B^{(\alpha)}_{\mu}(k)\right)^{-1/(2k)}$ diverges (resp.\ converges), then the
solution is unique (resp.\ non-unique solutions may exist). As in~\cite{sixdeniers99}, the
possible convergence of~$S$ can be tested by applying the logarithmic
test~\cite{prudnikov}. We obtain
\begin{equation}
  \lim_{k \to \infty} \frac{\ln\left[\left(B^{(\alpha)}_{\mu}(k)\right)^{-1/(2k)}\right]}
  {\ln k} = - \left(\frac{\lambda}{2} - \alpha\right).
\end{equation}
Hence $S$ converges or diverges according to whether $- \left(\frac{\lambda}{2} -
\alpha\right) < -1$ or $- \left(\frac{\lambda}{2} - \alpha\right) > -1$. We conclude
that except in those cases where $\lambda$ is odd and $\alpha = \frac{\lambda-1}{2}$,
or $\lambda$ is even and $\alpha = \frac{\lambda}{2}$,
other solutions than those found hereabove may exist. To construct them, we might
employ methods similar to those used in~\cite{sixdeniers99}, but we shall pursue this
subject no further, leaving it for a separate publication.\par
%
%---------------------------------------------------------------------------------------------------------
% 
As a consequence of the unity resolution relation~(\ref{eq:resol-mu}), we can express
any CS $|z; \mu; \alpha\rangle$, defined in ${\cal F}_{\mu}$, in terms of any set of CS
$|z; \mu; \alpha'\rangle$ ($\alpha' = \alpha$ or $\alpha' \ne \alpha$), defined in the
same,
\begin{equation}
  |z; \mu; \alpha\rangle = \int d\rho^{(\alpha')}_{\mu} (z', z^{\prime *}) 
  |z'; \mu; \alpha'\rangle \langle z'; \mu; \alpha'|z; \mu; \alpha\rangle.  
\end{equation}
The reproducing kernel $\langle z'; \mu; \alpha'|z; \mu; \alpha\rangle$ is given
in~(\ref{eq:CS1-overlap}). More generally, an arbitrary element $|\psi_{\mu}\rangle$ of
${\cal F}_{\mu}$ can be written in terms of any set of CS $|z; \mu; \alpha\rangle$
defined in ${\cal F}_{\mu}$,
\begin{equation}
  |\psi_{\mu}\rangle = \int d\rho^{(\alpha)}_{\mu} (z, z^*) 
  \tilde{\psi}^{(\alpha)}_{\mu}(z, z^*)  |z; \mu; \alpha\rangle,  
\end{equation}
where
\begin{equation}
  \tilde{\psi}^{(\alpha)}_{\mu}(z, z^*) = \langle z; \mu; \alpha|\psi_{\mu}
  \rangle = \sum_{k=0}^{\infty} c_k(z, z^*; \mu; \alpha) (z^*)^k \langle k\lambda +
  \mu|\psi_{\mu}\rangle,  \label{eq:psi-tilde}
\end{equation}
and $c_k(z, z^*; \mu; \alpha)$ is defined in~(\ref{eq:CS1-coeff})
and~(\ref{eq:CS1-norm}).\par
%
%++++++++++++++++++++++++++++++++++++++++++++++++++++++++++++
%
\subsubsection{\boldmath Unity Resolution in $\cal F$}

In the Fock space $\cal F$, the number states $|k\lambda + \mu\rangle$, $k=0$,
1,~\dots, $\mu=0$, 1, \ldots,~$\lambda-1$, form a complete set and their resolution of
unity is given by $\sum_{\mu=0}^{\lambda-1} \sum_{k=0}^{\infty} |k\lambda +
\mu\rangle \langle k\lambda + \mu| = \sum_{\mu=0}^{\lambda-1} P_{\mu} = I$.\par
%
%-----------------------------------------------------------------------------------------------------------
% 
The only set of CS $|z; \mu; \alpha\rangle$, $\mu=0$, 1, \ldots,~$\lambda - \alpha -
1$, defined in the whole space $\cal F$ corresponds to $\alpha=0$. Since two such
states belonging to different subspaces ${\cal F}_{\mu}$ and ${\cal F}_{\mu'}$ are
orthogonal (see~(\ref{eq:CS1-overlap})), it directly follows from~(\ref{eq:resol-mu}) that
this set of CS gives rise to a resolution of unity in~$\cal F$ with a set of positive
measures $d\rho^{(0)}_{\mu}(z, z^*)$, $\mu=0$, 1, \ldots,~$\lambda-1$,
\begin{equation}
  \sum_{\mu=0}^{\lambda-1} \int d\rho^{(0)}_{\mu}(z, z^*) |z; \mu; 0\rangle
  \langle z; \mu; 0| = I.  \label{eq:CS1-resol}  
\end{equation}
Taking (\ref{eq:SGA-CS}) into account, we recover the result previously obtained for the
annihilation-operator CS of the $C_{\lambda}$-extended oscillator SGA~\cite{cq00b}.\par
%
%----------------------------------------------------------------------------------------------------------
%
An arbitrary element $|\psi\rangle = \sum_{\mu=0}^{\lambda-1} |\psi_{\mu}\rangle$
of~$\cal F$, where $|\psi_{\mu}\rangle \in {\cal F}_{\mu}$, $\mu=0$, 1,
\ldots,~$\lambda-1$, can be expanded in terms of the CS $|z; \mu; 0\rangle$, $\mu=0$,
1, \ldots,~$\lambda - 1$,
\begin{equation}
  |\psi\rangle = \sum_{\mu=0}^{\lambda-1} \int d\rho^{(0)}_{\mu} (z, z^*) 
  \tilde{\psi}^{(0)}_{\mu}(z, z^*)  |z; \mu; 0\rangle,  
\end{equation}
where $\tilde{\psi}^{(0)}_{\mu}(z, z^*)$ is defined in~(\ref{eq:psi-tilde}).\par
%
%xxxxxxxxxxxxxxxxxxxxxxxxxxxxxxxxxxxxxxxxxxxxxxxxxxxxxxxxxxxxxxxxxxxxxxxxxxxx
%
\subsection{\boldmath Unity Resolution Relation for the Coherent States $|z\rangle$}

As mentioned at the end of Subsection~3.2, for values of $\lambda$ greater than two,
the eigenstates $|z\rangle$ of the $C_{\lambda}$-extended oscillator annihilation
operator $a$ may be considered as a generalization of paraboson CS. For the latter,
Sharma {\sl et al.}~\cite{sharma} demonstrated that a diagonal coherent state
resolution of unity, similar to~(\ref{eq:resol-mu}), does not exist, but found instead a
nondiagonal one.\par
%
%-------------------------------------------------------------------------------------------------------
% 
In the present Section, we will extend their approach to the case of the
$C_{\lambda}$-extended oscillator CS and prove that they satisfy a unity resolution
relation, which may be written in either form
\begin{equation}
  \sum_{\mu=0}^{\lambda-1} \int d\rho_{\mu}(z, z^*) |z_{\mu}\rangle
  \langle z_{\mu}| = I  \label{eq:CS2-resol1}
\end{equation}
or 
\begin{equation}
  \sum_{\mu=0}^{\lambda-1} \int d\sigma_{\mu}(z, z^*) |z\rangle
  \left\langle z e^{2\pi {\rm i} \mu/\lambda} \right| = I.  \label{eq:CS2-resol2}
\end{equation}
Here the integration is carried out over the whole complex plane,
\begin{eqnarray}
  d\rho_{\mu}(z, z^*) & = & {\cal N}(|z|) h_{\mu}(t) d^2z, \qquad 
         d\sigma_{\mu}(z, z^*) = {\cal N}(|z|) g_{\mu}(t) d^2z, \label{eq:CS2-drho} \\
  h_{\mu}(t) & = & \lambda^{\lambda} \left(\prod_{\nu=1}^{\mu} \bbeta_{\nu}\right)
         t^{\lambda - \mu - 1} h^{(0)}_{\mu}(t^{\lambda}), \label{eq:hmu} \\
  g_{\mu}(t) & = & \frac{1}{\lambda} \sum_{\nu=0}^{\lambda-1} e^{2\pi {\rm i}
         \mu \nu/\lambda} h_{\nu}(t), \\
  |z_{\mu}\rangle & = & \left(\frac{N^{(0)}_{\mu}(|z|^{\lambda})}
         {{\cal N}(|z|)}\right)^{1/2} \left(\prod_{\nu=1}^{\mu}
         \bbeta_{\nu}\right)^{-1/2} \left(\frac{z}{\sqrt{\lambda}}\right)^{\mu}
         \left| z^{\lambda}; \mu; 0\right\rangle, \label{eq:zmu}    
\end{eqnarray}
where $t = |z|^2/\lambda$, $d^2 z = |z| d|z| d \phi$, and $h^{(0)}_{\mu}(y)$ is
defined in~(\ref{eq:halphamu}). It should be noted that 
\begin{equation}
  |z\rangle = \sum_{\mu=0}^{\lambda-1} |z_{\mu}\rangle  \label{eq:CS2-exp-bis}
\end{equation}
and that the states $|z_{\mu}\rangle \in {\cal F}_{\mu}$, $\mu=0$, 1,
\ldots,~$\lambda-1$, are orthogonal to each other, but are not properly normalized.\par
%
%--------------------------------------------------------------------------------------------------------
%
To prove (\ref{eq:CS2-resol1}), let us start from~(\ref{eq:CS1-resol}), where $z$ is
replaced by $\omega = z^{\lambda}$. By taking~(\ref{eq:CS1-norm})
and~(\ref{eq:CS1-drho}) into account, it can be rewritten as 
\begin{equation}
  \sum_{\mu=0}^{\lambda-1} \int d^2\omega\, h^{(0)}_{\mu}(|\omega|^2/
  \lambda^{\lambda}) N^{(0)}_{\mu}(|\omega|) |\omega; \mu; 0\rangle \langle \omega;
  \mu; 0| = I,  \label{eq:CS1-resol-bis}
\end{equation}
where $d^2\omega = |\omega| d|\omega| d\varphi$. We now make the change of
variable $\omega = z^{\lambda}$, $d^2\omega = \lambda^2 |z|^{2\lambda-2} d^2z$,
and also allow for the fact that $\omega$ covers the complex plane $\lambda$ times
when $z$ covers it once. Equation~(\ref{eq:CS1-resol-bis}) becomes
\begin{equation}
  \lambda \sum_{\mu=0}^{\lambda-1} \int d^2z\, |z|^{2\lambda-2}
  h^{(0)}_{\mu}(|z|^{2\lambda}/\lambda^{\lambda}) N^{(0)}_{\mu}(|z|^{\lambda})
  |z^{\lambda}; \mu; 0\rangle \langle z^{\lambda}; \mu; 0| = I,  
\end{equation}
which directly leads to~(\ref{eq:CS2-resol1}) after expressing $|z^{\lambda}; \mu;
0\rangle$ and $\langle z^{\lambda}; \mu; 0|$ in terms of $|z_{\mu}\rangle$ and
$\langle z_{\mu}|$, respectively, and taking definitions~(\ref{eq:CS2-drho})
and~(\ref{eq:hmu}) into account.\par
%
%------------------------------------------------------------------------------------------------------
%
To transform~(\ref{eq:CS2-resol1}) into~(\ref{eq:CS2-resol2}), we note that
from~(\ref{eq:zmu}) and~(\ref{eq:CS2-exp-bis}), it follows that
\begin{equation}
  \left|z e^{2\pi {\rm i} \mu/\lambda}\right\rangle = \sum_{\nu=0}^{\lambda-1}
  e^{2\pi {\rm i} \mu \nu/\lambda} |z_{\nu}\rangle.
\end{equation}
Inverting this equation leads to
\begin{equation}
  |z_{\nu}\rangle = \frac{1}{\lambda} \sum_{\mu=0}^{\lambda-1}
  e^{-2\pi {\rm i} \mu \nu/\lambda} \left|z e^{2\pi {\rm i} \mu/\lambda}\right\rangle.
  \label{eq:znu-exp}
\end{equation}
By inserting~(\ref{eq:znu-exp}) and its Hermitian conjugate into~(\ref{eq:CS2-resol1}),
we obtain
\begin{equation}
  \frac{1}{\lambda^2} \sum_{\mu, \mu', \nu=0}^{\lambda-1} e^{2\pi {\rm i}
  (\mu' -\mu) \nu/\lambda} \int d^2z\, {\cal N}(|z|) h_{\nu}(t) \left|z e^{2\pi {\rm i}
  \mu/\lambda}\right\rangle \left\langle z e^{2\pi {\rm i} \mu'/\lambda}\right| = I.  
\end{equation}
It is now straightforward to get~(\ref{eq:CS2-resol2}) by changing $z$ into $z' e^{-2\pi
{\rm i} \mu/\lambda}$, $\mu' - \mu$ into $\mu'' {\rm mod} \lambda$, and by
performing the summation over~$\mu$.\par
%
%--------------------------------------------------------------------------------------------------------
% 
We have therefore shown that the diagonal coherent state resolution of unity valid for
the set of CS $\{\, |z; \mu; 0\rangle | \mu=0, 1, \ldots, \lambda-1\, \}$ is entirely
equivalent to a nondiagonal one for the CS $|z\rangle$. Via a change of variable $z \to
\omega$ and a redefinition of the component states $|z^{\lambda}; \mu; 0\rangle
\to |z_{\mu}\rangle \to \left|z e^{2\pi {\rm i} \mu/\lambda}\right\rangle$, the set
of positive weight functions $\{\,h^{(0)}_{\mu}(y) | \mu=0, 1, \ldots, \lambda-1\, \}$,
has been converted into a set of linear combinations $\{\, g_{\mu}(t) | \mu=0, 1,
\ldots, \lambda-1\, \}$ with complex coefficients of well-behaved positive functions.
Although this type of weight functions is unusual, the probabilistic interpretation of the
weight is not lost provided one keeps the above-mentioned equivalence in mind.\par
%
%----------------------------------------------------------------------------------------------------
%
As in Subsubsection 4.1.2, any vector of~$\cal F$ can be expanded in terms of the CS
$|z\rangle$,
\begin{equation}
  |\psi\rangle = \sum_{\mu=0}^{\lambda-1} \int d\sigma_{\mu}(z, z^*) \tilde{\psi}
  \left(z e^{2\pi {\rm i} \mu/\lambda}, z^* e^{-2\pi {\rm i} \mu/\lambda}\right) 
  |z\rangle, 
\end{equation}
where
\begin{equation}
  \tilde{\psi}(z, z^*) = \langle z|\psi\rangle.
\end{equation}
Equivalently, we may use~(\ref{eq:CS2-resol1}) to express $|\psi\rangle$ as
\begin{equation}
  |\psi\rangle = \sum_{\mu=0}^{\lambda-1} \int d\rho_{\mu}(z, z^*)
  \tilde{\psi}_{\mu}(z, z^*) |z_{\mu}\rangle,   
\end{equation}
where
\begin{equation}
  \tilde{\psi}_{\mu}(z, z^*) = \langle z_{\mu}|\psi \rangle. 
\end{equation}
\par
%
%--------------------------------------------------------------------------------------------------------
%
{}For $\lambda=2$, the unity resolution relations~(\ref{eq:CS2-resol1})
and~(\ref{eq:CS2-resol2}) become
\begin{equation}
  \int d\rho_0(z, z^*) |z_0\rangle \langle z_0| + \int d\rho_1(z, z^*) |z_1\rangle 
  \langle z_1| = I 
\end{equation}
and
\begin{equation}
  \int d\sigma_0(z, z^*) |z\rangle \langle z| + \int d\sigma_1(z, z^*) |z\rangle 
  \langle -z| = I, 
\end{equation}
where
\begin{equation}
  |z_0\rangle = \case{1}{2} (|z\rangle + |-z\rangle), \qquad |z_1\rangle = \case{1}{2}
  (|z\rangle - |-z\rangle). 
\end{equation}
In such a case, ${\cal N}(|z|)$, $h_{\mu}(t)$, and $g_{\mu}(t)$ can be rewritten in
terms of modified Bessel functions $I_{\nu}$ and $K_{\nu}$ as
\begin{eqnarray}
  {\cal N}(|z|) & = & 2^{\bbeta_1 - 1} \Gamma(\bbeta_1) |z|^{2(1 - \bbeta_1)} \left[
        I_{\bbeta_1 - 1}(|z|^2) + I_{\bbeta_1}(|z|^2)\right], \\
  h_0(t) & = & 4t\, h^{(0)}_0(t^2) = \left[2^{\bbeta_1 - 1} \pi \Gamma(\bbeta_1)
        \right]^{-1} |z|^{2\bbeta_1} K_{\bbeta_1 - 1}(|z|^2), \\
  h_1(t) & = & 4\bbeta_1 h^{(0)}_1(t^2) = \left[2^{\bbeta_1 - 1} \pi
        \Gamma(\bbeta_1) \right]^{-1} |z|^{2\bbeta_1} K_{\bbeta_1}(|z|^2), \\
  g_0(t) & = & \case{1}{2} [h_0(t) + h_1(t)] = \left[2^{\bbeta_1} \pi
        \Gamma(\bbeta_1)\right]^{-1} |z|^{2\bbeta_1} \left[K_{\bbeta_1 - 1}(|z|^2) +
        K_{\bbeta_1}(|z|^2)\right], \\
  g_1(t) & = & \case{1}{2} [h_0(t) - h_1(t)] = \left[2^{\bbeta_1} \pi
        \Gamma(\bbeta_1)\right]^{-1} |z|^{2\bbeta_1} \left[K_{\bbeta_1 - 1}(|z|^2) -
        K_{\bbeta_1}(|z|^2)\right],   
\end{eqnarray}
thus giving for $d\rho_{\mu}(z, z^*)$ and $d\sigma_{\mu}(z, z^*)$ the results
previously obtained for paraboson CS~\cite{sharma},
\begin{eqnarray}
  d\rho_{\mu}(z, z^*) & = & \frac{|z|^2}{\pi} \left[I_{\bbeta_1 - 1}(|z|^2) +
  I_{\bbeta_1}(|z|^2)\right] K_{\bbeta_1 - 1 + \mu}(|z|^2) d^2z, \\
  d\sigma_{\mu}(z, z^*) & = & \frac{|z|^2}{2\pi} \left[I_{\bbeta_1 - 1}(|z|^2) +
  I_{\bbeta_1}(|z|^2)\right] \left[K_{\bbeta_1 - 1}(|z|^2) + (-1)^{\mu}
  K_{\bbeta_1}(|z|^2)\right] d^2z. 
\end{eqnarray}
\par
%
%=============================================================
%
\section{BARGMANN REPRESENTATIONS}

\setcounter{equation}{0}

It is well known that for the harmonic oscillator, it is possible to find a realization of the
Fock space wherein any vector is described by an entire function: this is the so-called
Bargmann representation~\cite{bargmann} associated with the harmonic oscillator CS.
Similar realizations of other Lie algebras in Hilbert spaces of entire analytic functions have
also been found in connection with various types of CS~\cite{perelomov86,zhang}. Here
we shall construct such realizations for the CS considered in the previous Sections.\par
%
%xxxxxxxxxxxxxxxxxxxxxxxxxxxxxxxxxxxxxxxxxxxxxxxxxxxxxxxxxxxxxxxxxxxxxxxxxxxx
%
\subsection{\boldmath Bargmann Representations Corresponding to the Coherent States
$|z; \mu; \alpha\rangle$}

\subsubsection{\boldmath Bargmann Representations of ${\cal F}_{\mu}$}

As reviewed in Sec.~2, any subspace ${\cal F}_{\mu}$ of $\cal F$ carries a unirrep of
the $C_{\lambda}$-extended oscillator SGA, characterized by the eigenvalue $c_{\mu}$
of $C$ and by the lowest eigenvalue $\left(\mu + \gamma_{\mu} + \frac{1}{2}\right)/
\lambda = (\bbeta_{\mu} + \bbeta_{\mu+1})/2$ of $J_0$. Furthermore, in
Subsubsec.~4.1.1, it was proved that for any $\alpha \le \lambda - \mu - 1$, the states
$|z; \mu; \alpha\rangle$ form a complete (in fact, an overcomplete) set in ${\cal
F}_{\mu}$ for some appropriate choice of the algebra parameters if $z$ runs over the
complex plane for $\alpha \le [(\lambda - 1)/2]$ or over the unit disc for $\alpha =
[\lambda/2]$ and $\lambda$ even. We shall now proceed to show that with any such
set, parametrized by a given $\alpha$, we can associate a realization of ${\cal
F}_{\mu}$ as a space ${\cal B}^{(\alpha)}_{\mu}$ of entire analytic functions, wherein
the generators of the $C_{\lambda}$-extended oscillator SGA become differential
operators.\par
%
%-----------------------------------------------------------------------------------------------------------
% 
{}For such a purpose, it is convenient to replace the normalized CS $|z; \mu;
\alpha\rangle$ by unnormalized ones (denoted by a round bracket instead of an angular
one),
\begin{equation}
  |z; \mu; \alpha) \equiv \left[N^{(\alpha)}_{\mu}(|z|)\right]^{1/2} |z; \mu;
  \alpha\rangle,  
\end{equation}
and to substitute $z^*$ for $z$. Then any vector $|\psi_{\mu}\rangle \in {\cal
F}_{\mu}$ can be realized by the entire function
\begin{eqnarray}
  \psi^{(\alpha)}_{\mu}(z) & = & (z^*; \mu; \alpha | \psi\rangle \nonumber \\
  & =& \sum_{k=0}^{\infty} \left(\frac{\prod_{\nu=\mu+1}^{\mu+\alpha}
         (\bbeta_{\nu})_k}{k! \left(\prod_{\nu=1}^{\mu} (\bbeta_{\nu} + 1)_k\right)
         \left(\prod_{\nu=\mu+\alpha+1}^{\lambda-1} (\bbeta_{\nu})_k\right)}
         \right)^{1/2} \nonumber \\
  && \mbox{} \times \langle k\lambda + \mu | \psi_{\mu}\rangle 
         \left(\lambda^{-(\lambda - 2\alpha)/2} z\right)^k,  \label{eq:B1-psi}
\end{eqnarray}
related to $\tilde{\psi}^{(\alpha)}_{\mu}(z, z^*)$, defined in~(\ref{eq:psi-tilde}), by
\begin{equation}
  \tilde{\psi}^{(\alpha)}_{\mu}(z, z^*) = \left[N^{(\alpha)}_{\mu}(|z|)\right]^{-1/2}
  \psi^{(\alpha)}_{\mu}(z^*).   
\end{equation}
\par
%
%----------------------------------------------------------------------------------------------------------
%
The functions~(\ref{eq:B1-psi}) are the elements of a Hilbert space ${\cal
B}^{(\alpha)}_{\mu}$. Its scalar product is defined by
\begin{equation}
  \left(\psi^{\prime(\alpha)}_{\mu}, \psi^{(\alpha)}_{\mu}\right) = \int d^2z
\,  h^{(\alpha)}_{\mu}(y) \left[\psi^{\prime(\alpha)}_{\mu}(z)\right]^* 
  \psi^{(\alpha)}_{\mu}(z), \qquad y = |z|^2/ \lambda^{\lambda - 2\alpha},
  \label{eq:B1-sp}  
\end{equation}
where the integral is extended over the complex plane or the unit disc according to the
value of $\alpha$. For any function $\psi^{(\alpha)}_{\mu}(z) \in {\cal
B}^{(\alpha)}_{\mu}$, the condition $\left(\psi^{(\alpha)}_{\mu},
\psi^{(\alpha)}_{\mu}\right) < \infty$ is satisfied.\par
%
%-------------------------------------------------------------------------------------------------------
%
The functions
\begin{eqnarray}
  \varphi^{(\alpha)}_{\mu,k}(z) & = & (z^*; \mu; \alpha | k\lambda + \mu\rangle
         \nonumber \\
  & =& \left(\frac{\prod_{\nu=\mu+1}^{\mu+\alpha}
         (\bbeta_{\nu})_k}{k! \left(\prod_{\nu=1}^{\mu} (\bbeta_{\nu} + 1)_k\right)
         \left(\prod_{\nu=\mu+\alpha+1}^{\lambda-1} (\bbeta_{\nu})_k\right)}
         \right)^{1/2} \left(\lambda^{-(\lambda - 2\alpha)/2} z\right)^k,  
\end{eqnarray}
representing the number-state vectors $|k\lambda + \mu\rangle$, $k=0$, 1,~\ldots,
form an orthonormal basis of ${\cal B}^{(\alpha)}_{\mu}$. Equation~(\ref{eq:B1-psi})
can therefore be rewritten as
\begin{equation}
   \psi^{(\alpha)}_{\mu}(z) = \sum_{k=0}^{\infty} \langle k\lambda + \mu | \psi_{\mu}
  \rangle  \varphi^{(\alpha)}_{\mu,k}(z).  \label{eq:B1-psi-exp}
\end{equation}
\par
%
%-------------------------------------------------------------------------------------------------------
%
Any operator $X$ defined in ${\cal F}_{\mu}$ is represented in ${\cal
B}^{(\alpha)}_{\mu}$ by some differential operator ${\cal X}^{(\alpha)}_{\mu}$,
defined by
\begin{equation}
  (z^*; \mu; \alpha | X | \psi_{\mu}\rangle = {\cal X}^{(\alpha)}_{\mu}
  \psi^{(\alpha)}_{\mu}(z) 
\end{equation}
for any $| \psi_{\mu}\rangle \in {\cal F}_{\mu}$. For the number operator $N$, we get
\begin{equation}
  {\cal N}^{(\alpha)}_{\mu} = \lambda z \frac{\partial}{\partial z} + \mu.
  \label{eq:B1-N}
\end{equation}
\par
%
%-------------------------------------------------------------------------------------------------------------
%
Similarly, for the generators $J_+$, $J_-$, and $J_0$ of the $C_{\lambda}$-extended
oscillator SGA, we obtain
\begin{eqnarray}
  {\cal J}^{(\alpha)}_{+\mu} & = & \lambda^{\alpha-1} z 
         \prod_{\nu=\mu+1}^{\mu+\alpha} \left(z \frac{\partial}{\partial z} +
         \bbeta_{\nu}\right),  \label{eq:B1-J+} \\
  {\cal J}^{(\alpha)}_{-\mu} & = & \lambda^{\lambda-\alpha-1} \left[ 
         \prod_{\nu=1}^{\mu} \left(z \frac{\partial}{\partial z} + \bbeta_{\nu} + 1\right) 
         \right] \left[\prod_{\nu=\mu+\alpha+1}^{\lambda-1} \left(z
         \frac{\partial}{\partial z} + \bbeta_{\nu}\right)\right] \frac{\partial}{\partial z}, 
         \label{eq:B1-J-} \\
  {\cal J}^{(\alpha)}_{0\mu} & = & z \frac{\partial}{\partial z} + \frac{1}{2} 
         (\bbeta_{\mu} + \bbeta_{\mu+1}).  \label{eq:B1-J0}
\end{eqnarray}
In deriving these equations, use has been made of the expansion~(\ref{eq:B1-psi-exp})
and of the known action of $N$, $\left(\ap\right)^{\lambda}$, $a^{\lambda}$, and
$H_0$ on $|k\lambda + \mu\rangle$ (see (\ref{eq:alg-action}), (\ref{eq:CS1-inter2}),
(\ref{eq:CS1-inter1}) and~(\ref{eq:E}), respectively). It is straightforward to check that
the differential operators (\ref{eq:B1-J+})--(\ref{eq:B1-J0}) fulfil the commutation
relations~(\ref{eq:SGA-com}) characterizing the $C_{\lambda}$-extended oscillator SGA.
In Appendix~C, it is proved that they also satisfy the Hermiticity
properties~(\ref{eq:SGA-Herm}) with respect to the scalar product~(\ref{eq:B1-sp})
of~${\cal B}^{(\alpha)}_{\mu}$.\par
%
%---------------------------------------------------------------------------------------------------------
%
It should be noted that in the $\alpha=0$ case, ${\cal J}^{(0)}_{+\mu}$ is essentially
the operator that acts on ${\cal B}^{(0)}_{\mu}$as a multiplication by $z$,
whereas ${\cal J}^{(0)}_{-\mu}$ is a $\lambda$th-order differential operator. If in
addition $\lambda=2$, we get back the well-known realization of su(1,1) connected with
the Barut-Girardello CS~\cite{barut}, corresponding to the unirrep characterized by
$\frac{1}{2} (\bbeta_{\mu} + \bbeta_{\mu+1}) = \frac{1}{2} (\bbeta_1 + \mu)$ and
$c_{\mu}$, 
\begin{equation}
  {\cal J}^{(0)}_{+\mu} = \frac{1}{2} z, \qquad {\cal J}^{(0)}_{-\mu} = 2 \left(z
  \frac{\partial}{\partial z} + \bbeta_1 + \mu\right) \frac{\partial}{\partial z}, \qquad
  {\cal J}^{(0)}_{0\mu} = z \frac{\partial}{\partial z} + \frac{1}{2} \left(\bbeta_1 +
  \mu\right).  \label{eq:B1-BG} 
\end{equation}
\par
%
%----------------------------------------------------------------------------------------------------------
%
Similarly, for $\lambda=2$, $\alpha=1$, and $\mu=0$, ${\cal J}^{(1)}_{-0}$ is the
operator that acts on ${\cal B}^{(1)}_0$ as a derivation $\partial/\partial z$. This
corresponds to the realization of su(1,1) connected with the Perelomov
CS~\cite{perelomov86}, corresponding to the unirrep specified by $\frac{1}{2}
\bbeta_1$ and $c_0$,
\begin{equation}
  {\cal J}^{(1)}_{+0} = z \left(z \frac{\partial}{\partial z} + \bbeta_1\right),
  \qquad {\cal J}^{(1)}_{-0} = \frac{\partial}{\partial z}, \qquad
  {\cal J}^{(1)}_{00} = z \frac{\partial}{\partial z} + \frac{1}{2} \bbeta_1. 
  \label{eq:B1-Pere} 
\end{equation}
\par
%
%+++++++++++++++++++++++++++++++++++++++++++++++++++++++++++
%
\subsubsection{\boldmath Bargmann Representation of $\cal F$}

{}For $\alpha=0$, we may realize the whole Fock space ${\cal F} =
\sum_{\mu=0}^{\lambda-1} \oplus {\cal F}_{\mu}$ as a Hilbert space ${\cal B}^{(0)} =
\sum_{\mu=0}^{\lambda-1} \oplus {\cal B}^{(0)}_{\mu}$ of entire analytic functions. It
is then convenient to introduce vector CS~\cite{zhang, deenen}, defined as row vectors
\begin{equation}
  ||z; 0)) = \Bigl(|z; 0; 0), |z; 1; 0), \ldots, |z; \lambda-1; 0)\Bigr),  \label{eq:CS1-row}
\end{equation}
the corresponding bras being column vectors
\begin{equation}
  ((z; 0|| = \left(\begin{array}{c}
          (z; 0; 0| \\
          (z; 1; 0| \\
          \vdots \\
          (z; \lambda-1; 0|
  \end{array}\right).  \label{eq:CS1-column}
\end{equation}
The unity resolution relation in $\cal F$, given in~(\ref{eq:CS1-resol}), can be rewritten as
\begin{equation}
  \int d^2z\,  ||z; 0)) h^{(0)}(y) ((z; 0|| = I,  
\end{equation}
where $h^{(0)}(y) = {\rm diag}\left(h^{(0)}_0 (y), h^{(0)}_1(y), \ldots,
h^{(0)}_{\lambda-1}(y)\right)$ is a diagonal $\lambda \times \lambda$ matrix.\par
%
%----------------------------------------------------------------------------------------------------------
%
Any vector $|\psi\rangle = \sum_{\mu=0}^{\lambda-1} |\psi_{\mu}\rangle \in {\cal F}$
is realized by a column of entire functions
\begin{equation}
  \psi^{(0)}(z) = ((z^*; 0 || \psi\rangle = \left(\begin{array}{c}
         \psi^{(0)}_0(z) \\
         \psi^{(0)}_1(z) \\
         \vdots \\
         \psi^{(0)}_{\lambda-1}(z)  
  \end{array}\right),  \label{eq:B2-psi}
\end{equation}
and the scalar product in ${\cal B}^{(0)}$ reads
\begin{equation}
  \left(\psi^{\prime(0)}, \psi^{(0)}\right) = \int d^2z\, \left[\psi^{\prime(0)}(z)\right]
  ^{\dagger} h^{(0)}(y) \psi^{(0)}(y).  \label{eq:B2-sp}
\end{equation}
\par
%
%-----------------------------------------------------------------------------------------------------------
%
Any operator $X$ defined in $\cal F$ is represented in ${\cal B}^{(0)}$ by an
operator-valued $\lambda \times \lambda$ matrix ${\cal X}^{(0)}$, defined by
\begin{equation}
  ((z^*; 0||X|\psi\rangle = {\cal X}^{(0)} \psi^{(0)}(z).  \label{eq:B2-X}
\end{equation}
\par
%
%-----------------------------------------------------------------------------------------------
%
If $X$ does not produce transitions between different subspaces ${\cal F}_{\mu}$, then
${\cal X}^{(0)}$ is a diagonal matrix. Such is the case for the projection operators
$P_{\mu}$, which are represented by
\begin{equation}
  {\cal P}_{\mu} = e_{\mu+1,\mu+1},  \label{eq:B2-Pmu}
\end{equation}
where $e_{\mu \nu}$ denotes the matrix with entry one in row $\mu$ and column
$\nu$, and zeros everywhere else. The diagonal matrix ${\cal X}^{(0)}$ may then be
written as ${\cal X}^{(0)} = \sum_{\mu} {\cal X}^{(0)}_{\mu} {\cal P}_{\mu}$, where
${\cal X}^{(0)}_{\mu}$ is the operator representing in ${\cal B}^{(0)}_{\mu}$ the
restriction of $X$ to ${\cal F}_{\mu}$. In particular, for the generators of the
$C_{\lambda}$-extended oscillator SGA, we find
\begin{equation}
  {\cal J}^{(0)}_q = \sum_{\mu} {\cal J}^{(0)}_{q\mu} {\cal P}_{\mu}, \qquad q = +, -,
  0,  \label{eq:B2-J} 
\end{equation}
where ${\cal J}^{(0)}_{q\mu}$, $q=+, -, 0$, are given in
(\ref{eq:B1-J+})--(\ref{eq:B1-J0}). Similarly, the number operator $N$ is represented by
\begin{equation}
  {\cal N}^{(0)} = \sum_{\mu} {\cal N}^{(0)}_{\mu} {\cal P}_{\mu}, \label{eq:B2-N}
\end{equation}
where ${\cal N}^{(0)}_{\mu}$ is defined in~(\ref{eq:B1-N}).\par
%
%--------------------------------------------------------------------------------------------------------
%
On the contrary, operators giving rise to transitions between different subspaces ${\cal
F}_{\mu}$ are represented by nondiagonal operator-valued matrices. For the creation
and annihilation operators of the $C_{\lambda}$-extended oscillator, for instance, we
obtain the matrices
\begin{equation}
  {\cal A}^{\dagger (0)} = \left(\begin{array}{ccccc}
     0 & 0 & \ldots & 0 & z/\sqrt{\lambda^{\lambda-1} \bbeta_1 \bbeta_2 \ldots
          \bbeta_{\lambda-1}} \\[0.2cm]
     \sqrt{\lambda \bbeta_1} & 0 & \ldots & 0 & 0  \\[0.2cm]
     0 & \sqrt{\lambda \bbeta_2} & \ldots & 0 & 0  \\[0.2cm]
     \vdots & \vdots & \ddots & \vdots & \vdots  \\[0.2cm]
     0 & 0 & \ldots & \sqrt{\lambda \bbeta_{\lambda-1}} & 0
  \end{array}\right),  \label{eq:B2-aplus}
\end{equation}
\begin{equation}
  {\cal A}^{(0)} = \left(\begin{array}{ccccc}
     \ss 0 & \ss\sqrt{\lambda/\bbeta_1} \left(z \frac{\partial}{\partial z} +
          \bbeta_1\right) & \ss 0 & \ldots & \ss 0 \\[0.2cm]
     \ss 0 & \ss 0 & \ss \sqrt{\lambda/\bbeta_2} \left(z \frac{\partial}{\partial z} +
          \bbeta_2\right) & \ldots & \ss 0  \\[0.2cm]
     \vdots & \vdots & \vdots & \ddots & \vdots  \\[0.2cm]
     \ss 0 & \ss 0 & \ss 0 & \ldots & \ss \sqrt{\lambda/\bbeta_{\lambda-1}} \left(z
          \frac{\partial}{\partial z} + \bbeta_{\lambda-1}\right)  \\[0.2cm]
     \ss \sqrt{\lambda^{\lambda+1} \bbeta_1 \bbeta_2 \ldots \bbeta_{\lambda-1}}\,
          \frac{\partial}{\partial z} & \ss 0 & \ss 0 & \ldots & \ss 0
  \end{array}\right),  \label{eq:B2-a}
\end{equation}
respectively. It is straightforward to check that the operator-valued
matrices~(\ref{eq:B2-Pmu}) and (\ref{eq:B2-N})--(\ref{eq:B2-a}) satisfy the
commutation relations~(\ref{eq:alg-def}) of the $C_{\lambda}$-extended oscillator
algebra. The more difficult check of their Hermiticity properties with respect to the scalar
product~(\ref{eq:B2-sp}) of ${\cal B}^{(0)}$ is reviewed in Appendix~C.\par
%
%----------------------------------------------------------------------------------------------------------
%
{}From (\ref{eq:B2-N})--(\ref{eq:B2-a}), it is easy to recover (\ref{eq:B2-J}) by using
the relations
\begin{equation}
  {\cal J}^{(0)}_+ = \frac{1}{\lambda} \left({\cal A}^{\dagger(0)}\right)^{\lambda},
  \qquad {\cal J}^{(0)}_- = \frac{1}{\lambda} \left({\cal A}^{(0)}\right)^{\lambda}, 
  \qquad {\cal J}^{(0)}_0 = \frac{1}{\lambda} \left[{\cal N}^{(0)} + \frac{1}{2}
\sum_{\mu} (\beta_{\mu} + \beta_{\mu+1} + 1) {\cal P}_{\mu}\right],
\end{equation}
corresponding to~(\ref{eq:SGA-gen}).\par
%
%xxxxxxxxxxxxxxxxxxxxxxxxxxxxxxxxxxxxxxxxxxxxxxxxxxxxxxxxxxxxxxxxxxxxxxxxxxxx
%
\subsection{\boldmath Bargmann Representation Corresponding to the Coherent States
$|z\rangle$}

In $\cal F$, we may alternatively use the unnormalized CS
\begin{equation}
  |z) \equiv [{\cal N}(|z|)]^{1/2} |z\rangle = \sum_{\mu} |z_{\mu}), \qquad |z_{\mu})
  \equiv [{\cal N}(|z|)]^{1/2} |z_{\mu}\rangle. 
\end{equation}
Employing a vector notation again, let us define
\begin{equation}
  ||z)) = \Bigl(|z_0), |z_1), \ldots, |z_{\lambda-1})\Bigr), \qquad ((z|| = \left(
  \begin{array}{c}
       (z_0| \\
       (z_1| \\
       \vdots \\
       (z_{\lambda-1}|
  \end{array}\right),  \label{eq:CS2-row-column}
\end{equation}
which are related to the row and column vectors~(\ref{eq:CS1-row})
and~(\ref{eq:CS1-column}) through the relations
\begin{equation}
  ||z)) = ||\omega; 0)) D(z), \qquad ((z|| = D(z^*) ((\omega; 0||.
\end{equation}
Here $D(z)$ is a $\lambda \times \lambda$ diagonal matrix, given by
\begin{eqnarray}
  D(z) & = & {\rm diag}(D_0(z), D_1(z), \ldots, D_{\lambda-1}(z)), \nonumber \\
  D_{\mu}(z) & = & 
  \left(\prod_{\nu=1}^{\mu} \bbeta_{\nu}\right)^{-1/2} \left(\frac{z}{\sqrt{\lambda}}
  \right)^{\mu}, \qquad \mu=0, 1, \ldots, \lambda-1.
\end{eqnarray}
The unity resolution relation in $\cal F$, given in~(\ref{eq:CS2-resol1}), can be rewritten
as
\begin{equation}
  \int d^2z ||z)) h(t) ((z|| = I,
\end{equation}
where $h(t) = {\rm diag}(h_0(t), h_1(t), \ldots, h_{\lambda-1}(t))$.\par
%
%------------------------------------------------------------------------------------------------------
%
By using the new vector CS~(\ref{eq:CS2-row-column}), we get a new realization of~$\cal
F$ as a space~$\cal B$ of entire analytic functions. Any vector $|\psi\rangle \in {\cal F}$
is realized by a column of entire functions
\begin{equation}
  \psi(z) = ((z^*||\psi\rangle = D(z) \psi^{(0)}(\omega),
\end{equation}
where $\psi^{(0)}(\omega)$ is defined in~(\ref{eq:B2-psi}). The scalar product in~$\cal
B$ reads
\begin{equation}
  \left(\psi', \psi\right) = \int d^2z\, [\psi'(z)]^{\dagger} h(t) \psi(z).
\end{equation}
\par
%
%----------------------------------------------------------------------------------------------------------
%
Any operator $X$ defined in $\cal F$ is represented in~$\cal B$ by an operator-valued
matrix $\cal X$, defined by
\begin{equation}
  ((z^*||X|\psi\rangle = {\cal X} \psi(z),
\end{equation}
and given by
\begin{equation}
  {\cal X} = D(z) {\cal X}^{(0)} D^{-1}(z)
\end{equation}
in terms of ${\cal X}^{(0)}$ of Eq.~(\ref{eq:B2-X}).\par
%
%---------------------------------------------------------------------------------------------------------
%
{}From (\ref{eq:B1-J+})--(\ref{eq:B1-J0}) and (\ref{eq:B2-J})--(\ref{eq:B2-a}), it is
straightforward to derive the following results:
\begin{eqnarray}
  {\cal J}_q & = & \sum_{\mu} {\cal J}_{q\mu} {\cal P}_{\mu}, \qquad q=+, -, 0, \\
  {\cal J}_{+\mu} & = & \frac{1}{\lambda} z^{\lambda}, \\
  {\cal J}_{-\mu} & = & \frac{1}{\lambda} \left[\prod_{\nu=\mu+1}^{\lambda-1}
        \left(\frac{\partial}{\partial z} + \frac{\beta_{\nu}}{z}\right)\right]
        \frac{\partial}{\partial z} \left[\prod_{\nu=1}^{\mu}
        \left(\frac{\partial}{\partial z} + \frac{\beta_{\nu}}{z}\right)\right] , \\
  {\cal J}_{0\mu} & = & \frac{1}{\lambda} \left(z \frac{\partial}{\partial z} +
        \frac{1}{2} (\beta_{\mu} + \beta_{\mu+1} + 1)\right), \\
  {\cal N} & = & \sum_{\mu} {\cal N}_{\mu} {\cal P}_{\mu}, \qquad {\cal N}_{\mu} =
        z \frac{\partial}{\partial z}, \\
  {\cal A}^{\dagger} & = & \left(\begin{array}{ccccc}
          0 & 0 & \ldots & 0 & z \\[0.2cm]
          z & 0 & \ldots & 0 & 0 \\[0.2cm]
          0 & z & \ldots & 0 & 0 \\[0.2cm]
          \vdots & \vdots & \ddots & \vdots & \vdots \\[0.2cm]
          0 & 0 & \ldots & z & 0
        \end{array}\right), \\[0.2cm]
  {\cal A} & = & \left(\begin{array}{ccccc}
          0 & \frac{\partial}{\partial z} + \frac{\beta_1}{z} & 0 & \ldots & 0 \\[0.2cm]
          0 & 0 & \frac{\partial}{\partial z} + \frac{\beta_2}{z} & \ldots & 0 \\[0.2cm]
          \vdots & \vdots & \vdots & \ddots & \vdots \\[0.2cm]
          0 & 0 & 0 & \ldots & \frac{\partial}{\partial z} + \frac{\beta_{\lambda-1}}{z}
                \\[0.2cm]
          \frac{\partial}{\partial z} & 0 & 0 & \ldots & 0
        \end{array}\right). 
\end{eqnarray}
In particular, for $\lambda=2$, we get back known results for paraboson
CS~\cite{sharma}:
\begin{equation}
  {\cal N} = \left(\begin{array}{cc}
          z \frac{\partial}{\partial z} & 0 \\[0.2cm]
          0 & z \frac{\partial}{\partial z}
       \end{array}\right), \qquad
  {\cal A}^{\dagger} = \left(\begin{array}{cc}
          0 & z \\[0.2cm]
          z & 0
       \end{array}\right), \qquad
  {\cal A} = \left(\begin{array}{cc}
          0 & \frac{\partial}{\partial z} + \frac{\beta_1}{z}\\[0.2cm]
          \frac{\partial}{\partial z} & 0
       \end{array}\right).
\end{equation}
%
%=============================================================
% 
\section{PHYSICAL APPLICATIONS}

\setcounter{equation}{0}

In this Section, we shall investigate the statistical and squeezing properties of the
$C_{\lambda}$-extended oscillator CS $|z; \mu; \alpha\rangle$ and $|z\rangle$ with
special emphasis on the comparison with those of the standard harmonic oscillator, which
are retrieved for vanishing parameters~$\alpha_{\mu}$.\par
%
%xxxxxxxxxxxxxxxxxxxxxxxxxxxxxxxxxxxxxxxxxxxxxxxxxxxxxxxxxxxxxxxxxxxxxxxxxxx
%
\subsection{Photon Statistics}

For the conventional harmonic oscillator CS~\cite{glauber, klauder63, sudarshan}, the
probability distribution of the photon number is Poissonian. Experimentally, however, the
photon number statistics of real lasers is not exactly Poissonian~\cite{perina}.
Furthermore, various nonlinear interactions give rise to deviations from the Poisson
distribution~\cite{kimble, short}. A measure of such deviations is the Mandel
parameter~\cite{mandel}
\begin{equation}
  Q = \frac{(\Delta N_b)^2 - \langle N_b \rangle}{\langle N_b \rangle}, \qquad
  (\Delta N_b)^2 \equiv \langle N_b^2 \rangle - \langle N_b \rangle^2,
  \label{eq:mandel} 
\end{equation}
which vanishes for the Poisson distribution, is positive for a super-Poissonian distribution
(bunching effect), and negative for a sub-Poissonian distribution (antibunching effect). 
As mentioned in Sec.~1, in the present case we have $N_b = N$.\par
%
%++++++++++++++++++++++++++++++++++++++++++++++++++++++++++++
%
\subsubsection{\boldmath Statistical Properties of the Coherent States $|z; \mu;
\alpha\rangle$}

By using the Bargmann representation ${\cal N}^{(\alpha)}_{\mu}$ of $N$, given
in~(\ref{eq:B1-N}), the averages of $N$ and $N^2$ in the states $|z; \mu;
\alpha\rangle$ can be expressed in terms of derivatives of the normalization coefficients
$N^{(\alpha)}_{\mu}(|z|)$,
\begin{eqnarray}
  \langle z; \mu; \alpha | N^r |z; \mu; \alpha\rangle & = & [N^{(\alpha)}_{\mu}(|z|)]^{-1}
         \left(\lambda z^* \frac{\partial}{\partial z^*} + \mu\right)^r  
         N^{(\alpha)}_{\mu}(|z|) \nonumber \\
  & = & [N^{(\alpha)}_{\mu}(|z|)]^{-1}
         \left(\lambda y \frac{\partial}{\partial y} + \mu\right)^r  
         N^{(\alpha)}_{\mu}(|z|),  \label{eq:Nr}
\end{eqnarray}
where $r=1$, 2, and $y = |z|^2/\lambda^{\lambda - 2\alpha}$.\par
%
%-----------------------------------------------------------------------------------------------------------
%
{}For $\lambda=2$, $\alpha=1$ (and $\mu=0$), i.e., for the Perelomov su(1,1) CS given
in~(\ref{eq:Pere-CS}), the normalization coefficient $N^{(1)}_0(|z|)$ takes a very simple
form. The same is true for the Mandel parameter, which is independent of the
parameter~$\alpha_0$ and is given by
\begin{equation}
  Q = \frac{1+y}{1-y},
\end{equation}
where $y = |z|^2$. The distribution is then super-Poissonian as $1 < Q < \infty$ for $0 <
y < 1$.\par
%
%-----------------------------------------------------------------------------------------------------------
%
{}For the remaining CS $|z; \mu; \alpha\rangle$, the Mandel parameter can be written as
\begin{eqnarray}
  Q & = & \lambda \left[1 - \bbeta_{\lambda-1} - \left(\prod_{\nu=1}^{\alpha}
        \bbeta_{\nu}\right) \left(\prod_{\nu=\alpha+1}^{\lambda-1} \bbeta_{\nu} 
        \right)^{-1} y\, \Phi^{\lambda-1}_0(y; 0; \alpha) + \bbeta_{\lambda-1}
        \Phi^{\lambda-2}_{\lambda-1}(y; 0; \alpha)\right] \nonumber \\
  & & \mbox{} - 1 \qquad {\rm if\ } \mu=0, \nonumber \\
  & = & \left[\lambda^{-1} - \bbeta_1 + \bbeta_1 \Phi^0_1(y; 1; \alpha)\right]^{-1}
        \Biggl\{ \left(\bbeta_1 - \lambda^{-1}\right) \left[1 + \lambda \bbeta_1 
        \Phi^0_1(y; 1; \alpha)\right] \nonumber \\
  && \mbox{} - \lambda \bbeta_1^2 \left[\Phi^0_1(y; 1; \alpha)\right]^2 + \lambda
        \left(\prod_{\nu=2}^{\alpha+1} \bbeta_{\nu}\right)
        \left(\prod_{\nu=\alpha+2}^{\lambda-1} \bbeta_{\nu} \right)^{-1} y\,
        \Phi^{\lambda-1}_1(y; 1; \alpha)\Biggr\} \qquad {\rm if\ } \mu=1, \nonumber \\
  & = & \left[\mu \lambda^{-1} - \bbeta_{\mu} + \bbeta_{\mu} \Phi^{\mu-1}_{\mu}(y;
        \mu; \alpha)\right]^{-1} \Biggl\{ \bbeta_{\mu} - \mu \lambda^{-1} + \lambda
        \bbeta_{\mu} \left(\bbeta_{\mu} - \bbeta_{\mu-1} - \lambda^{-1}\right)
        \Phi^{\mu-1}_{\mu}(y; \mu; \alpha) \nonumber \\
  && \mbox{} - \lambda \bbeta_{\mu}^2 \left[\Phi^{\mu-1}_{\mu}(y; \mu;
        \alpha)\right]^2 + \lambda \bbeta_{\mu-1} \bbeta_{\mu}
        \Phi^{\mu-2}_{\mu}(y; \mu; \alpha)\Biggr\} \qquad {\rm if\ } \mu=2, 3, \ldots,
        \lambda - 1,  \label{eq:CS1-Q} 
\end{eqnarray}
where
\begin{equation}
  \Phi^{\nu'}_{\nu}(y; \mu; \alpha) = \frac{{}_\alpha F_{\lambda-\alpha-1} \left(
  \bbeta^{(\nu')}_{\mu+1}, \ldots, \bbeta^{(\nu')}_{\mu+\alpha};
  \bbeta^{(\nu')}_1, \ldots, \bbeta^{(\nu')}_{\mu}, \bbeta^{(\nu')}_{\mu+\alpha+1},
  \ldots, \bbeta^{(\nu')}_{\lambda-1}; y\right)}{{}_\alpha F_{\lambda-\alpha-1} \left(
  \bbeta^{(\nu)}_{\mu+1}, \ldots, \bbeta^{(\nu)}_{\mu+\alpha};
  \bbeta^{(\nu)}_1, \ldots, \bbeta^{(\nu)}_{\mu}, \bbeta^{(\nu)}_{\mu+\alpha+1},
  \ldots, \bbeta^{(\nu)}_{\lambda-1}; y\right)}  \label{eq:Phi} 
\end{equation}
and
\begin{eqnarray}
  \bbeta^{(\nu)}_{\mu} & = & \bbeta_{\mu} + 1 \qquad {\rm if\ } \mu \le \nu,
        \nonumber \\
  & = & \bbeta_{\mu} \qquad {\rm if\ } \mu > \nu. \label{eq:beta-bar} 
\end{eqnarray}
\par
%
%==========================================================
%
{}For $\alpha = 0$, Eq.~(\ref{eq:Phi}) reduces to
\begin{equation}
  \Phi^{\nu'}_{\nu}(y; \mu; 0) = \frac{N^{(0)}_{\nu'}(|z|)}{N^{(0)}_{\nu}(|z|)},
\end{equation}
which is independent of~$\mu$. We then get back the results previously derived for the
CS~(\ref{eq:SGA-CS}) of the $C_{\lambda}$-extended oscillator SGA (see Eq.~(36)
of~\cite{cq00b}). For such states, it has been shown that the parameters
$\alpha_{\mu}$ have a very strong effect on~$Q$. For instance, the super-Poissonian
distribution obtained for $\mu=0$ when $\alpha_{\mu} = 0$ (or $\bbeta_{\mu} =
\mu/\lambda$), $\mu = 0$, 1, \ldots,~$\lambda-1$, is changed into a sub-Poissonian
one when the $\alpha_{\mu}$'s are varied in an appropriate way, e.g., for $\alpha_0 <
0$ corresponding to $\bbeta_1 < \lambda^{-1}$.\par
%
%--------------------------------------------------------------------------------------------------------
%
{}For $\alpha=1$, let us consider the case where $\lambda=3$, corresponding to the
CS~(\ref{eq:CS1-ex1}) and~(\ref{eq:CS1-ex2}).\par
%
%-------------------------------------------------------------------------------------------------------
% 
{}For $\mu=0$, $Q$ goes to 2 for both $y \to 0$ and $y \to \infty$, and any parameter
values. Since the confluent hypergeometric function ${}_1F_1(a+n; a; y)$, $n \in \N$,
reduces to an exponential multiplied by an $n$th-degree polynomial~\cite{erdelyi}, it is
easy to get simple analytical formulae for $\bbeta_1 = \bbeta_2 + n$ (or $\alpha_1 = -
1 - 3n$). For instance
\begin{eqnarray}
  Q & = & 2 \qquad {\rm if\ } \bbeta_1 = \bbeta_2, \nonumber \\
  & = &  2 - \frac{3y}{(y + \bbeta_2)(y + \bbeta_2 + 1)} \qquad {\rm if\ } \bbeta_1 =
          \bbeta_2 + 1.
\end{eqnarray}
In the latter case, $Q \le 2$ for any value of~$|z|$ and its minimum, $Q_{min} = 2 - 3
\left(\sqrt{\bbeta_2} + \sqrt{\bbeta_2+1}\right)^{-2}$, decreases from 2 to $-1$
when $\bbeta_2$ decreases from $+ \infty$ to~0. For other values of~$\bbeta_1$, a
numerical study confirms this trend by showing that $Q \ge 2$ or $Q \le 2$ on the whole
range of $|z|$ according to whether $\bbeta_1 < \bbeta_2$ or $\bbeta_1 > \bbeta_2$.
In particular, as shown in Fig.~3a, the strong super-Poissonian character of the
distribution obtained for vanishing $\alpha_{\mu}$ parameters (i.e., $\bbeta_1 = 1/3$,
$\bbeta_2 = 2/3$) can be changed into a sub-Poissonian one over a limited range of
$|z|$ values provided $\alpha_1$ is negative enough (or $\bbeta_2$ small enough).\par
%
%-------------------------------------------------------------------------------------------------------
%
{}For $\mu=1$, $Q$ goes to $-1$ or~2 for any parameter values according to whether
$y \to 0$ or $y \to \infty$. Simple analytical formulae can be obtained for $\bbeta_2 =
\bbeta_1 + 1 + n$ (or $\alpha_1 = 2 + 3n$), $n \in \N$. For instance
\begin{equation}
  Q = 2 - \frac{3}{3y+1} \qquad {\rm if\ } \bbeta_2 = \bbeta_1 + 1.  \label{eq:Q-ex}
\end{equation}
In such a case, $Q$ is independent of~$\bbeta_1$ and increases from $-1$ to~2. For
other values of~$\bbeta_2$, a numerical study shows that if $\bbeta_2 < \bbeta_1 +
1$, $Q$ has a maximum greater than~2, whereas if $\bbeta_2 > \bbeta_1 +1$, its
behaviour is rather similar to that of~(\ref{eq:Q-ex}). Hence, as shown in Fig.~3b, large
negative values of~$\alpha_1$ (or small values of~$\bbeta_2$) again drastically change
$Q$ as compared with the case of vanishing $\alpha_{\mu}$ parameters.\par
%
%+++++++++++++++++++++++++++++++++++++++++++++++++++++++++
%
\subsubsection{\boldmath Statistical Properties of the Coherent States $|z\rangle$}

The averages of $N$ and $N^2$ in the CS $|z\rangle$ are easily calculated from those
determined in the previous Subsubsection since such states are linear combinations of the
CS $|\omega; \mu; 0\rangle$, where $\omega = z^{\lambda}$. The result for $Q$ reads
\begin{equation}
  Q = \frac{S_2(|z|)}{S_1(|z|)} - \lambda \frac{S_1(|z|)}{{\cal N}(|z|)},
\end{equation}
where
\begin{eqnarray}
  S_1(|z|) & = & \sum_{\mu=0}^{\lambda-1} N^{(0)}_{\mu}(|\omega|) (t + \mu
       \lambda^{-1} - \bbeta_{\mu}) \frac{t^{\mu}}{\prod_{\nu=1}^{\mu}
       \bbeta_{\nu}},  \label{eq:S1} \\
  S_2(|z|) & = & \sum_{\mu=0}^{\lambda-1} N^{(0)}_{\mu}(|\omega|) \Bigl[\mu
       (\mu-1) \lambda^{-1} - (2\mu-1) \bbeta_{\mu} + \lambda \bbeta_{\mu}^2 + 
       (2\mu + 1 - \lambda \bbeta_{\mu} - \lambda \bbeta_{\mu+1}) t \nonumber \\
  && \mbox{} + \lambda t^2 \Bigr] \frac{t^{\mu}}{\prod_{\nu=1}^{\mu}
       \bbeta_{\nu}}, 
\end{eqnarray}
and $t = |z|^2/\lambda$, $|\omega| = |z|^{\lambda}$.\par
%
%------------------------------------------------------------------------------------------------------
%
{}For $\lambda=2$, $Q$ can be rewritten in terms of Bessel modified functions
$I_{\nu}$ as
\begin{equation}
  Q = (1 - 2\bbeta_1) [2t - (1 - 2\bbeta_1) R(t)]^{-1} [2t - 2(2t + \bbeta_1) R(t) -
  (1 - 2\bbeta_1) R^2(t)],
\end{equation}
where
\begin{equation}
  R(t) = \frac{I_{\bbeta_1}(2t)}{ I_{\bbeta_1-1}(2t) + I_{\bbeta_1}(2t)} \qquad
  (t = |z|^2/2).  \label{eq:R}
\end{equation}
For $\bbeta_1 = 1/2$ (or $\alpha_0 = 0$), we obtain $Q=0$, as it should be, since the
CS~$|z\rangle$ then reduce to the conventional harmonic oscillator CS. In general, we find
that $Q$ behaves as $(2\bbeta_1 - 1) |z|^2/(2\bbeta_1)$ if $|z| \ll 1$, and as
$(2\bbeta_1 - 1)/(2|z|^2)$ if $|z| \gg 1$. As shown in Fig.~4a, the distribution is
super-Poissonian or sub-Poissonian according to whether $\alpha_0 > 0$ (i.e., $\bbeta_1
> 1/2$) or $\alpha_0 < 0$ (i.e., $\bbeta_1 < 1/2$).\par
%
%------------------------------------------------------------------------------------------------------
%
{}For higher values of~$\lambda$, the behaviour of~$Q$ becomes more complicated as
it now depends on more than one parameter. For $\lambda=3$ and $|z| \ll 1$, for
instance, $Q$ behaves as $(2\bbeta_1 - \bbeta_2) |z|^2/(3\bbeta_1 \bbeta_2)$ if
$\bbeta_2 \ne 2\bbeta_1$, and as $(3\bbeta_1 - 1) |z|^4/(18 \bbeta_1^2)$ if
$\bbeta_2 = 2\bbeta_1 \ne 2/3$. Hence values of~$\bbeta_1$ smaller (resp.\
greater) than both $\bbeta_2/2$ and $1/3$ favour sub-Poissonian (resp.\
super-Poissonian) distributions. In terms of the $\alpha_{\mu}$ parameters, this means
$\alpha_0 < 0$ and $\alpha_1 > \alpha_0$ (resp.\ $\alpha_0 > 0$ and 
$\alpha_1 < \alpha_0$). However, as shown in Fig.~4b, intermediate values
of~$\bbeta_1$ may lead to an oscillating behaviour of~$Q$.\par
%
%xxxxxxxxxxxxxxxxxxxxxxxxxxxxxxxxxxxxxxxxxxxxxxxxxxxxxxxxxxxxxxxxxxxxxxxxxxxxx
% 
\subsection{Squeezing Properties}

{}For conventional photon operators, the electromagnetic field components $x_b$
and~$p_b$ are given by
\begin{equation}
  x_b = \frac{1}{\sqrt{2}} \left(b + \bp\right), \qquad p_b = \frac{1}{{\rm i}\sqrt{2}}
  \left(b - \bp\right). 
\end{equation}
Their variances $(\Delta x_b)^2 = \langle x_b^2\rangle - \langle x_b\rangle^2$ and
$(\Delta p_b)^2 = \langle p_b^2\rangle - \langle p_b\rangle^2$ in any state satisfy
the conventional uncertainty relation
\begin{equation}
  (\Delta x_b)^2 (\Delta p_b)^2 \ge \case{1}{4},
\end{equation}
the lower bound being attained by the vacuum state, for which $(\Delta x_b)^2_0 =
(\Delta p_b)^2_0 = 1/2$. The quadrature $x_b$ (resp.\ $p_b$) is said to be squeezed
in a given state if $(\Delta x_b)^2 < (\Delta x_b)^2_0$ (resp.\ $(\Delta p_b)^2 <
(\Delta p_b)^2_0$) or, in other words, if the ratio $X \equiv (\Delta x_b)^2/ (\Delta
x_b)^2_0$ (resp.\ $P \equiv (\Delta p_b)^2/ (\Delta p_b)^2_0$) is less than
one.\par
%
%----------------------------------------------------------------------------------------------------
% 
The $C_{\lambda}$-extended oscillator creation and annihilation operators $\ap$, $a$
may be regarded as describing ``dressed'' photons, which may have some applications in
phenomenological models explaining non-intuitive observable phenomena~\cite{solomon}.
It is therefore also interesting to consider deformed electromagnetic field components
$x$ and $p$, given by
\begin{equation}
  x = \frac{1}{\sqrt{2}} \left(a + \ap\right), \qquad p = \frac{1}{{\rm i}\sqrt{2}}
  \left(a - \ap\right),  \label{eq:x-p} 
\end{equation}
where 
\begin{equation}
  a = b \left(\frac{F(N_b)}{N_b}\right)^{1/2}, \qquad \ap =
  \left(\frac{F(N_b)}{N_b}\right)^{1/2} \bp.  \label{eq:a-b} 
\end{equation}
Their variances satisfy the general uncertainty relation
\begin{equation}
  (\Delta x)^2 (\Delta p)^2 \ge \case{1}{4} |\langle [x,p] \rangle|^2 = \case{1}{4}
  |\langle [a, \ap] \rangle|^2.  \label{eq:UR}
\end{equation}
A state, for which there is equality in~(\ref{eq:UR}), is said to satisfy the minimum
uncertainty property. If the vacuum state satisfies the minimum uncertainty property,
then the deformed quadrature $x$ (resp.\ $p$) is
said to be squeezed in a given state if $X \equiv (\Delta x)^2/ (\Delta x)^2_0$
(resp.\ $P \equiv (\Delta p)^2/ (\Delta p)^2_0$) is less than one.\par
%
%----------------------------------------------------------------------------------------------------
%
We shall now proceed to study both types of squeezing properties for the CS $|z; \mu;
\alpha\rangle$ and~$|z\rangle$.\par
%
%++++++++++++++++++++++++++++++++++++++++++++++++++++++++++
%   
\subsubsection{\boldmath Squeezing Properties of the Coherent States $|z; \mu;
\alpha\rangle$}

In any state belonging to ${\cal F}_{\mu}$, the variances of the deformed quadratures
satisfy the uncertainty relation
\begin{equation}
  (\Delta x)^2 (\Delta p)^2 \ge \frac{\lambda^2}{4} (\bbeta_{\mu+1} -
  \bbeta_{\mu})^2, \label{eq:CS1-UR}
\end{equation}
where the right-hand side may be less than the conventional value 1/4~\cite{cq00b}.
They can be expressed as
\begin{equation}
  (\Delta x)^2 = \langle H_0\rangle + \delta_{\lambda,2} \langle J_+ + J_-\rangle,
  \qquad (\Delta p)^2 = \langle H_0\rangle - \delta_{\lambda,2} \langle J_+ +
  J_-\rangle.  \label{eq:CS1-variance}
\end{equation}
Since in ${\cal F}_{\mu}$ the role of the vacuum state is played by the number state
$|\mu\rangle = |0; \mu; \alpha\rangle$, and $\langle H_0\rangle \ge \langle
H_0\rangle_0$, it is obvious that there cannot be any squeezing for $\lambda \ge
3$.$^1$\par
%
%------------------------------------------------------------------------------------------------------
%
{}For $\lambda=2$ and $\alpha=0$, we obtain
\begin{eqnarray}
  (\Delta x)^2 & = & \langle N\rangle + \bbeta_1 + {\rm Re}\,z, \qquad (\Delta p)^2 =
  \langle N\rangle + \bbeta_1 - {\rm Re}\,z, \label{eq:CS1-var} \\
  (\Delta x)^2_0 & = & (\Delta p)^2_0 = \bbeta_1 + \mu = \bbeta_{\mu+1} +
  \bbeta_{\mu},  \label{eq:CS1-var0}
\end{eqnarray}
where $\langle N\rangle$ is given by
\begin{eqnarray}
  \langle N \rangle & = & \frac{2y}{\bbeta_1} \Phi^1_0(y;0;0) \qquad {\rm if\ } \mu=0,
         \nonumber \\
  & = & 1 + 2 \bbeta_1 \left[\Phi^0_1(y;1;0) - 1\right] \qquad {\rm if\ } \mu=1.
\end{eqnarray}
Comparing (\ref{eq:CS1-UR}) and~(\ref{eq:CS1-var0}), we see that $|\mu\rangle$
satisfies the minimum uncertainty property only for $\mu=0$, to which we shall restrict
ourselves.\par
%
%------------------------------------------------------------------------------------------------------
%
{}From (\ref{eq:CS1-var}) and~(\ref{eq:CS1-var0}), it is obvious that $X$ and $P$ are
related to each other by the transformation ${\rm Re}\, z \to - {\rm Re}\, z$ and that
the maximum squeezing in~$x$ is achieved for real, negative values of~$z$. So we only
consider $X$ for such values. As shown in Fig.~5a, a large squeezing effect over the
whole range of real, negative values of~$z$ is obtained for $\alpha_0 > 0$ (i.e.,
$\bbeta_1 > 1/2$).\par
%
%------------------------------------------------------------------------------------------------------
%
{}For the conventional quadratures, the counterpart of~(\ref{eq:CS1-variance}) is
\begin{eqnarray}
  (\Delta x_b)^2 = \langle N\rangle + \frac{1}{2} + \delta_{\lambda,2}\, {\rm Re}\, z
  \left\langle \left(\frac{(N+1)(N+2)}{F(N+1) F(N+2)}\right)^{1/2}\right\rangle, \\
  (\Delta p_b)^2 = \langle N\rangle + \frac{1}{2} - \delta_{\lambda,2}\, {\rm Re}\, z
  \left\langle \left(\frac{(N+1)(N+2)}{F(N+1) F(N+2)}\right)^{1/2}\right\rangle,
\end{eqnarray}
where use has been made of~(\ref{eq:a-b}). There is no squeezing for $\lambda \ge 3$
again.\par
%
%-------------------------------------------------------------------------------------------------------
%
{}For $\lambda=2$ and $\mu=\alpha=0$, we obtain from (\ref{eq:F}),
(\ref{eq:alg-action}), (\ref{eq:CS1-exp}), and~(\ref{eq:CS1-norm}),
\begin{equation}
  \left\langle \left(\frac{(N+1)(N+2)}{F(N+1) F(N+2)}\right)^{1/2}\right\rangle = \left[
  {}_0F_1(\bbeta_1; y)\right]^{-1} \sum_{k=0}^{\infty} \left(\frac{2k+1}{2k +
  2\bbeta_1}\right)^{1/2} \frac{y^k}{k! (\bbeta_1)_k},
\end{equation}
where $y = |z|^2/4$. The previous remarks about $X$ and $P$ remain valid. For real,
negative values of $z$, the behaviour of $X$ in terms of $- {\rm Re}\, z$ is shown in
Fig.~5b. Comparing with Fig.~5a, we observe more or less similar squeezing properies to
those corresponding to dressed photons.\par
%
%++++++++++++++++++++++++++++++++++++++++++++++++++++++++++
%
\subsubsection{\boldmath Squeezing Properties of the Coherent States $|z\rangle$}

{}From (\ref{eq:Ham}) and (\ref{eq:x-p}), it follows that in any state belonging to~$\cal
F$, the variances of the deformed quadratures can be expressed as
\begin{eqnarray}
  (\Delta x)^2 & = & \langle H_0\rangle - \langle a\rangle \langle \ap\rangle +
        \case{1}{2} \left[(\Delta a)^2 + (\Delta \ap)^2\right], \\
  (\Delta p)^2 & = & \langle H_0\rangle - \langle a\rangle \langle \ap\rangle -
        \case{1}{2} \left[(\Delta a)^2 + (\Delta \ap)^2\right].
\end{eqnarray}
\par
%
%-------------------------------------------------------------------------------------------------------
% 
{}For the CS $|z\rangle$, $\langle a \rangle = z$, $\langle \ap \rangle = z^*$, and
$(\Delta a)^2 = (\Delta \ap)^2 = 0$, so that $(\Delta x)^2 = (\Delta p)^2 = \langle
H_0\rangle - |z|^2$. Furthermore, since $\left[a, \ap\right] = \left\{a, \ap\right\} - 2
\ap a$, we also find $\frac{1}{2} \left\langle \left[a, \ap\right]\right\rangle = \langle
H_0\rangle - |z|^2$. We therefore conclude that the CS~$|z\rangle$ satisfy the
minimum uncertainty property and that in such states
\begin{equation}
  (\Delta x)^2 = (\Delta p)^2 = \frac{\lambda}{2}\, [{\cal N}(|z|)]^{-1} \sum_{\mu=0}
  ^{\lambda-1} (\bbeta_{\mu+1} - \bbeta_{\mu}) N^{(0)}_{\mu}(|\omega|) 
  \frac{t^{\mu}}{\prod_{\nu=1}^{\mu} \bbeta_{\nu}},  \label{eq:CS2-variance}
\end{equation}
where $t = |z|^2/\lambda$, $|\omega| = |z|^{\lambda}$. For the vacuum
state~$|0\rangle$, Eq.~(\ref{eq:CS2-variance}) reduces to
\begin{equation}
  (\Delta x)^2_0 = (\Delta p)^2_0 = \frac{\lambda}{2} \bbeta_1, 
\end{equation}
which is less than the conventional value 1/2 for $\bbeta_1 < \lambda^{-1}$ (i.e.,
$\alpha_0 < 0$). The variances of the deformed quadratures $x$ and $p$ are
simultaneously lower than their common vacuum value if $X = P < 1$.\par
%
%---------------------------------------------------------------------------------------------------------
%
{}For $\lambda=2$ corresponding to the paraboson CS~\cite{sharma}, $X = P$ can be
rewritten in terms of Bessel modified functions $I_{\nu}$ as 
\begin{equation}
  X = P = 1 + \frac{1- 2\bbeta_1}{\bbeta_1} R(t),
\end{equation}
where $R(t)$ is defined in~(\ref{eq:R}). They behave as $1 + (1- 2\bbeta_1) |z|^2/(2
\bbeta_1^2)$ or $(2\bbeta_1)^{-1}$ according to whether $|z| \ll 1$ or $|z|\gg 1$. In
these limits, we therefore get $X = P < 1$ if $\bbeta_1 > 1/2$ (i.e., $\alpha_0 > 0$). In
Fig.~6a, it is shown that this result is actually valid over the whole range of $|z|$
values.\par
%
%--------------------------------------------------------------------------------------------------------
%
{}For $\lambda \ge 3$, $X$ and $P$ behave as $1 + (\bbeta_2 - 2\bbeta_1)
|z|^2/(\lambda \bbeta_1^2)$ or $(\lambda \bbeta_1)^{-1}$ according to whether $|z| \ll
1$ or $|z|\gg 1$. Values of $\bbeta_1$ greater than both $1/\lambda$ and
$\bbeta_2/2$ (i.e., $\alpha_0 > 0$ and $\alpha_1 < \alpha_0$) therefore favour small
values of $X=P$. This is confirmed numerically, as shown in Fig.~6b for $\lambda=3$. It
should be noted however that there is a tendency for the behaviour of $X=P$ to become
more complicated than in the $\lambda=2$ case.\par
%
%----------------------------------------------------------------------------------------------------
%
{}For the conventional quadratures, the equality of the variances obtained
in~(\ref{eq:CS2-variance}) does not remain true. We obtain instead
\begin{eqnarray}
  (\Delta x_b)^2 & = & \langle N\rangle + \frac{1}{2} - |z|^2 \left\langle
        \left(\frac{(N+1)(N+2)}{F(N+1) F(N+2)}\right)^{1/2}\right\rangle \nonumber \\
  && \mbox{} + 2 ({\rm Re}\, z)^2 \left[\left\langle \left(\frac{(N+1)(N+2)}{F(N+1)
        F(N+2)}\right)^{1/2}\right\rangle - \left\langle
        \left(\frac{N+1}{F(N+1)}\right)^{1/2}\right\rangle^2\right],  
        \label{eq:CS2-variance-bis}
\end{eqnarray}
and $(\Delta p_b)^2$ given by a similar formula with ${\rm Re}\, z$ replaced by ${\rm
Im}\, z$. Here
\begin{equation}
  \langle N\rangle = \lambda [{\cal N}(|z|)]^{-1} S_1(|z|),
\end{equation}
\begin{eqnarray}
  \left\langle\left(\frac{N+1}{F(N+1)}\right)^{1/2}\right\rangle & = & [{\cal
        N}(|z|)]^{-1} \sum_{\mu=0}^{\lambda-1} \frac{t^{\mu}}{\prod_{\nu=1}^{\mu}
        \bbeta_{\nu}} \sum_{k=0}^{\infty} \left(\frac{k\lambda + \mu + 1}{k\lambda +
        \lambda \bbeta_{\mu+1}}\right)^{1/2} \nonumber \\
  && \mbox{} \times \frac{t^{k\lambda}}{k! \left(\prod_{\nu=1}^{\mu} (\bbeta_{\nu}
        + 1)_k\right) \left(\prod_{\nu=\mu+1}^{\lambda-1} (\bbeta_{\nu})_k\right)}, 
\end{eqnarray}
\begin{eqnarray}
  \left\langle\left(\frac{(N+1)(N+2)}{F(N+1)(F(N+2)}\right)^{1/2}\right\rangle & = &
        [{\cal N}(|z|)]^{-1} \sum_{\mu=0}^{\lambda-1}
        \frac{t^{\mu}}{\prod_{\nu=1}^{\mu} \bbeta_{\nu}} \nonumber \\
  && \mbox{} \times \sum_{k=0}^{\infty}
        \left(\frac{(k\lambda + \mu + 1)(k\lambda + \mu + 2)}{(k\lambda +
        \lambda \bbeta_{\mu+1})(k\lambda + \lambda \bbeta_{\mu+2})}\right)^{1/2}
        \nonumber \\
  && \mbox{} \times \frac{t^{k\lambda}}{k! \left(\prod_{\nu=1}^{\mu} (\bbeta_{\nu}
        + 1)_k\right) \left(\prod_{\nu=\mu+1}^{\lambda-1} (\bbeta_{\nu})_k\right)}, 
\end{eqnarray}
where $S_1(|z|)$ is given in~(\ref{eq:S1}) and $t = |z|^2/\lambda$. Since $X$ and $P$
are related to each other by the transformation ${\rm Re}\, z \leftrightarrow {\rm Im}\,
z$, it is enough to study the former.\par
%
%---------------------------------------------------------------------------------------------------
%
{}For small $|z|$ values, the coefficient of $2({\rm Re}\, z)^2$ on the right-hand side
of~(\ref{eq:CS2-variance-bis}) behaves as $\left(\sqrt{2\bbeta_1} -
1\right)/(2\bbeta_1)$ if $\lambda=2$, and as $\left(\sqrt{2\bbeta_1} -
\sqrt{\bbeta_2}\right)\Big/\left(\lambda \bbeta_1 \sqrt{\bbeta_2}\right)$ if
$\lambda\ge 3$. Hence, for such values, the maximum squeezing in~$x_b$ is achieved
when $z$ is real (resp.\ imaginary) if $\bbeta_1 < 1/2$ (resp.\ $\bbeta_1 > 1/2$) in
the former case and if $\bbeta_1 < \bbeta_2/2$ (resp.\ $\bbeta_1 > \bbeta_2/2$) in
the latter.\par
%
%---------------------------------------------------------------------------------------------------------
%
A numerical study does confirm this trend. In Fig.~7, $X$ is plotted against ${\rm Re}\,
z$ or ${\rm Im}\, z$ for $\lambda=2$ and some values of $\bbeta_1$ smaller or
greater than 1/2 (i.e., $\alpha_0 < 0$ or $\alpha_0 > 0$), respectively. In Fig.~8, the
same is done for $\lambda=3$ and some values of $(\bbeta_1, \bbeta_2)$ such that
$\bbeta_1 < \bbeta_2/2$ or $\bbeta_1 > \bbeta_2/2$ (i.e., $\alpha_0 < \alpha_1$
or $\alpha_0 > \alpha_1$). We conclude that a substantial squeezing effect can be
obtained for well-chosen parameters.\par
%
%============================================================
%  
\section{CONCLUSION}

In the present paper, we introduced two new types of CS associated with the
$C_{\lambda}$-extended oscillator. The first ones, which actually form a family of~CS
whose members are labelled by some index~$\alpha$, include as special cases both the
Barut-Girardello and the Perelomov su(1,1) CS for $\lambda=2$, as well as the
annihilation-operator CS of the $C_{\lambda}$-extended oscillator SGA for higher
$\lambda$ values. The second ones, which are eigenstates of the
$C_{\lambda}$-extended oscillator annihilation operator, extend to higher $\lambda$
values the paraboson~CS, to which they reduce for $\lambda=2$.\par
%
%---------------------------------------------------------------------------------------------------------
% 
We showed that all these states satisfy a unity resolution relation in the
$C_{\lambda}$-extended oscillator Fock space (or in some subspace thereof). It should
be stressed that it is this property that makes them qualify as generalized~CS and
distinguishes them from the plethora of putative~CS found in the literature. To arrive at
this result for some of the CS considered here, it has been essential to restrict the range
of the $C_{\lambda}$-extended oscillator parameters in such a way that the limit
$\alpha_{\mu} \to 0$, corresponding to the standard harmonic oscillator, is excluded.
The $C_{\lambda}$-extended oscillator parameters therefore play a regularizing role,
similar to that of the parameter $\alpha_0$ of the Calogero-Vasiliev algebra for the
Perelomov su(1,1)~CS. It is indeed worth reminding that for the latter, the unity
resolution relation invalid for the standard harmonic oscillator (i.e., for $\alpha_0 = 0$ or
$\bbeta_1 = 1/2$) becomes true for sufficiently large values of $\alpha_0$ (i.e., for
$\alpha_0 > 1$ or $\bbeta_1 > 1$).\par
%
%----------------------------------------------------------------------------------------------------------
% 
As a by-product of our study, we obtained Bargmann representations of the
$C_{\lambda}$-extended oscillator Fock space. This allowed us to solve, albeit in the
complex plane, the yet open problem of finding realizations of the
$C_{\lambda}$-extended oscillator algebra in terms of differential operators.\par
%
%-----------------------------------------------------------------------------------------------------
%
{}Furthermore, we investigated some characteristics of our CS relevant to quantum
optics, such as their statistical and squeezing properties, for a wide range of parameters
and from both viewpoints of real and dressed photons. The nonclassical features of these
states for some parameter ranges were clearly demonstrated.\par
%
%------------------------------------------------------------------------------------------------------
%
The remarkable properties of the new CS defined in this paper provide a useful tool for
theoretical investigation of model systems. Although their generation by nonlinear
systems remains to be studied, we believe that these states may play an important role
in quantum optics.\par
%
%============================================================
% 
\section*{APPENDIX A: PROOF OF EQS. (\ref{eq:g}) AND (\ref{eq:pos-cond})}

\renewcommand{\theequation}{A.\arabic{equation}}
\setcounter{equation}{0}

In this Appendix, we prove that Eq.~(\ref{eq:g}) provides us with a positive solation
of~(\ref{eq:mellin}) for any $r > 0$ and any $0 \le \alpha \le \lambda - \mu - 1$, if
condition~(\ref{eq:pos-cond}) is fulfilled.\par
%
%------------------------------------------------------------------------------------------------------
%
{}From (\ref{eq:g-zero}), it follows that Eq.~(\ref{eq:g}) is valid for $\alpha=0$ without
any restriction. We will now show that if it is valid for $\alpha - 1$ under a
condition similar to~(\ref{eq:pos-cond}), then the same is true for $\alpha$. For such a
purpose, let us factorize $g^*(s)$, defined in~(\ref{eq:mellin}), as
\begin{equation}
  g^*(s) = g^*_1(s) g^*_2(s), \qquad g^*_1(s) = \frac{\prod_{\nu=1}^{r +\alpha-1}
  \Gamma(s + b_{\nu})}{\prod_{\nu=1}^{\alpha-1} \Gamma(s + a_{\nu})}, \qquad
  g^*_2(s) = \frac{\Gamma(s + b_{r+\alpha})}{\Gamma(s +a_{\alpha})}.
  \label{eq:fact-A}
\end{equation}
From the induction hypothesis and tables of inverse Mellin transforms~\cite{prudnikov},
we get
\begin{equation}
  g_1(y) = G^{r + \alpha-1, 0}_{\alpha-1, r + \alpha-1} \left(y\, \Bigg| 
      \begin{array}{l}
         a_1, a_2, \ldots, a_{\alpha-1}\\
         b_1, b_2, \ldots, b_{r + \alpha-1}
      \end{array}\right)  \label{eq:g1}
\end{equation}
and
\begin{equation}
  g_2(y) = \left\{\begin{array}{ll}
       \frac{y^{b_{r+\alpha}} (1-y)^{a_{\alpha} - b_{r+\alpha} - 1}}{\Gamma(a_{\alpha}
            - b_{r+\alpha})} & {\rm if\ }0 < y < 1 \\[0.2cm]
       0 & {\rm if\ }1 < y < \infty 
  \end{array}\right.,  \label{eq:g2}
\end{equation}
respectively. Equation~(\ref{eq:g1}) is valid if the set $\{b_1, b_2, \ldots,
b_{r+\alpha-1}\}$ contains some subset of $\alpha-1$ elements $\{b_{i_1}, b_{i_2},
\ldots, b_{i_{\alpha-1}}\}$ such that
\begin{equation}
  a_1 > b_{i_1}, \qquad a_2 > b_{i_2}, \qquad \ldots, \qquad a_{\alpha-1} >
  b_{i_{\alpha-1}},  \label{eq:g1-cond}
\end{equation}
while Eq.~(\ref{eq:g2}) requires
\begin{equation}
  a_{\alpha} > b_{r+\alpha}.  \label{eq:g2-cond}
\end{equation}
\par
%
%-----------------------------------------------------------------------------------------------------
%
The second form of $g(y)$ in~(\ref{eq:convolution}) leads to the expression
\begin{equation}
  g(y) = \frac{y^{b_{r+\alpha}}}{\Gamma(a_{\alpha} - b_{r+\alpha})}\int_y^{\infty} 
  t^{-a_{\alpha}} (t-y)^{a_{\alpha} - b_{r+\alpha} - 1} 
  G^{r + \alpha-1, 0}_{\alpha-1, r + \alpha-1} \left(t\, \Bigg| 
      \begin{array}{l}
         a_1, a_2, \ldots, a_{\alpha-1}\\
         b_1, b_2, \ldots, b_{r + \alpha-1}
      \end{array}\right) dt,  \label{eq:int-A}
\end{equation}
which is a positive function if conditions~(\ref{eq:g1-cond}) and~(\ref{eq:g2-cond}) are
satisfied. The change of variable $u = t/y$ transforms (\ref{eq:int-A}) into
\begin{equation}
  g(y) = [\Gamma(a_{\alpha} - b_{r+\alpha})]^{-1} \int_1^{\infty} 
  u^{-a_{\alpha}} (u-1)^{a_{\alpha} - b_{r+\alpha} - 1} 
  G^{r + \alpha-1, 0}_{\alpha-1, r + \alpha-1} \left(yu\, \Bigg| 
      \begin{array}{l}
         a_1, a_2, \ldots, a_{\alpha-1}\\
         b_1, b_2, \ldots, b_{r + \alpha-1}
      \end{array}\right) du,  
\end{equation}
which is nothing else than the right-hand side of~(\ref{eq:g}) when use is made of
Eq.~(7.811.3) of~\cite{gradshteyn}.\par
%
%-------------------------------------------------------------------------------------------------------
%
To extend conditions~(\ref{eq:g1-cond}) and~(\ref{eq:g2-cond}) to the more general
ones given in~(\ref{eq:pos-cond}), it is enough to note that $g^*(s)$, and hence $g(y)$,
are symmetric under any permutation of the $a_{\nu}$'s or the $b_{\nu}$'s. In other
words, considering other factorizations than that assumed in~(\ref{eq:fact-A}) would lead
to the general condition~(\ref{eq:pos-cond}), which completes the proof.\par
%
%===========================================================
%
\section*{APPENDIX B: PROOF OF EQS. (\ref{eq:halpha0}), (\ref{eq:halpha0-inter}), AND
(\ref{eq:halpha0-cond})}

\renewcommand{\theequation}{B.\arabic{equation}}
\setcounter{equation}{0}

In this Appendix, we prove that Eqs.~(\ref{eq:halpha0}) and (\ref{eq:halpha0-inter})
provide us with a positive solution of~(\ref{eq:moment}) for any even $\lambda$,
$\alpha = \lambda/2 \ge 2$, and $\mu=0$, if condition~(\ref{eq:halpha0-cond}) is
fulfilled. For such a purpose, it is enough to consider the inversion problem
of~(\ref{eq:mellin}) for $r=0$, any $\alpha \ge 2$, and $\mu=0$.\par
%
%---------------------------------------------------------------------------------------------------------
%
Let us start with the $\alpha=2$ case and factorize $g^*(s)$ as
\begin{equation}
  g^*(s) = g^*_1(s) g^*_2(s), \qquad g^*_i(s) = \frac{\Gamma(s + b_i)}{\Gamma(s +
  a_i)} \qquad i=1, 2.  \label{eq:fact-B}
\end{equation}
The inverse Mellin transforms of the factors are given by~\cite{prudnikov}
\begin{equation}
  g_i(y) = \left\{\begin{array}{ll}
       \frac{y^{b_i} (1-y)^{a_i - b_i - 1}}{\Gamma(a_i - b_i)} & {\rm if\ }0 < y < 1
           \\[0.2cm]
       0 & {\rm if\ }1 < y < \infty 
  \end{array}\right.,  
\end{equation}
provided
\begin{equation}
  a_i > b_i \qquad i=1, 2.  \label{eq:g1-g2-cond}
\end{equation}
\par
%
%-------------------------------------------------------------------------------------------------
%
The second form of $g(y)$ in~(\ref{eq:convolution}) leads to the expression
\begin{equation}
  g(y) = \frac{y^{b_2}}{\Gamma(a_1-b_1) \Gamma(a_2-b_2)} \int_y^1 t^{b_1-a_2}
  (1-t)^{a_1-b_1-1} (t-y)^{a_2-b_2-1} dt,  \label{eq:int-B}
\end{equation}
if $0 < y < 1$ and to zero if $1 < y < \infty$. The right-hand side of~(\ref{eq:int-B}) is
clearly a positive function on (0, 1) if condition~(\ref{eq:g1-g2-cond}) is fulfilled.
Substituting $u = (1-t)/(1-y)$ into~(\ref{eq:int-B}) transforms it into
\begin{equation}
  g(y) = \frac{y^{b_2} (1-y)^{a_1+a_2-b_1-b_2-1}}{\Gamma(a_1-b_1)
  \Gamma(a_2-b_2)} \int_0^1 u^{a_1-b_1-1} (1-u)^{a_2-b_2-1} [1 -
  (1-y)u]^{b_1-a_2} du, 
\end{equation}
which is nothing else than
\begin{equation}
  g(y) = \frac{y^{b_2} (1-y)^{a_1+a_2-b_1-b_2-1}}{\Gamma(a_1+a_2-b_1-b_2)}
  {}_2F_1(a_1-b_1, a_2-b_1; a_1+a_2-b_1-b_2; 1-y),  \label{eq:gB-1} 
\end{equation}
when use is made of Eq.~(3.197.3) of~\cite{gradshteyn}.\par
%
%------------------------------------------------------------------------------------------------------
%
Had we considered the other possible factorization of $g^*(s)$, obtained
from~(\ref{eq:fact-B}) by permuting $b_1$ and $b_2$ while leaving $a_1$, $a_2$
unchanged, we should have obtained
\begin{equation}
  g(y) = \frac{y^{b_1} (1-y)^{a_1+a_2-b_1-b_2-1}}{\Gamma(a_1+a_2-b_1-b_2)}
  {}_2F_1(a_1-b_2, a_2-b_2; a_1+a_2-b_1-b_2; 1-y),  \label{eq:gB-2} 
\end{equation}
if
\begin{equation}
  a_1 > b_2, \qquad a_2 > b_1.  \label{eq:g1-g2-cond-bis}
\end{equation}
However the right-hand side of~(\ref{eq:gB-2}) can be rewritten in the
form~(\ref{eq:gB-1}) by using Eq.~(9.131.1) of~\cite{gradshteyn}. We therefore
conclude that Eq.~(\ref{eq:gB-1}) applies in both cases~(\ref{eq:g1-g2-cond})
and~(\ref{eq:g1-g2-cond-bis}), thereby leading to~(\ref{eq:h20}).\par
%
%------------------------------------------------------------------------------------------------------
%
{}For $\alpha > 2$, the generalization of~(\ref{eq:gB-1}) reads
\begin{eqnarray}
  g(y) & = & \frac{y^{b_{\alpha}}}{\Gamma(a_1-b_1) \left[\prod_{p=1}^{\alpha-2}
         \Gamma(a_{p+1}-b_p)\right]} \nonumber \\
  && \mbox{} \times  \sum_{n_1 n_2 \ldots n_{\alpha-2}} \frac{\prod_{p=1}^{\alpha-2}
         \left[\Gamma(a_{p+1}-b_p+n_p) \Gamma(\xi_{\alpha,p}(\bn))\right]}
         {\left[\prod_{p=1}^{\alpha-3} \Gamma\left(\eta_{\alpha,p}(\bn)\right)\right]
         \Gamma\left(\zeta_{\alpha}(\bn)\right) \left(\prod_{p=1}^{\alpha-2} n_p!
         \right)} \nonumber \\
  && \mbox{} \times (1-y)^{\zeta_{\alpha}(\bn) - 1} {}_2F_1(a_{\alpha} - b_{\alpha-1},
         \eta_{\alpha, \alpha-2}(\bn); \zeta_{\alpha}(\bn); 1-y),  \label{eq:gB-3}
\end{eqnarray}
where
\begin{eqnarray}
  \xi_{\alpha,p}(\bn) & = & \sum_{q=1}^p (a_q - b_q + n_q), \nonumber \\
  \eta_{\alpha,p}(\bn) & = & \sum_{q=1}^{p+1} (a_q - b_q) + \sum_{q=1}^p n_q,
         \nonumber \\
  \zeta_{\alpha}(\bn) & = & \sum_{p=1}^{\alpha} (a_p - b_p) + \sum_{p=1}^{\alpha-2}
         n_p. \label{eq:gB-inter}
\end{eqnarray}
The positivity condition on $g(y)$ is now that there exists some permutation $(i_1, i_2,
\ldots, i_{\alpha})$ of $(1, 2, \ldots, \alpha)$ such that
\begin{equation}
  a_1 > b_{i_1}, \qquad a_2 > b_{i_2}, \qquad \ldots, \qquad a_{\alpha} >
b_{i_{\alpha}}.
  \label{eq:gB-cond}
\end{equation}
The proof of (\ref{eq:gB-3}), (\ref{eq:gB-inter}), and~(\ref{eq:gB-cond}) is by induction
over~$\alpha$, using the factorization~(\ref{eq:fact-A}) of $g^*(s)$ for $r=0$ and the
convolution property~(\ref{eq:convolution}). The intermediate steps are the same as in
the $\alpha=2$ case, once the hypergeometric function ${}_2F_1(a_{\alpha-1} -
b_{\alpha-2}, \eta_{\alpha-1, \alpha-3}(\bn); \zeta_{\alpha-1}(\bn); (1-y)u)$ has been
expanded into powers of $(1-y)u$. Finally, Eqs.~(\ref{eq:halpha0}),
(\ref{eq:halpha0-inter}), and~(\ref{eq:halpha0-cond}) are directly obtained
from~(\ref{eq:gB-3}), (\ref{eq:gB-inter}), and~(\ref{eq:gB-cond}) after permuting the
$b_{\nu}$'s in a convenient way and substituting (\ref{eq:a}) and~(\ref{eq:b}).\par
%
%===========================================================
%
\section*{APPENDIX C: HERMITICITY PROPERTIES OF SOME OPERATOR REALIZATIONS IN
BARGMANN SPACE}

\renewcommand{\theequation}{C.\arabic{equation}}
\setcounter{equation}{0}

In this Appendix, we consider the Hermiticity properties of some operators or some
operator-valued matrices with respect to the scalar product~(\ref{eq:B1-sp}) of ${\cal
B}^{(\alpha)}_{\mu}$ or (\ref{eq:B2-sp}) of ${\cal B}^{(0)}$, respectively.\par
%
%----------------------------------------------------------------------------------------------------------
%
Let us start with the operators ${\cal J}^{(\alpha)}_{+\mu}$, ${\cal
J}^{(\alpha)}_{-\mu}$, and ${\cal J}^{(\alpha)}_{0\mu}$, defined in
(\ref{eq:B1-J+})--(\ref{eq:B1-J0}), and prove that they satisfy relations similar
to~(\ref{eq:SGA-Herm}) with respect to~(\ref{eq:B1-sp}). Considering the first relation
in~(\ref{eq:SGA-Herm}), we have to show that for any two functions
$\psi^{(\alpha)}_{\mu}(z)$,
$\psi^{\prime(\alpha)}_{\mu}(z) \in {\cal B}^{(\alpha)}_{\mu}$,
\begin{equation}
  \int d^2z \, h^{(\alpha)}_{\mu}(y) \left[{\cal J}^{(\alpha)}_{+\mu}
  \psi^{\prime(\alpha)}_{\mu}(z)\right]^* \psi^{(\alpha)}_{\mu}(z) = \int d^2z \,
  h^{(\alpha)}_{\mu}(y) \left[\psi^{\prime(\alpha)}_{\mu}(z)\right]^*
  {\cal J}^{(\alpha)}_{-\mu} \psi^{(\alpha)}_{\mu}(z).  \label{eq:C-result1}   
\end{equation}
\par
%
%------------------------------------------------------------------------------------------------------
%
By substituting~(\ref{eq:B1-J-}) into the right-hand side of~(\ref{eq:C-result1}),
integrating by parts, and re-ordering some operators, we obtain
\begin{eqnarray}
  \lefteqn{\int d^2z \, h^{(\alpha)}_{\mu}(y)
        \left[\psi^{\prime(\alpha)}_{\mu}(z)\right]^*
        {\cal J}^{(\alpha)}_{-\mu} \psi^{(\alpha)}_{\mu}(z)} \nonumber \\
  & = & - \lambda^{\lambda-\alpha-1} \int d^2z \left\{\frac{\partial}{\partial z} \left[
        \prod_{\nu=1}^{\mu} \left(- z \frac{\partial}{\partial z} + \bbeta_{\nu}\right)
        \right] \left[\prod_{\nu=\mu+\alpha+1}^{\lambda-1} \left(- z
        \frac{\partial}{\partial z} + \bbeta_{\nu} - 1\right)\right]
        h^{(\alpha)}_{\mu}(y)\right\} \nonumber \\
  && \mbox{} \times \left[\psi^{\prime(\alpha)}_{\mu}(z)\right]^*
        \psi^{(\alpha)}_{\mu}(z).  \label{eq:C-int1}   
\end{eqnarray}
Since
\begin{equation}
  \frac{\partial}{\partial z} h^{(\alpha)}_{\mu}(y) = \lambda^{2\alpha-\lambda} z 
  \frac{d}{dy} h^{(\alpha)}_{\mu}(y), \qquad z \frac{\partial}{\partial z}
  h^{(\alpha)}_{\mu}(y) = y \frac{d}{dy} h^{(\alpha)}_{\mu}(y),  \label{eq:C-inter}   
\end{equation}
Eq.~(\ref{eq:C-int1}) becomes
\begin{eqnarray}
  \lefteqn{\int d^2z \, h^{(\alpha)}_{\mu}(y)
        \left[\psi^{\prime(\alpha)}_{\mu}(z)\right]^*
        {\cal J}^{(\alpha)}_{-\mu} \psi^{(\alpha)}_{\mu}(z)} \nonumber \\
  & = & (-1)^{\lambda-\alpha} \lambda^{\alpha-1} \int d^2z
        \left\{z \frac{d}{dy} \left[\prod_{\nu=1}^{\mu} \left(y \frac{d}{dy} -
        \bbeta_{\nu}\right)\right] \left[\prod_{\nu=\mu+\alpha+1}^{\lambda-1} \left(y
        \frac{d}{dy} - \bbeta_{\nu} + 1\right)\right] h^{(\alpha)}_{\mu}(y)\right\}
        \nonumber \\
  && \mbox{} \times \left[\psi^{\prime(\alpha)}_{\mu}(z)\right]^*
        \psi^{(\alpha)}_{\mu}(z).  
\end{eqnarray}
\par
%
%------------------------------------------------------------------------------------------------------
%
By proceeding similarly, the left-hand side of~(\ref{eq:C-result1}) is transformed into
\begin{eqnarray}
  \lefteqn{\int d^2z \, h^{(\alpha)}_{\mu}(y)
        \left[{\cal J}^{(\alpha)}_{+\mu} \psi^{\prime(\alpha)}_{\mu}(z)\right]^*
         \psi^{(\alpha)}_{\mu}(z)} \nonumber \\
  & = & (-1)^{\alpha} \lambda^{\alpha-1} \int d^2z
        \left\{z \left[\prod_{\nu=\mu+1}^{\mu+\alpha} \left(y \frac{d}{dy} -
        \bbeta_{\nu} + 2\right)\right] h^{(\alpha)}_{\mu}(y)\right\}
        \left[\psi^{\prime(\alpha)}_{\mu}(z)\right]^* \psi^{(\alpha)}_{\mu}(z).  
\end{eqnarray}
\par
%
%------------------------------------------------------------------------------------------------------
%
Equation~(\ref{eq:C-result1}) is therefore equivalent to the following differential equation
for $h^{(\alpha)}_{\mu}(y)$,
\begin{eqnarray}
  &&\left\{(-1)^{\alpha} \left[\prod_{\nu=\mu+1}^{\mu+\alpha} \left(y 
        \frac{d}{dy} - \bbeta_{\nu} + 2\right)\right] - (-1)^{\lambda-\alpha} \frac{d}{dy} 
        \left[\prod_{\nu=1}^{\mu} \left(y \frac{d}{dy} - \bbeta_{\nu}\right)\right]
        \right.\nonumber \\
  && \mbox{} \times  \left.\left[\prod_{\nu=\mu+\alpha+1}^{\lambda-1} \left(y 
        \frac{d}{dy} - \bbeta_{\nu} + 1\right)\right]\right\}h^{(\alpha)}_{\mu}(y) = 0. 
\label{eq:C-diff1}
\end{eqnarray}
After multiplication by $y$, Eq.~(\ref{eq:C-diff1}) is nothing else than the differential
equation satisfied by the Meijer $G$-function given in (\ref{eq:halphamu})~\cite{erdelyi}.
This completes the proof of~(\ref{eq:C-result1}).\par
%
%----------------------------------------------------------------------------------------------------
%
The proof of the second relation in~(\ref{eq:SGA-Herm}) is easily carried out by using
only integrations by parts and relations such as~(\ref{eq:C-inter}).\par
%
%-----------------------------------------------------------------------------------------------------
%
Let us now turn ourselves to the operator-valued matrices ${\cal A}^{\dagger(0)}$ and
${\cal A}^{(0)}$, defined in~(\ref{eq:B2-aplus}) and~(\ref{eq:B2-a}), respectively, and
show that they are Hermitian conjugate of one another with respect to~(\ref{eq:B2-sp}),
i.e.,
\begin{equation}
  \sum_{\mu, \mu'=0}^{\lambda-1}\int d^2z \, h^{(0)}_{\mu}(y) \left[{\cal
  A}^{\dagger(0)}_{\mu\mu'} \psi^{\prime(0)}_{\mu'}(z)\right]^* \psi^{(0)}_{\mu}(z)
  = \sum_{\mu, \mu'=0}^{\lambda-1} \int d^2z \, h^{(0)}_{\mu}(y)
  \left[\psi^{\prime(0)}_{\mu}(z)\right]^* {\cal A}^{(0)}_{\mu\mu'}
  \psi^{(0)}_{\mu'}(z).  \label{eq:C-result2}   
\end{equation}
\par
%
%------------------------------------------------------------------------------------------------------
% 
Substituting the explicit expressions of ${\cal A}^{\dagger(0)}_{\mu\mu'}$ into the
left-hand side of~(\ref{eq:C-result2}) directly leads to
\begin{eqnarray}
  \lefteqn{\sum_{\mu, \mu'=0}^{\lambda-1}\int d^2z \, h^{(0)}_{\mu}(y) \left[{\cal
        A}^{\dagger(0)}_{\mu\mu'} \psi^{\prime(0)}_{\mu'}(z)\right]^*
        \psi^{(0)}_{\mu}(z)}\nonumber \\
  & = & \left[\lambda^{\lambda-1} \left(\prod_{\nu=1}^{\lambda-1} \bbeta_{\nu}\right)
        \right]^{-1/2} \int d^2z \, h^{(0)}_0(y) \left[z \psi^{\prime(0)}_{\lambda-1}(z)
        \right]^* \psi^{(0)}_0(z) \nonumber \\
  && \mbox{} +\sum_{\mu=0}^{\lambda-2} \sqrt{\lambda \bbeta_{\mu+1}} \int d^2z
        \, h^{(0)}_{\mu+1}(y) \left[\psi^{\prime(0)}_{\mu}(z)\right]^*
        \psi^{(0)}_{\mu+1}(z).   
\end{eqnarray}
\par
%
%-----------------------------------------------------------------------------------------------------
%
After substituting the explicit expressions of ${\cal A}^{(0)}_{\mu\mu'}$ into the
right-hand side of~(\ref{eq:C-result2}), integrating by parts, re-ordering some operators,
and using~(\ref{eq:C-inter}) for $\alpha=0$, we get
\begin{eqnarray}
  \lefteqn{\sum_{\mu, \mu'=0}^{\lambda-1} \int d^2z \, h^{(0)}_{\mu}(y)
       \left[\psi^{\prime(0)}_{\mu}(z)\right]^* {\cal A}^{(0)}_{\mu\mu'}
       \psi^{(0)}_{\mu'}(z)} \nonumber \\
  & = & - \left[\lambda^{1-\lambda} \left(\prod_{\nu=1}^{\lambda-1} 
       \bbeta_{\nu}\right) \right]^{1/2} \int d^2z \, \left[\frac{d}{dy}
       h^{(0)}_{\lambda-1}(y)\right] \left[z \psi^{\prime(0)}_{\lambda-1}(z)\right]^*
       \psi^{(0)}_0(z) \nonumber \\
  && \mbox{} - \sum_{\mu=0}^{\lambda-2} \sqrt{\frac{\lambda} {\bbeta_{\mu+1}}}
       \int d^2z \, \left[\left(y \frac{d}{dy} - \bbeta_{\mu+1} + 1\right)
       h^{(0)}_{\mu}(y)\right] \left[\psi^{\prime(0)}_{\mu}(z)\right]^*
       \psi^{(0)}_{\mu+1}(z).  
\end{eqnarray}
\par
%
%---------------------------------------------------------------------------------------------------------
% 
Equation~(\ref{eq:C-result2}) is therefore equivalent to the following system of
$\lambda$ differential equations for $h^{(0)}_{\mu}(y)$, $\mu=0$, 1,
\ldots,~$\lambda-1$,
\begin{eqnarray}
  \left(y \frac{d}{dy} - \bbeta_{\mu+1} + 1\right) h^{(0)}_{\mu}(y) & = & -
        \bbeta_{\mu+1} h^{(0)}_{\mu+1}(y), \qquad \mu=0, 1, \ldots, \lambda-2, \\
  \frac{d}{dy} h^{(0)}_{\lambda-1}(y) & = & - \left(\prod_{\nu=1}^{\lambda-1} 
       \bbeta_{\nu}\right)^{-1} h^{(0)}_0(y). 
\end{eqnarray}
By taking (\ref{eq:Aalphamu}) and~(\ref{eq:halphamu}) into account, this leads to the
system
\begin{eqnarray}
  && \left(y \frac{d}{dy} - \bbeta_{\mu+1} + 1\right) G^{\lambda 0}_{0 \lambda} (y | 0,
        \bbeta_1, \ldots, \bbeta_{\mu}, \bbeta_{\mu+1} - 1, \ldots, \bbeta_{\lambda-1}
        - 1) \nonumber \\
  && = - G^{\lambda 0}_{0 \lambda} (y | 0, \bbeta_1, \ldots, \bbeta_{\mu+1},
        \bbeta_{\mu+2} - 1, \ldots, \bbeta_{\lambda-1} - 1), \qquad \mu=0, 1, \ldots,
        \lambda-2,  \label{eq:C-diff2}
\end{eqnarray}
\begin{equation}
  \frac{d}{dy} G^{\lambda 0}_{0 \lambda} (y | 0, \bbeta_1, \ldots, \bbeta_{\lambda-1})
  = - G^{\lambda 0}_{0 \lambda} (y | 0, \bbeta_1 - 1, \ldots, \bbeta_{\lambda-1} - 1),  
  \label{eq:C-diff3}
\end{equation}
to be satisfied by Meijer $G$-functions.\par
%
%-----------------------------------------------------------------------------------------------------
%
To prove (\ref{eq:C-diff2}), we successively use Eq.~(5.3.1.9) of~\cite{erdelyi}, the
symmetry  of the $G$-function in its parameters, Eq.~(5.3.1.13) of the same, and the
symmetry again to rewrite its left-hand side as
\begin{eqnarray}
  && \left(y \frac{d}{dy} - \bbeta_{\mu+1} + 1\right) G^{\lambda 0}_{0 \lambda} (y | 0,
        \bbeta_1, \ldots, \bbeta_{\mu}, \bbeta_{\mu+1} - 1, \ldots, \bbeta_{\lambda-1}
        - 1) \nonumber \\
  && = - \left(u \frac{d}{du} + \bbeta_{\mu+1} - 1\right) G^{0 \lambda}_{\lambda 0} (u
        | 1, 1 - \bbeta_1, \ldots, 1 - \bbeta_{\mu}, 2 - \bbeta_{\mu+1}, \ldots,
        2 - \bbeta_{\lambda-1}) \nonumber \\
  && = - \left(u \frac{d}{du} + \bbeta_{\mu+1} - 1\right) G^{0 \lambda}_{\lambda 0} (u
        | 2 - \bbeta_{\mu+1}, 1, 1 - \bbeta_1, \ldots, 1 - \bbeta_{\mu}, 2 -
        \bbeta_{\mu+2}, \ldots, 2 - \bbeta_{\lambda-1}) \nonumber \\
  && = - G^{0 \lambda}_{\lambda 0} (u | 1 - \bbeta_{\mu+1}, 1, 1 - \bbeta_1, \ldots, 1 -
        \bbeta_{\mu}, 2 - \bbeta_{\mu+2}, \ldots, 2 - \bbeta_{\lambda-1}) \nonumber \\
  && = - G^{0 \lambda}_{\lambda 0} (u | 1, 1 - \bbeta_1, \ldots, 1 -
        \bbeta_{\mu+1}, 2 - \bbeta_{\mu+2}, \ldots, 2 - \bbeta_{\lambda-1}),
\end{eqnarray}
where $u = 1/y$. Applying Eq.~(5.3.1.9) of~\cite{erdelyi} then leads to the right-hand
side of~(\ref{eq:C-diff2}).\par
%
%------------------------------------------------------------------------------------------------------
%
Similarly, the left-hand side of~(\ref{eq:C-diff3}) can be transformed into
\begin{eqnarray}
  && \frac{d}{dy} G^{\lambda 0}_{0 \lambda} (y | 0, \bbeta_1, \ldots,
          \bbeta_{\lambda-1}) = - \frac{1}{y} u \frac{d}{du} G^{0 \lambda}_{\lambda 0} (u
          | 1, 1 - \bbeta_1, \ldots, 1 - \bbeta_{\lambda-1}) \nonumber \\ 
  &&  = - \frac{1}{y} G^{0 \lambda}_{\lambda 0} (u | 0, 1 - \bbeta_1, \ldots, 1 -
          \bbeta_{\lambda-1}) = - \frac{1}{y} G^{\lambda 0}_{0 \lambda} (y | 1, \bbeta_1
          , \ldots, \bbeta_{\lambda-1}),  
\end{eqnarray}
which gives the right-hand side of~(\ref{eq:C-diff3}) when using Eq.~(5.3.1.8)
of~\cite{erdelyi}.\par
%
%===========================================================
% 
\section*{ACKNOWLEDGMENTS}

The author is a Research Director of the National Fund for Scientific Research (FNRS),
Belgium.\par
%
%=============================================================
% 

%
%============================================================
%
\newpage
\section*{FOOTNOTES}

$^1$ This is only valid for the second-order squeezing. In~\cite{cq00b}, we actually
showed that there is a small fourth-order squeezing effect for $\lambda=4$, $\mu=0$,
$\alpha=0$.\par
%
%===========================================================
%
\newpage
\section*{FIGURE CAPTIONS}

\hspace{\parindent} 
{\bf FIG.\ 1.} The weight function $h^{(1)}_0(y)$ in terms of $y$ for
$\lambda=3$. The parameter values are $\bbeta_1 = 4/3$, $\bbeta_2 = 2/3$, i.e.,
$\alpha_0 = - \alpha_1 = 3$ (solid line), or $\bbeta_1 = \bbeta_2 = 4/3$, i.e., $\alpha_0
= 3$, $\alpha_1 = -1$ (dashed line).

\bigskip
{\bf FIG.\ 2.} The weight function $h^{(2)}_0(y)$ in terms of $y$ for $\lambda=4$
and $\bbeta_1 = 3/2$, i.e., $\alpha_0 = 5$. The other parameter values are: (a)
$\bbeta_2 = 3/2$, $\bbeta_3 = 5/4$, i.e., $\alpha_1 = -1$, $\alpha_2 = -2$ (solid line),
$\bbeta_2 = 7/4$, $\bbeta_3 = 5/4$, i.e., $\alpha_1 = 0$, $\alpha_2 = -3$ (dashed
line), or $\bbeta_2 = 2$, $\bbeta_3 = 5/4$, i.e., $\alpha_1 = 1$, $\alpha_2 = -4$
(dot-dashed line); (b) $\bbeta_2 = 1$, $\bbeta_3 = 3/4$, i.e., $\alpha_1 = -3$,
$\alpha_2 = -2$ (solid line), $\bbeta_2 = 5/4$, $\bbeta_3 = 3/4$, i.e., $\alpha_1 =
-2$, $\alpha_2 = -3$ (dashed line), or $\bbeta_2 = 3/2$, $\bbeta_3 = 3/4$, i.e.,
$\alpha_1 = -1$, $\alpha_2 = -4$ (dot-dashed line).

\bigskip
{\bf FIG.\ 3.} Mandel's parameter $Q$ as a function of $|z| \equiv r$ for the CS $|z; 1;
\mu\rangle$, $\lambda=3$, and various parameters: (a) $\mu=0$ and $\bbeta_1 =
1/3$, $\bbeta_2 = 2/3$, i.e., $\alpha_0 = \alpha_1 = 0$ (solid line), $\bbeta_1 =
1$, $\bbeta_2 = 1/10$, i.e., $\alpha_0 = 2$, $\alpha_1 = -37/10$ (dashed line),
$\bbeta_1 = 1$, $\bbeta_2 = 1/100$, i.e., $\alpha_0 = 2$, $\alpha_1 = -397/100$
(dotted line), or $\bbeta_1 = 1/10$, $\bbeta_2 = 2/3$, i.e., $\alpha_0 = - \alpha_1 =
-7/10$ (dot-dashed line); (b) $\mu=1$, $\bbeta_1 = 1/3$, i.e., $\alpha_0 = 0$, and
$\bbeta_2 = 2/3$, i.e., $\alpha_1 = 0$ (solid line), $\bbeta_2 = 1/10$, i.e., $\alpha_1
= -17/10$ (dashed line), $\bbeta_2 = 1/50$, i.e., $\alpha_1 = -97/50$ (dotted line),
or $\bbeta_2 = 2$, i.e., $\alpha_1 = 4$ (dot-dashed line). 

\bigskip
{\bf FIG.\ 4.} Mandel's parameter $Q$ as a function of $|z| \equiv r$ for the CS
$|z\rangle$ and various parameters: (a) $\lambda=2$ and $\bbeta_1 = 1/90$, i.e.,
$\alpha_0 = -44/45$ (solid line), $\bbeta_1 = 1/4$, i.e., $\alpha_0 = -1/2$ (dashed
line), $\bbeta_1 = 1$, i.e., $\alpha_0 = 1$ (dotted line), or $\bbeta_1 = 10$, i.e.,
$\alpha_0 =19$ (dot-dashed line); (b) $\lambda=3$ and $\bbeta_1 = 1/100$,
$\bbeta_2 = 2/3$, i.e., $\alpha_0 = -\alpha_1 = -97/100$ (solid line), $\bbeta_1 =
1/3$, $\bbeta_2 = 10$, i.e., $\alpha_0 = 0$, $\alpha_1 = 28$ (dashed line), $\bbeta_1
= 2/3$, $\bbeta_2 = 1/100$, i.e., $\alpha_0 = 1$, $\alpha_1 = -297/100$ (dotted
line), or $\bbeta_1 = 10$, $\bbeta_2 = 1/3$, i.e., $\alpha_0 = 29$, $\alpha_1 = -30$
(dot-dashed line). 

\bigskip
{\bf FIG.\ 5.} $X$ as a function of $- {\rm Re}\, z$ for the CS $|z; 0; 0\rangle$,
$\lambda=2$, ${\rm Im}\, z = 0$, and (a) the case of dressed photons, or (b) the case
of real photons. The parameter values are $\bbeta_1 = 1/2$, i.e., $\alpha_0 = 0$ (solid
lines), $\bbeta_1 = 3/10$, i.e., $\alpha_0 = -2/5$ (dashed lines), $\bbeta_1 = 1$, i.e.,
$\alpha_0 = 1$ (dotted lines), or $\bbeta_1 = 2$, i.e., $\alpha_0 = 3$ (dot-dashed
lines).

\bigskip
{\bf FIG.\ 6.} The ratio $X = (\Delta x)^2/(\Delta x)^2_0$ as a function of $|z| \equiv r$
for the CS $|z\rangle$ and various parameters: (a) $\lambda=2$ and $\bbeta_1 = 10$,
i.e., $\alpha_0 = 19$ (solid line), $\bbeta_1 = 2$, i.e., $\alpha_0 = 3$ (dashed line),
$\bbeta_1 = 3/4$, i.e., $\alpha_0 = 1/2$ (dotted line), or $\bbeta_1 = 3/10$, i.e.,
$\alpha_0 = -2/5$ (dot-dashed line); (b) $\lambda=3$ and $\bbeta_1 = 2$, $\bbeta_2
= 1/20$, i.e., $\alpha_0 = 5$, $\alpha_1 = -137/20$ (solid line), $\bbeta_1 = \bbeta_2
= 2/3$, i.e., $\alpha_0 = -\alpha_1 = 1$ (dashed line), $\bbeta_1 = 2$, $\bbeta_2 =
5$, i.e., $\alpha_0 = 5$, $\alpha_1 = 8$ (dotted line), or $\bbeta_1 = 1/4$, $\bbeta_2
= 1/8$, i.e., $\alpha_0 = -1/4$, $\alpha_1 = -11/8$ (dot-dashed line).    

\bigskip
{\bf FIG.\ 7.} The ratio $X = (\Delta x_b)^2/(\Delta x_b)^2_0$ for $\lambda=2$ as a
function of (a) ${\rm Re}\, z$ for ${\rm Im}\, z = 0$ and $\bbeta_1 = 1/4$, i.e.,
$\alpha_0 = -1/2$ (solid line), $\bbeta_1 = 1/10$, i.e., $\alpha_0 = -4/5$ (dashed line),
$\bbeta_1 = 1/40$, i.e., $\alpha_0 = -19/20$ (dotted line), or $\bbeta_1 = 1/100$,
i.e., $\alpha_0 = -49/50$ (dot-dashed line); (b) ${\rm Im}\, z$ for ${\rm Re}\, z = 0$
and $\bbeta_1 = 1$, i.e., $\alpha_0 = 1$ (solid line), $\bbeta_1 = 4$, i.e., $\alpha_0 =
7$ (dashed line), $\bbeta_1 = 10$, i.e., $\alpha_0 = 19$ (dotted line), or
$\bbeta_1 = 40$, i.e., $\alpha_0 = 79$ (dot-dashed line).   

\bigskip
{\bf FIG.\ 8.} The ratio $X = (\Delta x_b)^2/(\Delta x_b)^2_0$ for $\lambda=3$ as a
function of (a) ${\rm Re}\, z$ for ${\rm Im}\, z = 0$ and $\bbeta_1 = 1/10$,
$\bbeta_2 = 2/5$, i.e., $\alpha_0 = -7/10$, $\alpha_1 = -1/10$ (solid line),
$\bbeta_1 = 1/10$, $\bbeta_2 = 1$, i.e., $\alpha_0 = -7/10$, $\alpha_1 = 17/10$
(dashed line), $\bbeta_1 = 1$, $\bbeta_2 = 15$, i.e., $\alpha_0 = 2$, $\alpha_1 = 41$
(dotted line), or $\bbeta_1 = 5$, $\bbeta_2 = 50$, i.e., $\alpha_0 = 14$,
$\alpha_1 = 134$ (dot-dashed line); (b) ${\rm Im}\, z$ for ${\rm Re}\, z = 0$ and
$\bbeta_1 = \bbeta_2 = 1/10$, i.e., $\alpha_0 = -7/10$, $\alpha_1 = -1$ (solid
line), $\bbeta_1 = 1$, $\bbeta_2 = 1/10$, i.e., $\alpha_0 =2$, $\alpha_1 = -37/10$
(dashed line), $\bbeta_1 = 5$, $\bbeta_2 = 1/10$, i.e., $\alpha_0 = 14$, $\alpha_1 =
-157/10$ (dotted line), or $\bbeta_1 = 5$, $\bbeta_2 = 1$, i.e., $\alpha_0 = 14$,
$\alpha_1 = -13$ (dot-dashed line).     
%
%=============================================================
%
\newpage
\begin{picture}(160,100)
\put(35,0){\mbox{\scalebox{1.0}{\includegraphics{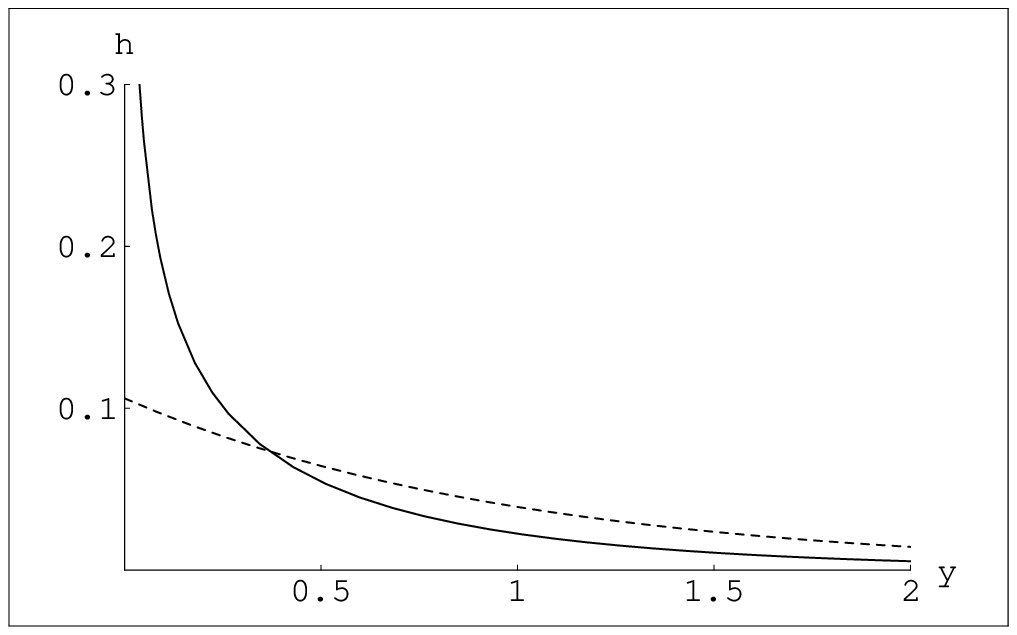}}}}
\end{picture}
\vspace{5cm}
\centerline{Figure 1}
%%
%%----------------------------------------------------------------------
%% 
\newpage
\begin{picture}(160,100)
\put(35,0){\mbox{\scalebox{1.0}{\includegraphics{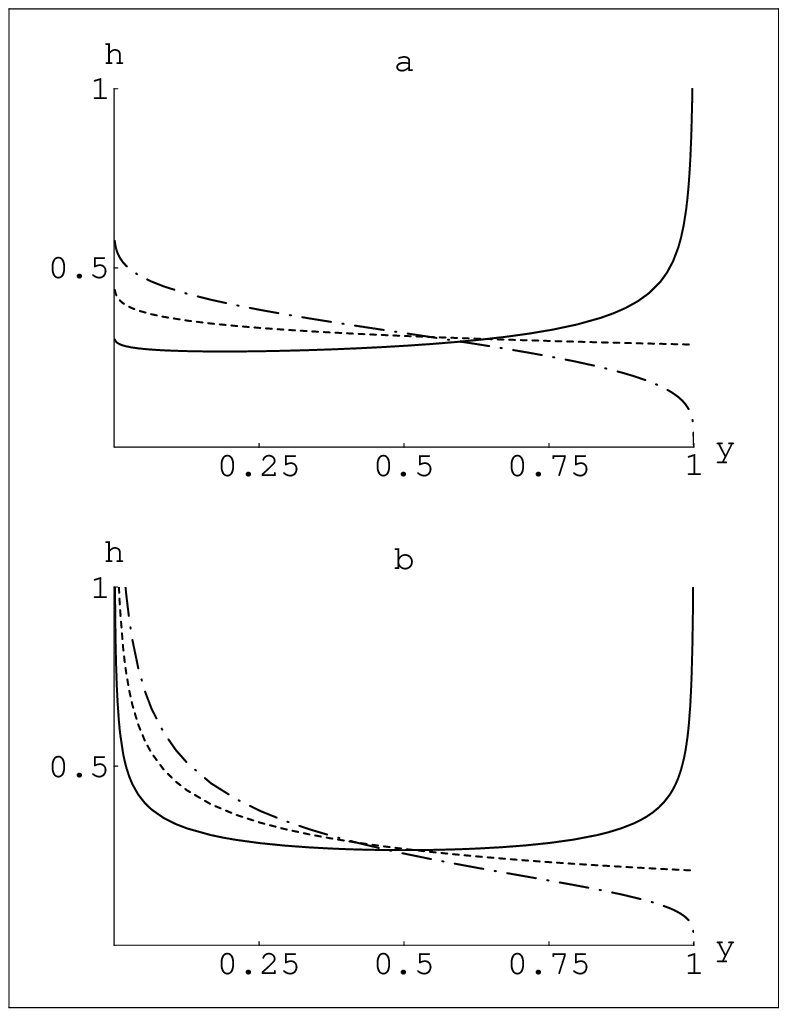}}}}
\end{picture}
\vspace{5cm}
\centerline{Figure 2}
%%
%%----------------------------------------------------------------------
%%
\newpage
\begin{picture}(160,100)
\put(35,0){\mbox{\scalebox{1.0}{\includegraphics{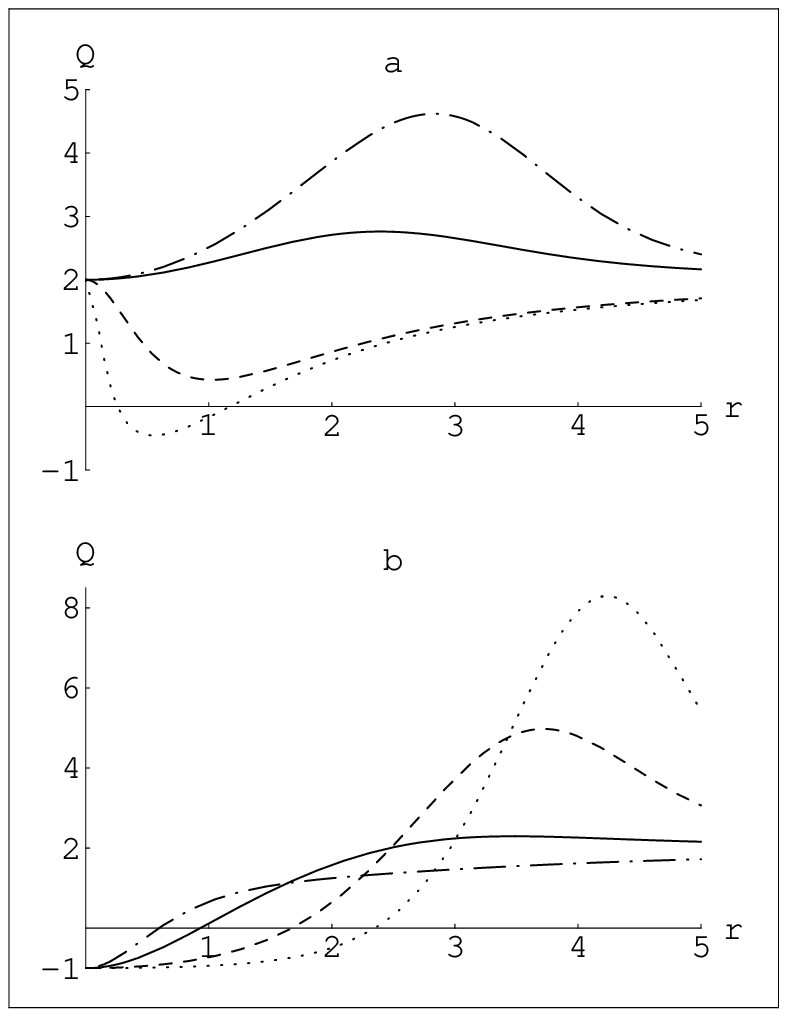}}}}
\end{picture}
\vspace{5cm}
\centerline{Figure 3}
%%
%%----------------------------------------------------------------------
%% 
\newpage
\begin{picture}(160,100)
\put(35,0){\mbox{\scalebox{1.0}{\includegraphics{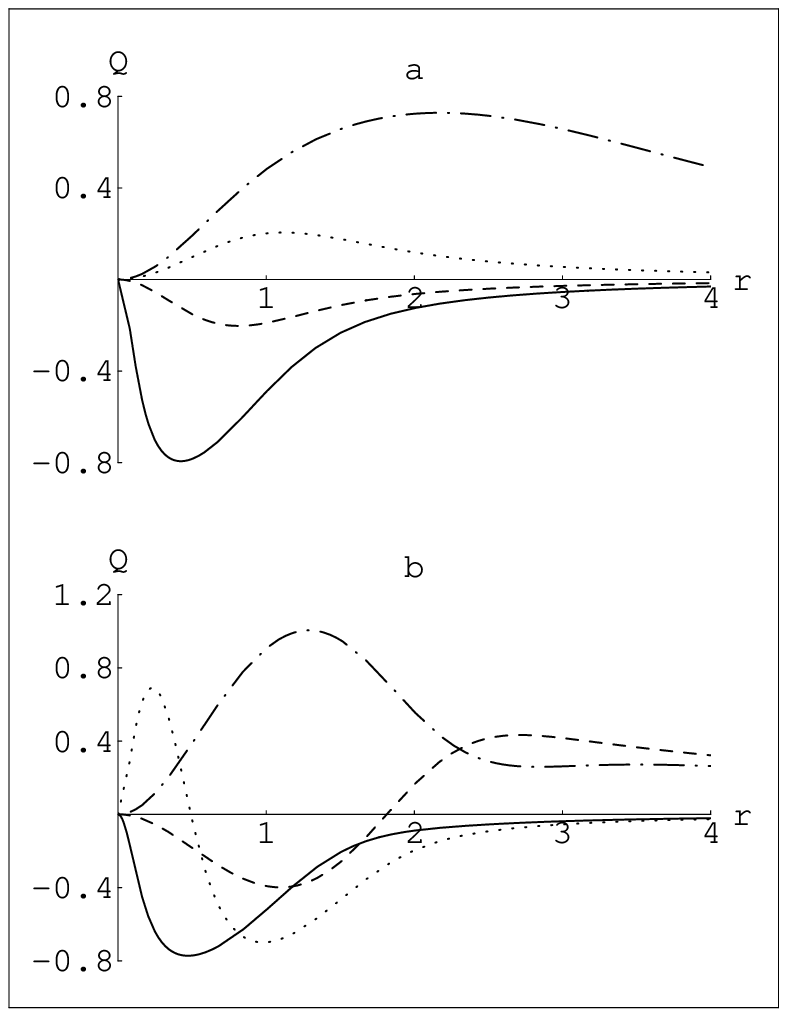}}}}
\end{picture}
\vspace{5cm}
\centerline{Figure 4}
%%
%%----------------------------------------------------------------------
%%
\newpage
\begin{picture}(160,100)
\put(35,0){\mbox{\scalebox{1.0}{\includegraphics{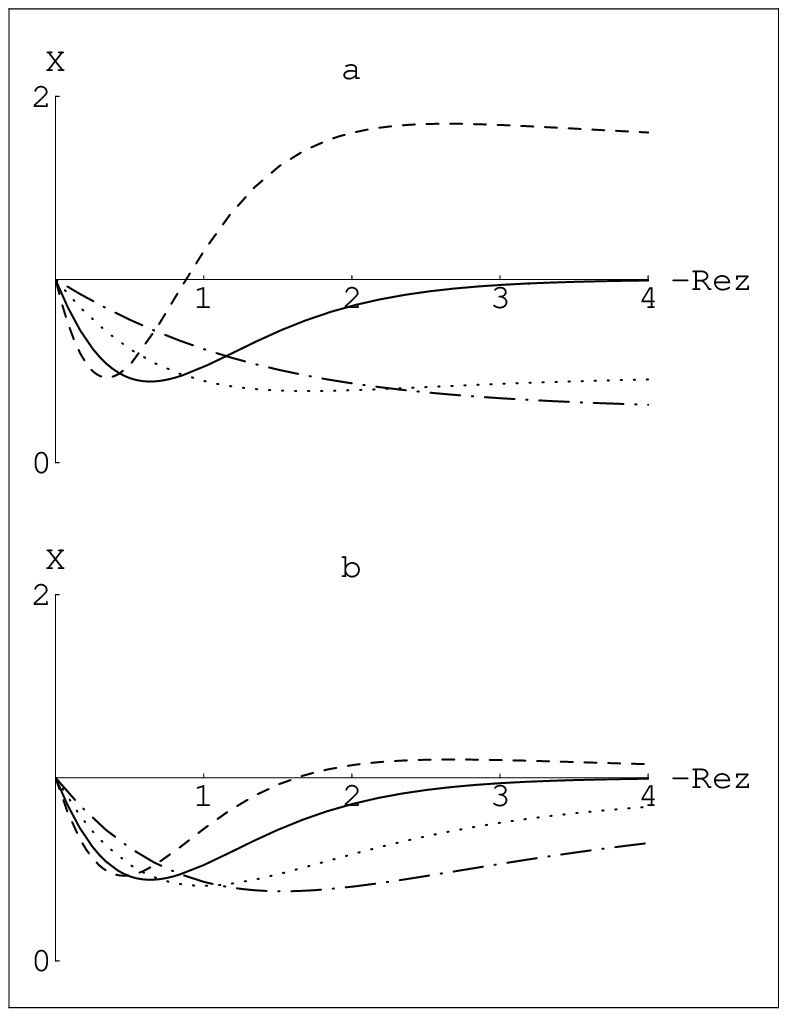}}}}
\end{picture}
\vspace{5cm}
\centerline{Figure 5}
%%
%%----------------------------------------------------------------------
%%
\newpage
\begin{picture}(160,100)
\put(35,0){\mbox{\scalebox{1.0}{\includegraphics{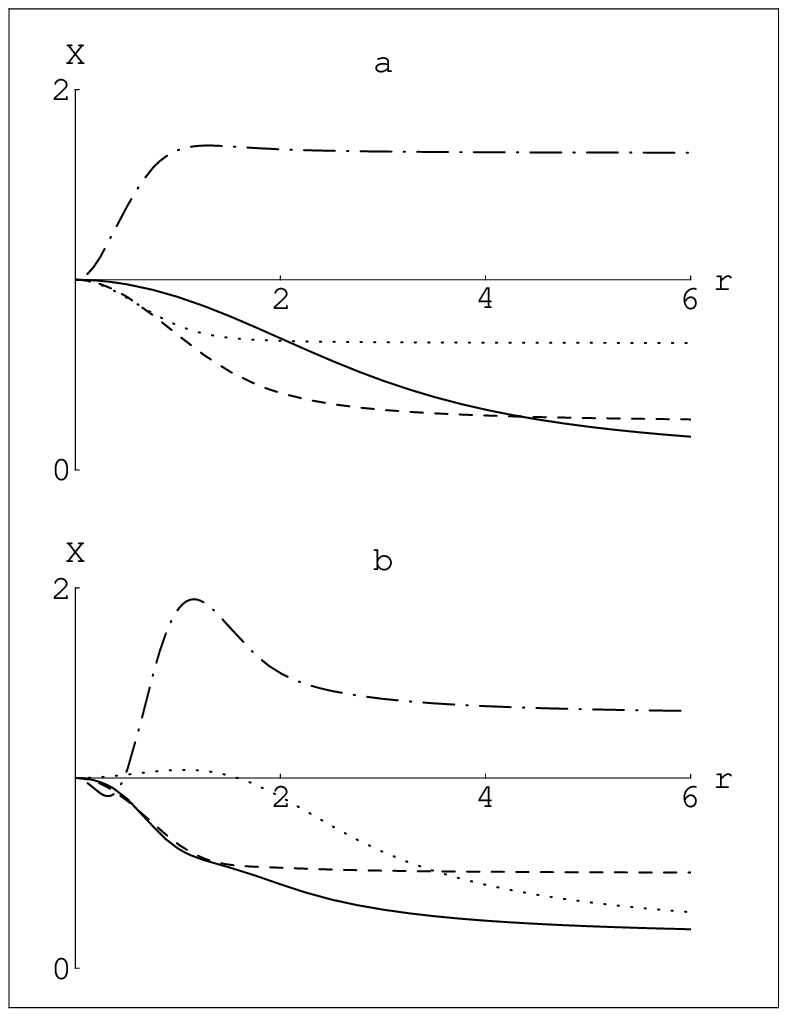}}}}
\end{picture}
\vspace{5cm}
\centerline{Figure 6}
%%
%%----------------------------------------------------------------------
%%
\newpage
\begin{picture}(160,100)
\put(35,0){\mbox{\scalebox{1.0}{\includegraphics{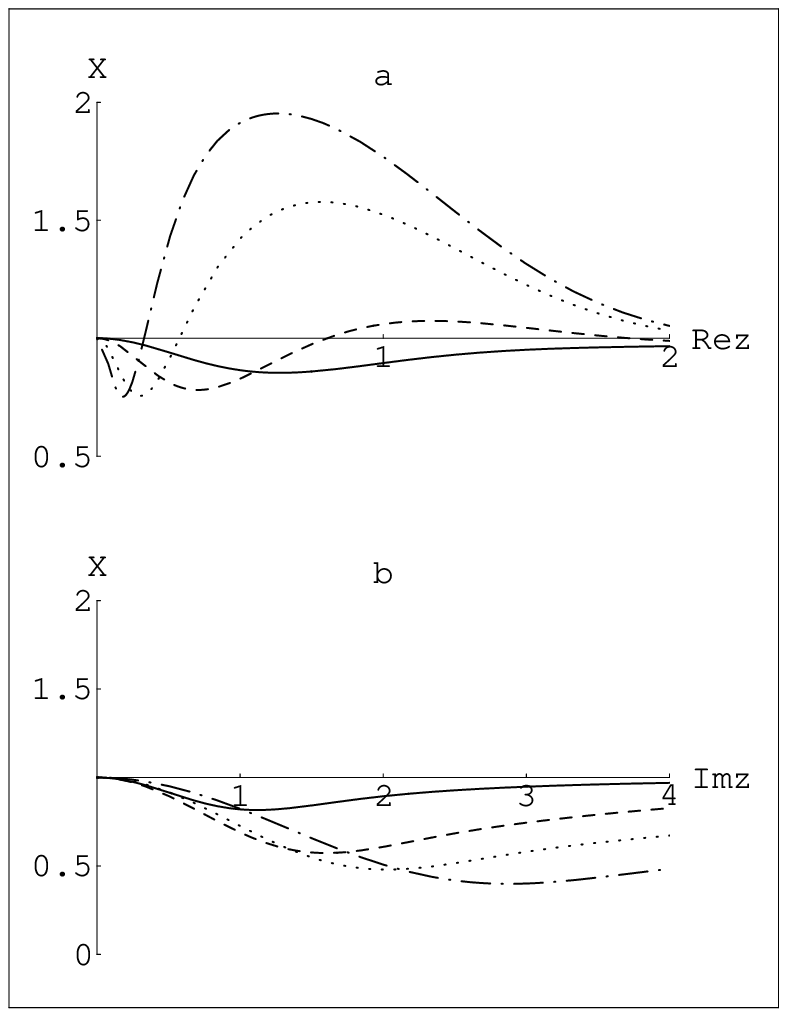}}}}
\end{picture}
\vspace{5cm}
\centerline{Figure 7}
%%
%%----------------------------------------------------------------------
%%
\newpage
\begin{picture}(160,100)
\put(35,0){\mbox{\scalebox{1.0}{\includegraphics{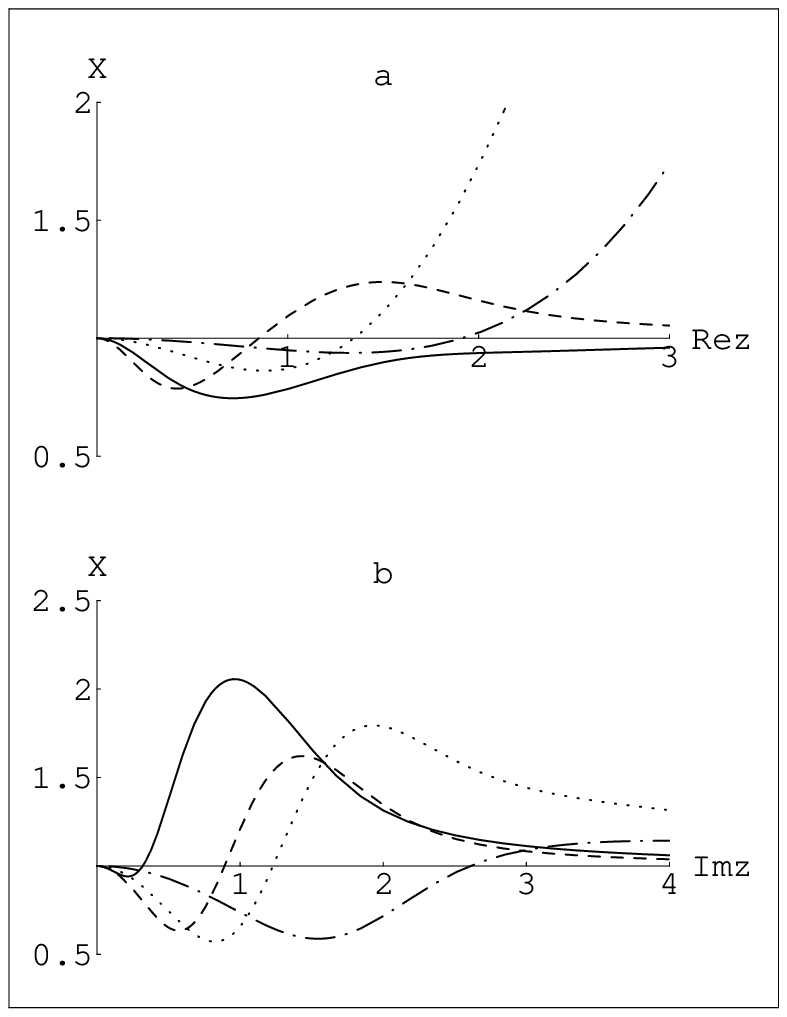}}}}
\end{picture}
\vspace{5cm}
\centerline{Figure 8}
%%
%%----------------------------------------------------------------------
%%  
%\newpage 
%\vspace*{14cm}
%\centerline{Figure 1}
%
%------------------------------------------------------------------------------------------------------------
%
%\newpage 
%\vspace*{\fill}
%\centerline{Figure 2}
%
%-----------------------------------------------------------------------------------------------------------
%
%\newpage 
%\vspace*{\fill}
%\centerline{Figure 3}
%
%------------------------------------------------------------------------------------------------------------
%
%\newpage 
%\vspace*{\fill}
%\centerline{Figure 4} 
%
%------------------------------------------------------------------------------------------------------------
%
%\newpage 
%\vspace*{\fill}
%\centerline{Figure 5}
%
%-----------------------------------------------------------------------------------------------------------
%
%\newpage 
%\vspace*{\fill}
%\centerline{Figure 6}
%
%------------------------------------------------------------------------------------------------------------
%
%\newpage 
%\vspace*{\fill}
%\centerline{Figure 7}
%
%------------------------------------------------------------------------------------------------------------
%
%\newpage 
%\vspace*{\fill}
%\centerline{Figure 8}

\end{document}